\documentclass[preprint,showpacs,titlepage,aps,prd,tightenlines,amsmath,byrevtex,nofootinbib]{revtex4}

\usepackage{graphicx}

\newcommand{\mn}{\Delta m_{21}^2}
\newcommand{\mt}{\Delta m_{31}^2}
\newcommand{\si}{s_{12}}
\newcommand{\sn}{s_{23}}
\newcommand{\st}{s_{13}}
\newcommand{\ci}{c_{12}}
\newcommand{\cn}{c_{23}}

\def\lsim{\raise0.3ex\hbox{$\;<$\kern-0.75em\raise-1.1ex
\hbox{$\sim\;$}}}
\def\gsim{\raise0.3ex\hbox{$\;>$\kern-0.75em\raise-1.1ex
\hbox{$\sim\;$}}}

\begin{document}


\title{Perturbation Theory of Neutrino Oscillation with 
Nonstandard Neutrino Interactions}

\author{Takashi Kikuchi} 
\email{t-kiku@phys.metro-u.ac.jp}

\author{Hisakazu Minakata}
\email{minakata@phys.metro-u.ac.jp}

\author{Shoichi Uchinami}
\email{uchinami@phys.metro-u.ac.jp}

\affiliation{Department of Physics, Tokyo Metropolitan University \\
1-1 Minami-Osawa, Hachioji, Tokyo 192-0397, Japan}

\date{February 5, 2009}

\vglue 1.4cm

\begin{abstract}

We discuss various physics aspects of neutrino oscillation with 
non-standard interactions (NSI). 
We formulate a perturbative framework by taking 
$\Delta m^2_{21} / \Delta m^2_{31} $, 
$s_{13}$, and the NSI elements $\varepsilon_{\alpha \beta}$ 
($\alpha, \beta = e, \mu, \tau$) as small expansion parameters 
of the same order $\epsilon$. 
Within the $\epsilon$ perturbation theory we obtain the $S$ matrix 
elements and the neutrino oscillation probability formula 
to second order (third order in $\nu_e$ related channels) in $\epsilon$. 
The formula allows us to estimate size of the contribution of any particular 
NSI element $\varepsilon_{\alpha \beta}$ to the oscillation probability 
in arbitrary channels, and gives a global bird-eye view of 
the neutrino oscillation phenomena with NSI. 
Based on the second-order formula we discuss how all the 
conventional lepton mixing as well as NSI parameters can be determined. 
Our results shows that while $\theta_{13}$, $\delta$, and the NSI 
elements in $\nu_e $ sector can in principle be determined, 
complete measurement of the NSI parameters in the $\nu_\mu - \nu_\tau$ 
sector is not possible by the rate only analysis. 
The discussion for parameter determination and the analysis based on 
the matter perturbation theory indicate that the parameter degeneracy prevails 
with the NSI parameters. In addition, a new solar-atmospheric variable 
exchange degeneracy is found. 
Some general properties of neutrino oscillation with and without NSI 
are also illuminated.

\end{abstract}


\maketitle

\section{ Introduction }

Neutrino masses and lepton flavor mixing \cite{MNS} discovered by 
the atmospheric \cite{SKatm}, the solar \cite{solar}, and the reactor 
neutrino \cite{KamLAND} experiments constitute still 
the uniques evidence for physics beyond the Standard Model. 
A possible next step would be a discovery of neutrino interactions 
outside the standard electroweak theory. 
Based on expectation of new physics at TeV scale such non-standard 
interactions (NSI) with matter possessed by neutrinos are proposed and 
extensively discussed 
\cite{wolfenstein,valle,guzzo,roulet,grossmann,berezhiani}. 
The experimental constraints on NSI are summarized in 
\cite{davidson}. See also \cite{DELPHI}.

Recognition of structure of neutrino masses and lepton flavor mixing, 
at least up to now, relies on neutrino flavor transformation 
\cite{MNS,pontecorvo,wolfenstein,MSW}, 
which we generically refer as neutrino oscillation in this paper. 
Quite naturally, there have been numerous theoretical analyses 
to understand the structure of the phenomena. 
In the context of long-baseline neutrino experiments, an exact expression 
of the oscillation probability is derived under the constant matter density approximation \cite{KTY}. 
To understand physics of neutrino oscillation, however, it is often more 
illuminating to have suitable approximation schemes. 
In the latter category, various perturbative formulations of three-flavor 
neutrino oscillation have been developed and proven to be quite 
useful in particular in the context of long-baseline 
accelerator and reactor experiments. 
They include one-mass scale dominance approximation in vacuum 
\cite{one-mass}, short-distance expansion in matter \cite{shortBL}, 
matter perturbation theory \cite{AKS,MNprd98}, and 
perturbation theory with the small expansion parameters 
$\Delta m^2_{21} / \Delta m^2_{31}$ and $\theta_{13}$ \cite{golden} 
(that are taken as of order $\epsilon$) 
which we call the $\epsilon$ perturbation theory in this paper. 
See, for example, 
\cite{yasuda1,freund,munich04,MSS04} for subsequent development of 
perturbation theory of neutrino oscillation.

When the effects of NSI are included, however, 
theoretical analysis of the system of neutrino flavor transformation 
does not appears to achieve the same level of completeness as that 
only with standard interactions (SI). 
Perturbative formulas of the oscillation probabilities with NSI have been 
derived under various assumptions 
\cite{concha1,confusion1,confusion2,ota1,yasuda2,kopp2,NSI-nufact}. 
Even some exact formulas are known \cite{NSI-nufact}. 
However, one cannot answer the questions such as: 
How large is the effects of $\epsilon_{\mu \tau}$ in the oscillation 
probability $P(\nu_{e} \rightarrow \nu_{\mu} )$? 
(See below for definition of NSI elements $\epsilon_{\alpha \beta}$.) 
How large is the effects of $\epsilon_{e e}$ in the oscillation 
probability $P(\nu_{e} \rightarrow \nu_{e} )$?
Which set of measurement is sufficient to determine all the 
NSI elements? 
References of neutrino oscillation and the sensitivity analyses 
with NSI are too numerous to quote here and may be found 
in bibliographies in the existing literatures, for example, in 
\cite{kopp2,NSI-nufact,NOVE08_mina}.

It is the purpose of this paper to fill the gap between understanding of 
neutrino oscillations with and without NSI. 
We try to do it by formulating the similar $\epsilon$ perturbation theory 
as in \cite{golden} but with including effects of NSI by assuming that 
NSI elements are of order $\sim \epsilon$. 
We derive the perturbative formula of the oscillation probability 
similar to the one in \cite{golden}, which we call, respectively, the 
NSI and the SI second-order formulas in this paper. 
The approximate formula will allow us to have a bird-eye view of 
the neutrino oscillations with NSI, and will enable us to answer 
the above questions.

The other limitation that is present in some foregoing analyses, 
which we want to overcome, is the assumption of single (or, a few) 
$\epsilon_{\alpha \beta}$ dominance. 
Upon identification or getting hint for possible NSI interactions 
it will become possible to express 
$\varepsilon_{\alpha \beta}$ in propagation as a function of 
couplings involved in the higher dimensional operators. 
When this situation comes it is likely that all (or at least most of) the NSI 
elements $\varepsilon_{\alpha \beta}$ exist in the Hamiltonian 
with comparable magnitudes. 
Therefore, the theoretical machinery we prepare for the analysis 
must include all the NSI elements at the same time.\footnote{
The similar comments also apply to the procedure by which the 
current constraints on NSI is derived (for example in \cite{davidson}) 
where the constraints are derived under the assumption of presence 
of a particular NSI element in each time, the point carefully mentioned 
by the authors themselves. 
}

More about necessity and usefulness of the $S$ matrix and the 
NSI second-order formula of all oscillation channels and with all 
NSI elements included; 
If we are to include the effects of NSI in production and detection processes 
it is necessary to sum up all the oscillation channels that can contribute. 
Hence, the formulas of all channels are necessary. 
In a previous paper it was uncovered that the so called 
$\theta_{13} - $NSI confusion \cite{confusion1,confusion2} can be 
resolved by a two-detector setting in neutrino factory experiments 
\cite{NSI-nufact}. 
Keeping the terms with the solar $\Delta m^2_{21}$ is shown to be 
crucial for resolving the confusion, and hence a full second-order 
formula is useful. 
In fact, the NSI second-order formula is surprisingly simple in its form, 
keeping the form of the original SI one with generalized variables, 
and the structure is even more transparent than those with first-order 
approximation of NSI.

With the NSI second order formula, we are able to discuss, for the first time, 
a strategy for simultaneous complete determination of the SI and NSI parameters. 
Through the course of discussions we indicate that, as in the system 
without NSI, the parameter degeneracy \cite{intrinsicD,MNjhep01,octant} 
exist in systems with NSI, but in a new form which involve both the SI 
and the NSI parameters. 
See Secs.~\ref{determination} and \ref{matter-perturbation}. 
Moreover, we will uncover a new type of degeneracy, the one exchanging 
the generalized solar and atmospheric variables in Sec.~\ref{degeneracy}.

Finally, we should mention about what will not be achieved in this paper 
even within the context of theoretical analysis. 
First of all, our perturbative formulation relies on the particular assumption 
on relative magnitudes of SI and NSI parameters, and 
we cannot say many for cases in which our assumptions are not valid. 
We discuss the effects of NSI while neutrinos propagate in matter, 
and its effects in production and detection of neutrinos are ignored. 
Therefore, this paper must be regarded as merely the first step toward 
complete treatment of neutrino oscillation with NSI.

\section{ Physics summary }

Because this paper has been developed into a long one, unfortunately, 
we think it convenient for readers, in particular experimentalists, to 
summarize the physics outputs of the perturbative treatment of 
neutrino oscillation with NSI. 
We highly recommend the readers to read this section first.

\subsection{New result in the standard three flavor mixing}

Though this paper aims at uncovering structure of neutrino oscillation 
with NSI, we have observed a new features of standard neutrino oscillation 
without NSI in Sec.~\ref{hesitation-theorem}, the property we call the 
``matter hesitation''. 
It states that in our perturbative framework 
the matter effect comes in into the oscillation probability only at the 
second order in the small expansion parameter $\epsilon$ in all the 
channels of neutrino oscillation.\footnote{
Though this property must be known in the community 
as the results of perturbative calculation, it appears to us that 
it did not receive enough attention so far. 
}
%
It is a highly nontrivial feature because we treat the matter effect 
as of order unity. The ``matter hesitation'' explains why it is so difficult to 
have a sufficiently large matter effect, e.g., to resolve the 
mass hierarchy, in many long-baseline neutrino oscillation experiments. 
It also has implications to neutrino oscillation with NSI as will be discussed in Sec.~\ref{implication}.

\subsection{Guide for experimentalists; importance of various NSI elements in each channel} 
\label{guide}

Experimentalists who want to hunt NSI in neutrino propagation 
may ask the following questions:

\begin{itemize}

\item 
We want to uncover the effect of $\varepsilon_{e \tau}$ 
(or $\varepsilon_{e \mu}$). 
What is the neutrino oscillation channel do you recommend to 
use for this purpose?

\item 
We plan to detect the effect of $\varepsilon_{\mu \tau} $. 
Which set of measurement do we need to prepare? 

\item 
We seek a complete determination of all the SI and the NSI parameters. 
What would be the global strategy to adopt?

\end{itemize}

\noindent
With the oscillation probability formulas given in Sec.~\ref{formula} 
we will try to answer these questions. 
Though we can offer only a partial answer to the last question above 
we can certainly give the answer to the first two questions within 
the framework of perturbation theory we use. 
In Table~\ref{order} the relative importance of the effects of each 
element $\varepsilon_{\alpha \beta}$ of NSI are tabulated as 
order of a small parameter $\epsilon$ that they first appear in 
each oscillation channel. We presume $\epsilon \sim10^{-2}$. 
Thus, our answer to the above questions based on the assumption 
that only the terms up to second order in $\epsilon$ are relevant 
would be (in order): 

\begin{itemize}

\item 
The neutrino oscillation channels in which only $\varepsilon_{e \tau}$ 
and $\varepsilon_{e \mu}$ come in and the other elements do not 
are the $\nu_{e}$ related ones, 
$\nu_{e} \rightarrow \nu_{e}$, 
$\nu_{e} \rightarrow \nu_{\mu}$, and 
$\nu_{e} \rightarrow \nu_{\tau}$. 
Obviously, the latter two appearance channels would be more 
interesting experimentally.
One can in principle determine them simultaneously with $\theta_{13}$ 
and $\delta$ by rate only analysis. 

\item 
Do measurement at the 
$\nu_{\mu} \rightarrow \nu_{\mu}$ channel to determine 
$\varepsilon_{\mu \tau}$. 
Adding $\nu_{\mu} \rightarrow \nu_{\tau}$ channel does not help. 
Effects of the NSI element are relatively large because they are 
first order in $\epsilon$, but the spectrum information is crucial to utilize 
this feature and to separate its effects from those of 
$\varepsilon_{\mu \mu} - \varepsilon_{\tau \tau}$. 
If an extreme precision is required you might want to supplement 
the measurement by the $\nu_{\mu}$ and $\nu_{\tau}$ appearance 
measurement above. 

\item 
We will show that, in fact, there is a difficulty in complete determination 
of all the NSI and SI parameters by the rate only analysis. 
The trouble occurs in the $\nu_{\mu} - \nu_{\tau}$ sector.
Even though we are allowed to assume perfect measurement of 
all the channels including the one with $\nu_{\tau}$ beam 
(which, of course, would not be practical), one of the three unknowns, 
$\varepsilon_{\mu \mu} - \varepsilon_{\tau \tau}$ and 
$\varepsilon_{\mu \tau}$ including its phase, cannot be determined 
if we rely on the rate only analysis. 
See Sec.~\ref{determination} for more details. 
Clearly, the spectrum information is the key to the potential of 
being able to determine all the SI and the NSI parameters, which should 
be taken into account in considering future facilities which search for NSI.

\end{itemize}

\noindent
With regard to the second point above, some remarks are in order; 
Usually, disappearance channels are disadvantageous in looking for 
a small effect such as $\theta_{13}$, because one has to make 
the statistical error smaller than the effect one wants to detect. 
In this respect, the NSI search in the $\nu_{\mu} - \nu_{\tau}$ sector 
is promising because it is the first order effect in $\epsilon$. 
In fact, rather high sensitivities for determining 
$\varepsilon_{\mu \tau}$ and 
$\varepsilon_{\mu \mu} - \varepsilon_{\tau \tau}$ observed 
in atmospheric \cite{fornengo} and future accelerator \cite{NSI-T2KK} 
neutrino analyses are benefited by this feature.

We must warn the readers that experimental observable will be affected by 
NSI effects in production and detection of neutrinos. 
Therefore, our comments in this subsection assumes that they are 
well under control and shown to be smaller than the NSI effects in 
propagation by near detector measurement with an extreme precision. 
It should be also emphasized that some of our comments rely on the 
second-order perturbative formula of the oscillation probability.

\begin{table}
\caption[aaa]{
Presented are the order in $\epsilon$ ($\sim 10^{-2}$) at which  
each type of $\varepsilon_{\alpha \beta}$ ($\alpha, \beta = e, \mu, \tau$) 
and  $a$ dependence 
($a$ is Wolfenstein's matter effect coefficient \cite{wolfenstein}) 
starts to come in into the expression of the 
oscillation probability in $\epsilon$ perturbation theory.\footnote{
To second order in $\epsilon$ the sensitivity to $\varepsilon_{\mu \mu}$ 
and $\varepsilon_{\tau \tau}$ is through the form 
$\varepsilon_{\mu \mu} - \varepsilon_{\tau \tau}$, and 
hence no sensitivity to the individual $\varepsilon$'s. 
Generally, the diagonal $\varepsilon$'s appear in a form of difference 
in the oscillation probabilities because an over-all phase is an unobservable.
}
%
The last column is for the $a$ dependence in the standard oscillation without NSI.
The order of $\epsilon$ indicated in parentheses implies the one 
for the maximal $\theta_{23}$ in which cancellation takes place in 
the leading order. 
See the text for the definition of $\epsilon$ perturbation theory and 
for more details. }
\vglue 0.5cm
\begin{tabular}{c|cccccccc}
\hline 
\hline
\ \ Channel \ \ 
& \ \ $\epsilon_{ee}$ \ \ 
& \ \ $\epsilon_{e\mu}$ \ \
& \ \ $\epsilon_{e\tau}$ \ \ 
& \ \ $\epsilon_{\mu \tau}$ \ \ 
& \ \ $\epsilon_{\mu \mu} $ \ \
& \ \ $\epsilon_{\tau \tau}$ \ \ 
& \ a dep. (NSI) \ 
& \ a dep. (SI) \ \\
\hline
\ $P(\nu_{e} \rightarrow \nu_{\alpha})$: $\alpha=e, \mu, \tau$  \ 
& $\epsilon^3$ 
& $\epsilon^2$ 
& $\epsilon^2$ 
& $\epsilon^3$ 
& $\epsilon^3$ 
& $\epsilon^3$ 
& $\epsilon^2$
& $\epsilon^2$ \\
\hline
$P(\nu_{\alpha} \rightarrow \nu_{\beta})$: $\alpha, \beta=\mu, \tau$ 
& $\epsilon^3$ 
& $\epsilon^2$ 
& $\epsilon^2$ 
& $\epsilon^1 $ 
& $\epsilon^1 (\epsilon^2) $
& $\epsilon^1 (\epsilon^2) $ 
& $\epsilon^1$
& $\epsilon^2$\\
\hline
\hline
\end{tabular}
\label{order}
\end{table}

\subsection{Some interesting or peculiar features of neutrino oscillation with NSI}
\label{peculiarity}

We list here some interesting features of neutrino oscillation with NSI 
which will be fully discussed in the following sections in this paper. 
Some of them are either unexpected, or might be showed up in 
previous analyses but without particular attention. 
A few points in them requires further investigation for full understanding. 

\begin{itemize}

\item 
One of the most significant feature in Table. 1 is that $\varepsilon_{e e}$ 
appears only at third order in $\epsilon$ in all oscillation channels. 
It will be shown in Sec.~\ref{implication} 
that this feature can be explained as a consequence of the matter  
hesitation mentioned earlier. 

\item 
It is interesting to observe from Table~\ref{order} that 
Wolfenstein's matter effect coefficient $a$ in the oscillation probability, 
shows up in first (second) order in $\epsilon$ in system 
with (without) NSI, which makes effects of matter density uncertainty 
larger in system with NSI. 
It occurs in the $\nu_{\mu} - \nu_{\tau}$ sector, and is easily understood 
as a consequence of ``tree level'' transition by the NSI element. 

\item 

The results in the last column in Table~\ref{order} indicates that 
sensitivity to $\varepsilon_{\mu \mu} - \varepsilon_{\tau \tau}$ will 
depend upon if $\theta_{23}$ is maximal or not. 
This feature is clearly seen, e.g., in \cite{NSI-T2KK}. 
Analysis to resolve the $\theta_{23}$ octant degeneracy 
similar to the one proposed for cases without NSI 
\cite{MSYIS03,resolve23,shaevitz,atm-method,T2KK2nd}, 
would be required for correct estimation of the sensitivity to NSI.

\end{itemize}

\section{Introducing the Effects of NSI in neutrino production, propagation and detection processes}

We consider NSI involving neutrinos of the type 
\begin{eqnarray}
{\cal L}_{\text{eff}}^{\text{NSI}} =
-2\sqrt{2}\, \varepsilon_{\alpha\beta}^{fP} G_F
(\overline{\nu}_\alpha \gamma_\mu P_L \nu_\beta)\,
(\overline{f} \gamma^\mu P f),
\label{LNSI}
\end{eqnarray}
where $G_F$ is the Fermi constant, and 
$f$ stands for the index running over fermion species in the earth, 
$f = e, u, d$, where 
$P$ stands for a projection operator and is either
$P_L\equiv \frac{1}{2} (1-\gamma_5)$ or $P_R\equiv \frac{1}{2} (1+\gamma_5)$. 
The current constraints on $\varepsilon$ parameters are summarized 
in \cite{davidson}.

Upon introduction of the NSI as in (\ref{LNSI}) it affects neutrino 
production, detection as well as propagation in matter 
\cite{grossmann,concha1,confusion2,ota1}. 
Therefore, we have to analyze the following ``grand transition amplitude'' 
from a parent $\Pi$ particles (which needs not to be pions) to the 
particular detection particle $N$ (which needs not be nucleons):\footnote{
One can talk about momentum reconstructed detected neutrinos instead 
of detected positrons, for example, but the reconstruction process must 
involve the effects of NSI. 
The expression in (\ref{grand-amplitude}) is just to symbolically indicate this point. 
}
%
\begin{eqnarray} 
T(E_{\Pi}, E_{N}) = \sum_{\alpha, \beta}
\int dE_{\nu \alpha} 
D( E_{\Pi}, E_{\nu \alpha}) 
S(\nu_{\alpha} \rightarrow \nu_{\beta}; E_{\nu \alpha}) 
R( E_{\nu \beta}, E_{N} ) 
\label{grand-amplitude}
\end{eqnarray}
where the sum over $\alpha$ and $\beta$ must be taken only 
if they are amenable to be produced by the decay, or to undergo the reaction. 
Here, we have assumed the particular decay process to produce neutrinos 
as $\Pi \rightarrow \nu_{\alpha} + X_{\alpha}$ with decay amplitude 
$D( E_{\Pi}, E_{\nu \alpha}) $ with the energies $E_{\Pi}$ and 
$E_{\nu \alpha}$ of parent and daughter particles, 
and the particular reaction 
$\nu_{\beta} + P_{TG} \rightarrow N_{\beta} + Y_{\beta}$ with 
reaction amplitude $R( E_{\nu \beta}, E_{N} )$ which produce 
$N_{\beta}$ particle with energy $E_{N}$. 
Here, $X_{\alpha}$ and $Y_{\beta}$ are meant to be some inclusive 
collections of particles and $P_{TG}$ denotes the target particle.
$S(\nu_{\alpha} \rightarrow \nu_{\beta})$ denotes the neutrino oscillation 
amplitude of the channel $\nu_{\alpha} \rightarrow \nu_{\beta}$. 
The observable quantity is of course $\vert T(E_{\Pi}, E_{N}) \vert^2$.

We assume that the coupling constant $\varepsilon_{\alpha\beta}$ 
possessed by NSI is small, 
$\sim \left( \frac{M_{W}}{M_{NP}} \right)^2$ where $M_{NP}$ 
is a new physics scale, 
so that we can organize perturbative treatment of the effects of NSI. 
$\varepsilon_{\alpha\beta}$ can be as small as $10^{-2}$ ($10^{-4}$)
for $M_{NP} = 1 (10)$ TeV, and is even smaller if higher 
dimension operators (higher than six) are required. 
We assume that all the $\varepsilon_{\alpha\beta}$ have similar order 
of magnitudes and denote the small number collectively as $\epsilon$. 
Under these assumptions we expect that the decay and the detection 
functions, and the oscillation probabilities can be expanded as 
\begin{eqnarray} 
D( E_{\Pi}, E_{\nu \alpha}) &=& 
D^{ (0) } + 
D^{ (1) } \epsilon + 
D^{ (2) } \epsilon^2 + ...
\nonumber \\
S(\nu_{\alpha} \rightarrow \nu_{\beta}; E_{\nu \alpha}) &=& 
S^{ (0) } (\nu_{\alpha} \rightarrow \nu_{\beta}) + 
S^{ (1) } (\nu_{\alpha} \rightarrow \nu_{\beta}) \epsilon + 
S^{ (2) } (\nu_{\alpha} \rightarrow \nu_{\beta}) \epsilon^2 + ...
\nonumber \\
R( E_{\nu \alpha}, E_{N} ) &=& 
R^{ (0) } + 
R^{ (1) } \epsilon + 
R^{ (2) } \epsilon^2 + ...
\label{expansion}
\end{eqnarray}
where we have suppressed the kinematical dependences in 
quantities in the right-hand-side of (\ref{expansion}). 
The first terms in (\ref{expansion}) are the one without NSI. 
Now, because of the smallness of $\epsilon \sim 10^{-2}$ (or smaller) 
we take the attitude that keeping terms up to second order in $\epsilon$ 
must be good enough to discuss the effects of NSI 
and eventually to estimate the sensitivity to NSI.\footnote{
As far as the appearance channels $\nu_e \rightarrow \nu_\mu$ and 
$\nu_e \rightarrow \nu_\tau$ are concerned the oscillation 
amplitudes start from first order in $\epsilon$, as we will see below. 
Therefore, only the first order corrections to $D$ and $R$ are 
relevant for the observable to order $\epsilon^2$. 
}

Unfortunately, even with the perturbative treatment this is a highly 
complicated system to analyze its full structure. 
It is possible that the types of NSI that contribute to production 
and detection processes are more numerous than the ones in the 
propagation process \cite{kopp2}. 
If this occurs the effects of NSI into production and detection processes 
could be qualitatively different from those in propagation. 
Therefore, the effects of NSI come into the decay and the reaction 
amplitudes generally in a model-dependent fashion, so that the flavor 
($\alpha, \beta$) dependence of NSI effects are also model-dependent. 
Also they do so in an energy dependent way so that integration 
over neutrino energy in (\ref{grand-amplitude}) is required for the full analysis. 
For an explicit example of how NSI enter into the decay and the reaction 
amplitudes as well as to the neutrino propagation in matter in 
concrete models, see for example the ``unitarity violation'' approach 
developed in \cite{gavela}.

In this paper, therefore, we confine ourselves to analysis of the structure 
of neutrino propagation with NSI, namely the terms with no effects of 
NSI in the decay and the reaction amplitudes in (\ref{grand-amplitude}). 
This is a particularly simple system (relatively speaking with the full one) 
in the sense that no unitarity violation comes in because it deals with 
propagation of three light neutrinos. 
Furthermore, it has no explicit model dependence once the effects of 
NSI is parametrized in the familiar way. 
See the Hamiltonian in (\ref{hamiltonian}). 
We should emphasize that limitation of our scope to the problem of 
neutrino propagation, in fact, allows us to dig out structure of neutrino 
oscillation with NSI in a transparent manner. 
Therefore, we think it a meaningful first step.

Our analysis can become the whole story provided that extremely stringent 
bounds on NSI effects in decay as well as detection reactions are 
placed by front detector measurement in future experiments. 
Otherwise, it covers only a leading (zeroth) order terms in NSI effect in 
decay and detection. 
When the first order corrections to them are taken into account 
what is needed is to compute the oscillation amplitude up to first order 
in $\epsilon$ to obtain the observable to order $\epsilon^2$. Hence, 
we present the results of $S$ matrix elements in Appendix~\ref{Smatrix-element}, not only the expression of the oscillation probabilities, 
for future use.

\section{ General Properties of Neutrino Oscillation with and without NSI }
\label{general}

Now, we analyze the structure and the properties of neutrino propagation 
in matter with NSI. 
We, however, sometimes go back to the system without NSI 
whenever it is illuminating. 
The results obtained in this section are exact, that is, they are valid without 
recourse to perturbation theory we will formulate in the next section. 
To discuss effects of NSI on neutrino propagation it is customary 
to introduce the $\varepsilon$ parameters, which are defined as
$\varepsilon_{\alpha\beta} \equiv \sum_{f,P} \frac{n_f}{n_e}
\varepsilon_{\alpha\beta}^{fP}$, where 
$n_f$ ($n_e$) denotes the $f$-type fermion (electron) number density 
along the neutrino trajectory in the earth. 
Then, the neutrino evolution equation can be written in flavor basis as 
\begin{eqnarray}
i \frac{ d }{ dx } \nu_{\alpha} = H_{ \alpha \beta } \nu_{ \beta } 
\hspace{0.6cm}
(\alpha, \beta = e, \mu, \tau). 
\label{evolution}
\end{eqnarray}
In the standard three-flavor neutrino scheme, Hamiltonian including NSI 
is given by 
\begin{eqnarray}
H= 
\frac{1}{2E} 
\left\{ 
U \left[
\begin{array}{ccc}
0 & 0 & 0 \\
0 & \Delta m^2_{21} & 0 \\
0 & 0 & \Delta m^2_{31} 
\end{array}
\right] U^{\dagger}
+ 
a(x) \left[
\begin{array}{ccc}
1 & 0 & 0 \\
0 & 0 & 0 \\
0 & 0 & 0
\end{array}
\right] 
\right.
\nonumber \\
&&\hspace*{-36mm} {} +
\left.
a(x) 
\left[
\begin{array}{ccc}
\varepsilon_{e e} & |\varepsilon_{e \mu}| e^{i \phi_{e \mu}} & |\varepsilon_{e \tau}| e^{i \phi_{e \tau}} \\
|\varepsilon_{e \mu}| e^{-i \phi_{e \mu}} & \varepsilon_{\mu \mu} & |\varepsilon_{\mu \tau}| e^{i \phi_{\mu \tau}} \\
|\varepsilon_{e \tau}| e^{-i \phi_{e \tau}} & |\varepsilon_{\mu \tau}| e^{-i \phi_{\mu \tau}} & \varepsilon_{\tau \tau} 
\end{array}
\right] 
\right\} 
\label{hamiltonian}
\end{eqnarray}
where $\Delta m^2_{ji} \equiv m^2_{j} - m^2_{i}$, and 
$a(x) \equiv 2\sqrt{2} G_F N_e(x) E$ is the coefficient which is 
related to the index of refraction of neutrinos in medium of electron 
number density $N_e(x)$~\cite{wolfenstein}, 
where $G_F$ is the Fermi constant and $E$ is the neutrino energy. 
The first two terms in (\ref{hamiltonian}) are the Standard Model 
interactions, whereas the last term denotes the non-standard 
neutrino interactions with matter. 
$U$ denotes the flavor mixing matrix, the 
Maki-Nakagawa-Sakata (MNS) matrix \cite{MNS}, in the lepton sector. 
In its standard form \cite{PDG} it reads 
\begin{eqnarray}
U = U_{23} U_{13} U_{12} = 
\left[
\begin{array}{ccc}
1 & 0 & 0 \\
0 & c_{23} & s_{23} \\
0 & - s_{23} & c_{23} \\
\end{array}
\right] 
\left[
\begin{array}{ccc}
c_{13} & 0 & s_{13} e^{- i \delta} \\
0 & 1 & 0 \\
- s_{13} e^{ i \delta} & 0 & c_{13} \\
\end{array}
\right] 
\left[
\begin{array}{ccc}
c_{12} & s_{12} & 0 \\
- s_{12} & c_{12} & 0 \\
0 & 0 & 1 \\
\end{array}
\right] 
\label{MNSmatrix}
\end{eqnarray}
where $\delta$ stands for the leptonic Kobayashi-Maskawa (KM) 
phase \cite{KM}.

Most of the formulas in this and the next sections 
(Secs.~\ref{general} and \ref{perturbation}) can be written in forms valid 
for arbitrary matter density profiles if the adiabatic approximation holds. 
We, however, present the ones derived under the constant matter density approximation because it makes the equations simpler, in particular, 
the perturbative formulas for the oscillation probabilities in Sec.~\ref{formula}. 
Unlike the case of the MSW solar neutrino solutions \cite{MSW} in which 
the matter density variation is the key to the problem, the constant density 
approximation to $N_e(x)$ in long-baseline experiments 
should serve as a reasonable first approximation.

The $S$ matrix describes possible flavor changes after traversing 
a distance $L$, 
\begin{eqnarray} 
\nu_{\alpha} (L) = S_{\alpha \beta} \nu_{\beta} (0), 
\label{def-S}
\end{eqnarray}
and the oscillation probability is given by 
\begin{eqnarray} 
P(\nu_{\beta} \rightarrow \nu_{\alpha}; L )= 
\vert S_{\alpha \beta} \vert^2. 
\label{def-P}
\end{eqnarray}
If the neutrino evolution is governed by the Schr\"odinger equation 
(\ref{evolution}), $S$ matrix is given as 
\begin{eqnarray} 
S = T \text{exp} \left[ -i \int^{L}_{0} dx H(x) \right] 
\label{Smatrix}
\end{eqnarray}
where $T$ symbol indicates the ``time ordering'' 
(in fact ``space ordering'' here). 
The right-hand-side of (\ref{Smatrix}) may be written as 
$e^{-i H L}$ for the case of constant matter density. 
For notational convenience, we denote the $S$ matrix elements as 
\begin{eqnarray}
S =
\left[
\begin{array}{ccc}
S_{ee} & S_{e \mu} & S_{e \tau} \\
S_{\mu e} & S_{\mu \mu} & S_{\mu \tau} \\
S_{\tau e} & S_{\tau \mu} & S_{\tau \tau} 
\end{array}
\right]. 
\label{def-Selements}
\end{eqnarray}

The primary purpose of this paper is to discuss the properties of 
neutrino oscillation in the standard three flavor system with NSI. 
But, we recollect the properties of neutrino oscillation 
without NSI whenever necessary, and treat both systems simultaneously 
or go back and forth between them to make our discussion transparent. 
By this way the properties of the neutrino oscillations can be 
better illuminated.

\subsection{Relations between neutrino oscillation amplitudes without NSI}

If NSI, the third term in (\ref{hamiltonian}), is absent the matter term 
(the second term in (\ref{hamiltonian})) has a symmetry; 
It is invariant under $U_{23}$ rotation which act on 
$\nu_\mu - \nu_\tau$ subspace.
Due to this symmetry the Hamiltonian can be conveniently written in the form  
\begin{eqnarray} 
H = U_{23} \tilde{H} U_{23}^{\dagger}, 
\label{tilde-hamiltonian}
\end{eqnarray}
and hence the $S$ matrix can be written as 
\begin{eqnarray} 
S(L) = U_{23} \tilde{S} (L) U_{23}^{\dagger}
\label{Smatrix-tilde}
\end{eqnarray}
as noticed in \cite{munich04} where 
$ \tilde{S} (L) = T \text{exp} \left[ -i \int^{L}_{0} dx \tilde{H} (x) \right] $. 
The point here is that $ \tilde{H} $ and $ \tilde{S} (L) $ do {\em not} 
contain $\theta_{23}$.

If we denote $ \tilde{S} (L) $ matrix elements in a form 
analogous to the one in (\ref{def-Selements}) $S$ matrix 
can be written as 
\begin{eqnarray}
\left[
\begin{array}{ccc}
\tilde{ S }_{ee} & c_{23} \tilde{ S }_{e \mu} + s_{23} \tilde{ S }_{e \tau} & - s_{23} \tilde{ S }_{e \mu} + c_{23} \tilde{ S }_{e \tau} \\
c_{23} \tilde{ S }_{\mu e} + s_{23} \tilde{ S }_{\tau e} & c^2_{23} \tilde{ S }_{\mu \mu} + s^2_{23} \tilde{ S }_{\tau \tau} + c_{23} s_{23} ( \tilde{ S }_{\mu \tau} + \tilde{ S }_{\tau \mu} ) & c^2_{23} \tilde{ S }_{\mu \tau} - s^2_{23} \tilde{ S }_{\tau \mu} + c_{23} s_{23} ( \tilde{ S }_{\tau \tau} - \tilde{ S }_{\mu \mu} ) \\
- s_{23} \tilde{ S }_{\mu e} + c_{23} \tilde{ S }_{\tau e} & c^2_{23} \tilde{ S }_{\tau \mu } - s^2_{23} \tilde{ S }_{ \mu \tau} + c_{23} s_{23} ( \tilde{ S }_{\tau \tau} - \tilde{ S }_{\mu \mu} ) & s^2_{23} \tilde{ S }_{\mu \mu} + c^2_{23} \tilde{ S }_{\tau \tau} - c_{23} s_{23} ( \tilde{ S }_{\mu \tau} + \tilde{ S }_{\tau \mu} ) 
\end{array}
\right]. 
\nonumber \\
\label{SbyS-tilde}
\end{eqnarray}
It should be noticed that $ S_{e e} = \tilde{S}_{e e}$ is independent of 
$\theta_{23}$. 
Therefore, the $S$ matrix elements obey relationships \cite{munich04} 
\begin{eqnarray} 
S_{e \tau} &=& S_{e \mu} ( c_{23} \rightarrow - s_{23}, s_{23} \rightarrow c_{23} ), 
\nonumber \\
S_{\tau \tau} &=& S_{\mu \mu} ( c_{23} \rightarrow - s_{23}, s_{23} \rightarrow c_{23} ), 
\nonumber \\
S_{\tau \mu} &=& - S_{\mu \tau} ( c_{23} \rightarrow - s_{23}, s_{23} \rightarrow c_{23} ). 
\label{relation}
\end{eqnarray}
%

\subsection{Relations between neutrino oscillation amplitudes with NSI}

The secret behind the relations between $S_{e \mu}$ and $S_{e \tau}$ 
and the others 
in (\ref{relation}) is that $\tilde{ H }$ is independent of $\theta_{23}$, 
or in other words, the invariance of $\tilde{ H }$ 
under the transformation 
$c_{23} \rightarrow - s_{23}, s_{23} \rightarrow c_{23} $. 
When NSI is introduced there exists the following additional term 
in $\tilde{ H }$: 
\begin{eqnarray}
&& \tilde{H}^{\text{NSI}} = 
U_{23}^{\dagger} 
\left[
\begin{array}{ccc}
\varepsilon_{e e} & \varepsilon_{e \mu} & \varepsilon_{e \tau} \\
\varepsilon_{e \mu}^* & \varepsilon_{\mu \mu} & \varepsilon_{\mu \tau} \\
\varepsilon_{e \tau}^* & \varepsilon_{\mu \tau}^* & \varepsilon_{\tau \tau} 
\end{array}
\right] 
U_{23} 
\equiv
\left[
\begin{array}{ccc}
\tilde{\varepsilon}_{e e} & \tilde{\varepsilon}_{e \mu} & \tilde{\varepsilon}_{e \tau} \\
\tilde{\varepsilon}_{e \mu}^* & \tilde{\varepsilon}_{\mu \mu} & \tilde{\varepsilon}_{\mu \tau} \\
\tilde{\varepsilon}_{e \tau}^* & \tilde{\varepsilon}_{\mu \tau}^* & \tilde{\varepsilon}_{\tau \tau} 
\end{array}
\right] 
\nonumber \\
&=& 
\left[
\begin{array}{ccc}
\varepsilon_{e e} & c_{23} \varepsilon_{e \mu} - s_{23} \varepsilon_{e \tau} & s_{23} \varepsilon_{e \mu} + c_{23} \varepsilon_{e \tau} \\
c_{23} \varepsilon_{e \mu}^* - s_{23} \varepsilon_{e \tau}^* & c^2_{23} \varepsilon_{\mu \mu} + s^2_{23} \varepsilon_{\tau \tau} - c_{23} s_{23} ( \varepsilon_{\mu \tau} + \varepsilon_{\mu \tau}^* ) & c^2_{23} \varepsilon_{\mu \tau} - s^2_{23} \varepsilon_{\mu \tau}^* + c_{23} s_{23} ( \varepsilon_{\mu \mu} - \varepsilon_{\tau \tau} ) \\
s_{23} \varepsilon_{e \mu}^* + c_{23} \varepsilon_{e \tau}^* & c^2_{23} \varepsilon_{\mu \tau}^* - s^2_{23} \varepsilon_{\mu \tau} + c_{23} s_{23} ( \varepsilon_{\mu \mu} - \varepsilon_{\tau \tau} ) & s^2_{23} \varepsilon_{\mu \mu} + c^2_{23} \varepsilon_{\tau \tau} + c_{23} s_{23} ( \varepsilon_{\mu \tau} + \varepsilon_{\mu \tau}^* ) 
\end{array}
\right] 
\nonumber \\
\label{tilde-H1NSI}
\end{eqnarray}
Because $\tilde{H}^{\text{NSI}}$ in (\ref{tilde-H1NSI}) {\em does} 
depend on $\theta_{23}$, the $S$ matrix relations as given in 
(\ref{relation}) do not hold. 
However, if we consider the extended transformation 
\begin{eqnarray} 
&& c_{23} \rightarrow - s_{23}, 
\hspace{1cm}
s_{23} \rightarrow c_{23}, 
\nonumber \\
&& \varepsilon_{e \mu} \rightarrow \varepsilon_{e \tau}, 
\hspace{1.3cm}
\varepsilon_{e \tau} \rightarrow - \varepsilon_{e \mu}, 
\nonumber \\
&& \varepsilon_{\mu \mu} \rightarrow \varepsilon_{\tau \tau}, 
\hspace{1.3cm}
\varepsilon_{\tau \tau} \rightarrow \varepsilon_{\mu \mu}, 
\nonumber \\
&& \varepsilon_{\mu \tau} \rightarrow - \varepsilon_{\mu \tau}^*, 
\hspace{1cm}
\varepsilon_{\mu \tau}^* \rightarrow - \varepsilon_{\mu \tau}, 
\label{transformation}
\end{eqnarray}
it is easy to show that $\tilde{H}^{\text{NSI}}$ is invariant under 
the transformation (\ref{transformation}). 
It means that the $S$ matrix relations (\ref{relation}) hold even with 
NSI provided that we extend the transformation to the ones in 
(\ref{transformation}). 
It not only implies the existence of useful relations between the $S$ 
matrix elements, but also serves as a powerful tool for consistency check 
of perturbative computation. 
We will see in Appendices.~\ref{Smatrix-element} and \ref{2nd-order-formula} 
that the computed results 
of the $S$ matrix elements, and hence the oscillation probabilities, 
do satisfy (\ref{relation}) with the extended transformation 
(\ref{transformation}).

As we will see in Sec.~\ref{mu-tau-sector}, the invariance under the 
extended transformation (\ref{transformation}) entails a remarkable 
feature that the terms which depend on $\varepsilon$'s in the 
$\nu_{\mu} - \nu_{\tau}$ sector in the oscillation probabilities 
$P(\nu_\mu \rightarrow \nu_\mu)$, 
$P(\nu_\mu \rightarrow \nu_\tau)$, and 
$P(\nu_\tau \rightarrow \nu_\tau)$
are all equal up to sign.

\subsection{ Phase reduction theorem}
\label{reduction-theorem}

Now, we present a general theorem on reduction of number of 
CP violating phases in system with NSI, which we call 
``phase reduction theorem'' for short. 
By looking into the results of perturbative computation \cite{ota1} 
it was observed that when the solar $\Delta m^2_{21}$ is switched off 
the oscillation probabilities with NSI depends on phases 
which come from NSI elements and $\delta$ in a particular manner, 
e.g., $\vert \varepsilon \vert e^{ i (\delta + \phi) }$. 
It was conjectured on physics ground that the property must hold 
in the exact expressions of the oscillation probabilities \cite{NSI-nufact}; 
With vanishing $\Delta m^2_{21}$ the system becomes effectively 
two flavor and hence the observable CP violating phase must be unique.

Here, we give a general proof of this property which is, in fact, 
very easy to do. We first notice a simple relation which holds 
in the absence of $\Delta m^2_{21}$, 
\begin{eqnarray} 
\hat{H} &\equiv& 
\left[ 
\begin{array}{ccc}
e^{i \delta} & 0 & 0 \\
0 & 1 & 0 \\
0 & 0 & 1
\end{array}
\right]
H
\left[ 
\begin{array}{ccc}
e^{ - i \delta}& 0 & 0 \\
0 & 1 & 0 \\
0 & 0 & 1
\end{array}
\right]
\nonumber \\
&=&
\Delta
\left[ 
\begin{array}{ccc}
s_{13}^2 & c_{13} s_{13} s_{23} &
c_{13} s_{13} c_{23} \\
c_{13} s_{13} s_{23} & c_{13}^2 s_{23}^2 &
c_{13}^2 c_{23} s_{23} \\
c_{13} s_{13} c_{23} & c_{13}^2 c_{23} s_{23} & 
c_{13}^2 c_{23}^2
\end{array}
\right] 
+ \frac{ a }{ 2 E } \left[ 
\begin{array}{ccc}
1 + \varepsilon_{e e} & |\varepsilon_{e \mu}| e^{i \chi} & |\varepsilon_{e \tau}| e^{i \omega} \\
|\varepsilon_{e \mu}| e^{-i \chi} & \varepsilon_{\mu \mu} & |\varepsilon_{\mu \tau}| e^{i \phi_{\mu \tau}} \\
|\varepsilon_{e \tau}| e^{-i \omega} & |\varepsilon_{\mu \tau}| e^{-i \phi_{\mu \tau}} & \varepsilon_{\tau \tau}
\end{array}
\right], 
\label{hatH}
\end{eqnarray}
where $\Delta \equiv \frac{ \Delta m^2_{31} }{ 2 E } $, 
$\chi \equiv \delta + \phi_{e \mu}$, and 
$\omega \equiv \delta + \phi_{e \tau}$. 
Then, if we use a new basis 
$\hat{\nu_{\alpha}} \equiv [ \text{diag} ( e^{i \delta}, 1, 1) ]_{\alpha \beta} \nu_{\beta}$, the evolution equation reads 
\begin{eqnarray}
i \frac{d}{dx}
\left[ 
\begin{array}{c}
\hat{\nu}_{e} \\
\hat{\nu}_{\mu} \\
\hat{\nu}_{\tau}
\end{array}
\right] 
=
\hat{H}
\left[ 
\begin{array}{c}
\hat{\nu}_{e} \\
\hat{\nu}_{\mu} \\
\hat{\nu}_{\tau}
\end{array}
\right]. 
\label{evolution2}
\end{eqnarray}
It is obvious from (\ref{evolution2}) that the system depends on only 
three phases $\chi = \delta + \phi_{e \mu}$, 
$\omega = \delta + \phi_{e \tau}$ and $\phi_{\mu \tau}$ out of four.
This particular combination of phases is, of course, depends upon 
the specific parametrization of the MNS matrix. 
The phase factor attached to the transformation matrix in (\ref{hatH}) 
does not affect the oscillation probability because it is an over-all phase.

The similar treatment with the same transformation as in (\ref{hatH}) 
($ \text{diag.} (1, 1, e^{-i \delta})$) 
can be used to prove that the phase reduction occurs if $\theta_{12}=0$ 
($\theta_{23}=0$) even though $\Delta m^2_{21} \neq 0$.\footnote{
We thank Hiroshi Nunokawa for calling our attention to this feature. 
}
If $\theta_{13}=0$ it is obvious that there are no effect of $\delta$.

This completes a general proof that number of CP violating phases 
is reduced by one when the solar $\Delta m^2_{21}$ is switched off, 
or one of the mixing angles vanishes. 
We emphasize that this property has implications to the real world; 
For example, the phenomenon of phase reduction occurs at the 
magic baseline, $\frac{a L}{4 E} = \pi$, in the perturbative formula 
to be obtained in Sec.~\ref{formula} even though 
$\Delta m^2_{21} \neq 0$.

\section{Perturbation Theory of Neutrino Oscillation }
\label{perturbation}

\subsection{$\epsilon$ Perturbation theory} 
\label{parameters}

To formulate perturbation theory one has to specify the expansion 
parameters. 
We take the following dimensionless parameters as small expansion 
parameters and assume that they are of the same order:\footnote{
We do not take $\frac{1}{ \sqrt{ 2 } } - s_{23}$ as an expansion parameter 
because a rather large range is currently allowed and the situation 
will not be changed even with the next generation experiments \cite{MSS04}. 
}
%
\begin{eqnarray} 
\frac{ \Delta m^2_{21} } { \Delta m^2_{31} } 
\sim s_{13} 
\sim \varepsilon_{\alpha \beta} 
\sim \epsilon 
\hspace{0.5cm}
(\alpha, \beta = e, \mu, \tau). 
\label{def-epsilon}
\end{eqnarray}
Whereas, we treat $\frac{ a }{ \Delta m^2_{31} } $ and 
$\frac{ \Delta m^2_{31} L } { 2 E} $ as of order unity. 
We collectively denote order of magnitude of the expansion parameters 
as $\epsilon$, and hence we call the perturbative framework the 
$\epsilon$ perturbation theory. 
In the absence of NSI our formulas of oscillation probabilities, 
of course, reduces to the Cervera {\it et al.} formula \cite{golden}, 
which we call the SI second-order formula in this paper. 
Correspondingly, we call our second-order probability formula the 
``NSI second-order formula''. 
It appears that in the standard case this perturbative framework 
accommodates the situation of relatively large $\theta_{13}$ within 
the Chooz bound \cite{CHOOZ}, and applicable to wide variety of 
experimental settings.

Another approach would be to just expand in terms of 
$\varepsilon_{\alpha \beta} $ which is assumed to be small without 
any correlation with other SI mixing parameters. 
If NSI elements are extremely small, much smaller than the SI expansion 
parameters, such first-order formulas of NSI would be sufficient. 
It would be the case of NSI search in the next generation experiments as 
discussed e.g., in \cite{kopp2}. 

On the contrary, it often occurs in deriving constraints on various 
NSI parameters that the bounds on the diagonal $\varepsilon$'s, 
$\varepsilon_{ee}$, $\varepsilon_{\mu \mu}$, and $\varepsilon_{\tau \tau}$, 
are sometimes milder than the ones on the off-diagonal $\varepsilon$'s 
by an order of magnitude. 
If it is the case, we may need to keep the higher order of the diagonal 
$\varepsilon$'s in $\epsilon$-perturbation theory, 
to e.g. $\epsilon^4$, in probabilities to analyze such situations. 
We try not to enter into this problem in our present treatment.

\subsection{ Formulating perturbative framework } 
\label{framework}

We follow the standard perturbative formulation to calculate the 
$S$ matrix and the neutrino oscillation probabilities \cite{AKS}. 
Yet, we present a simplified treatment which is suitable for 
higher order calculations. 
For convenience, we start by treating the system without NSI 
in this section. We use the tilde-basis 
$ \tilde{\nu} = U_{23}^{\dagger} \nu $ 
with Hamiltonian $\tilde{H}$ defined in (\ref{tilde-hamiltonian}). 
The tilde-Hamiltonian is decomposed as 
$ \tilde{H} = \tilde{H}_{0} + \tilde{H}_{1} $, where 
\begin{eqnarray} 
\tilde{H}_{0} (x) &=& 
\Delta 
\left[
\begin{array}{ccc}
r_{A} (x) & 0 & 0 \\
0 & 0 & 0 \\
0 & 0 & 1
\end{array}
\right] 
\label{H0}
\\
\tilde{H}_{1} &=& 
\Delta 
\left[
\begin{array}{ccc}
s^2_{13} & 0 & c_{13} s_{13} e^{ -i \delta} \\
0 & 0 & 0 \\
c_{13} s_{13} e^{ i \delta} & 0 & - s^2_{13} 
\end{array}
\right] 
+ 
\Delta r_{\Delta} \left[
\begin{array}{ccc}
s^2_{12} c^2_{13} & c_{12} s_{12} c_{13} & - s^2_{12} c_{13} s_{13} e^{ -i \delta} \\
c_{12} s_{12} c_{13} & c^2_{12} & - c_{12} s_{12} s_{13} e^{ - i \delta} \\
- s^2_{12} c_{13} s_{13} e^{ i \delta} & - c_{12} s_{12} s_{13} e^{ i \delta} & s^2_{12} s^2_{13} 
\end{array}
\right] 
\nonumber
\\
\label{H1tilde}
\end{eqnarray} 
where 
$\Delta \equiv \frac{ \Delta m^2_{31} }{2E} $
$r_{\Delta} \equiv \frac{ \Delta m^2_{21} }{ \Delta m^2_{31} } $, 
$r_{A} (x) \equiv \frac{ a(x) }{ \Delta m^2_{31} } $. 
Though our treatment can be easily generalized to cases 
with matter density variation as far as the adiabatic approximation holds, 
we present, for ease of presentation, the formulas with constant matter 
density approximation.

To calculate $\tilde {S} (L)$ we define $\Omega(x)$ as 
\begin{eqnarray} 
\Omega(x) = e^{i \tilde{H}_{0} x} \tilde{S} (x).
\label{def-omega}
\end{eqnarray}
$\Omega(x)$ obeys the evolution equation 
\begin{eqnarray} 
i \frac{d}{dx} \Omega(x) = H_{1} \Omega(x) 
\label{omega-evolution}
\end{eqnarray}
where 
\begin{eqnarray} 
H_{1} \equiv e^{i \tilde{H}_{0} x} \tilde{H}_{1} e^{-i \tilde{H}_{0} x} 
\label{def-H1}
\end{eqnarray}
Then, $\Omega(x)$ can be computed perturbatively as 
\begin{eqnarray} 
\Omega(x) = 1 + 
(-i) \int^{x}_{0} dx' H_{1} (x') + 
(-i)^2 \int^{x}_{0} dx' H_{1} (x') \int^{x'}_{0} dx'' H_{1} (x'') + 
\mathcal{O} ( \epsilon^3 ). 
\label{Omega-exp}
\end{eqnarray}
where the ``space-ordered'' form in (\ref{Omega-exp}) is essential 
because of the highly nontrivial spatial dependence in $H_{1}$. 
Collecting the formulas the $S$ matrix can be written as 
\begin{eqnarray} 
S(L) = U_{23} e^{- i \tilde{H}_{0} L} \Omega(L) U_{23}^{\dagger}
\label{Smatrix-tilde2}
\end{eqnarray}
Therefore, essentially we are left with perturbative computation of 
$\Omega(x)$ with use of (\ref{def-H1}) to calculate the $S$ matrix.\footnote{
Since $ \tilde{H}_{1} $ in (\ref{H1tilde}) contains order $\epsilon^2$ 
terms in addition to order $\epsilon$ terms the formal expression 
in (\ref{Omega-exp}) includes terms higher than $\mathcal{O} ( \epsilon^2 )$ 
which are meant to be ignored. The same statement applies to 
the computation to be carried out in Sec.~\ref{formula}. 
}

\subsection{Matter hesitation and unitarity}
\label{hesitation-theorem}

One of the usefulness of the $\epsilon$ perturbation theory is that it allows 
to prove the property  ``matter hesitation'', which is a characteristic feature 
of neutrino oscillation in matter without NSI in small $\theta_{13}$ regime. 
The matter hesitation refers to the property that the matter effect dependent 
terms in the neutrino oscillation probabilities 
$P(\nu_{\alpha} \rightarrow \nu_{\beta})$ ($\alpha, \beta = e, \mu, \tau$) 
are absent to first order in $\epsilon$. 
Namely, it hesitates to come in before computation goes to second order 
in $\epsilon$. 
Though its validity heavily relies on the particular perturbative framework 
we work in this paper, it explains why it is so difficult to detect 
the matter effect in many accelerator experiments.

In fact, it is easy to observe the property of matter hesitation; 
It directly follows from the structure of the $S$ matrix in (\ref{Smatrix-tilde2}) 
itself. We first note that in the tilde-basis $\tilde{H}_{1}$ is free from the 
mater effect and it exists only in $\tilde{H}_{0}$. 
Therefore, the matter effect dependence exists only in $e\mu$ 
and $e\tau$  (and their conjugate) elements in $H_{1}$ in (\ref{def-H1}), 
and they are of order $\epsilon$. 
The same statement follows for $\Omega$ in (\ref{Omega-exp}). 
Then, the matter effect dependence in 
$\tilde{S} \equiv e^{- i \tilde{H}_{0} L} \Omega$ 
is only in $e\mu$, $e\tau$ and $ee$ elements at first order 
in $\epsilon$. 
Notice that the final rotation in 23 space to obtain the $S$ matrix in 
(\ref{Smatrix-tilde2}) does not alter the property of the $\tilde{S}$ matrix. 
Therefore, no matter effect dependence appears 
in the oscillation probabilities to first order in $\epsilon$ in $ee$, 
$e\mu - e\tau$ and in the $\mu\tau$ sector  channels for different reasons: 
In $P(\nu_\mu \rightarrow \nu_\mu)$ and $P(\nu_\mu \rightarrow \nu_\tau)$ 
the matter effect is trivially absent to order $\epsilon$ because of no 
dependence in the $S$ matrix.
In $P(\nu_e \rightarrow \nu_\mu)$ and $P(\nu_e \rightarrow \nu_\tau)$ 
it comes in only at order $\epsilon^2$ because the $S$ matrix elements 
are of order $\epsilon$. 
In $P(\nu_e \rightarrow \nu_e)$ the matter effect is absent to order $\epsilon$ because it is contained in a phase factor of the $S$ matrix element. 
This completes the derivation of the matter hesitation, 
the property that matter effects comes in into the oscillation probability 
only at second order in $\epsilon$.

We stress that the absence of the matter effect in the oscillation probability 
to first order in $\epsilon$ is highly nontrivial, in particular in $S_{ee}$. 
Since the matter effect coefficient $r_{A} = \frac{ a }{ \Delta m^2_{31}}$ 
is zeroth order in $\epsilon$ it can affect the $S$ matrix 
in all orders of $\epsilon$. 
But, in fact, absence of matter effect to first order in $\epsilon$ in 
$ P(\nu _e \rightarrow \nu _e) $
can be understood by unitarity. 
For the most nontrivial channel, the relevant unitarity relation is 
\begin{eqnarray} 
1 - P(\nu _e \rightarrow \nu _e) = 
P(\nu _e \rightarrow \nu _\mu) + P(\nu _e \rightarrow \nu _\tau) 
\label{unitarity2}
\end{eqnarray}
Since 
$P(\nu _e \rightarrow \nu _\mu)$ and $P(\nu _e \rightarrow \nu _\tau)$ 
are at least of order $\epsilon^2$ as will be shown in Sec.~\ref{formula} 
and Appendix \ref{2nd-order-formula}, the matter dependent term in 
$ P(\nu _e \rightarrow \nu _e) $, which is involved in the 
left-hand-side in (\ref{unitarity2}), has to be second order, or higher, 
in $\epsilon$. 
It should be noticed that this argument is valid not only in systems 
with SI only but also in the one with NSI, the matter hesitation 
property for $ P(\nu _e \rightarrow \nu _e) $ in the presence of NSI. 
Also notice that the same argument does not go through for 
$1- P(\nu _\mu \rightarrow \nu _\mu) $ because 
$P(\nu _\mu \rightarrow \nu _\tau) $ can contain the terms lower 
than $\epsilon^2$. 
If fact, there exists the first order term in $\epsilon$ in 
$P(\nu _\mu \rightarrow \nu _\mu) $ and 
$P(\nu _\mu \rightarrow \nu _\tau) $ which are proportional to the 
matter effect coefficient $a$.

\subsection{Implication of matter hesitation to neutrino oscillation with NSI }
\label{implication}

There is a clear implication of the property of matter hesitation 
to the system with NSI; 
The terms with the NSI element $\varepsilon_{ee}$ must appear in the 
oscillation probability only at third-order in $\epsilon$ or higher. 
It is due to the special nature of $\varepsilon_{ee}$ that can be 
introduced as a renormalization factor of the matter effect coefficient $a$, 
$a \rightarrow a ( 1 + \varepsilon_{ee} )$.
%
%
Since the terms with $a$ are already of order $\epsilon^2$, 
the terms with $\varepsilon_{ee}$ must be at least of order $\epsilon^3$.

The reader should be puzzled by the above statement. 
One may argue quite naturally that there must exist a term with 
first order in $\varepsilon_{ee}$ in the survival probability 
$P(\nu_{e} \rightarrow \nu_{e})$. 
In fact, such a term does exist in the relevant $S$ matrix element 
as one can see in (\ref{See}): 
\begin{eqnarray}
S_{ee} &=& 
e^{- i r_{A} \Delta L} \Bigl\{ 
1 - i ( s^2_{12} r_{\Delta} + \varepsilon_{ee} r_{A} ) \Delta L \Big\} 
\label{See-1st}
\end{eqnarray}
Resolution of the puzzle, therefore, is that 
the first order term of $\varepsilon_{ee}$ cannot appears in 
the oscillation probability because it is purely imaginary, 
or a phase ignoring $\epsilon^2$ terms. 
However, to confirm the cancellation of second order term we must go 
beyond the present treatment by keeping the order $\epsilon^2$ terms 
in the $S$ matrix. It will be done in the next section.

A more general question is whether the matter hesitation 
can be generalized into the whole systems with NSI. 
We have already answered the question at the end of the previous subsection. 
This feature is to be verified by explicit computation in Sec.~\ref{formula}.

\section{NSI second-order Probability Formulas}
\label{formula}

Now, we present the expressions of the oscillation probabilities 
with NSI which is valid to second order in $\epsilon$. 
For ease of computation we use a slightly different basis which we call 
the double-tilde basis 
with Hamiltonian 
\begin{eqnarray} 
H = U_{23} U_{13} \tilde{ \tilde{H} } U_{13}^{\dagger} U_{23}^{\dagger}
\label{hamiltonian2}
\end{eqnarray}
and the corresponding $S$ matrix  
\begin{eqnarray} 
S(L) = U_{23} U_{13} \tilde{  \tilde{S} } (L)  U_{13}^{\dagger} U_{23}^{\dagger}
\label{Smatrix-dtilde}
\end{eqnarray}
where 
$ \tilde{ \tilde{S} } (L)  = T \text{exp} \left[ -i \int^{L}_{0} dx \tilde{ \tilde{H} } (x)  \right] $. 
The zeroth order and the perturbed part of the reduced Hamiltonian 
$ \tilde{ \tilde{H} }$ are given by 
\begin{eqnarray} 
 \tilde{ \tilde{H} }_{0}  &=& 
\Delta \left[
\begin{array}{ccc}
r_{A}  & 0 & 0 \\
0 & 0 & 0 \\
0 & 0 & 1
\end{array}
\right] 
\label{H0-tilde} \\
%
 \tilde{  \tilde{H} }_{1} &=& 
\Delta \left\{
r_{\Delta}  \left[
\begin{array}{ccc}
s^2_{12} & c_{12} s_{12}  & 0 \\
c_{12} s_{12}  & c^2_{12}  & 0 \\
0 & 0 & 0 
\end{array}
\right] 
+ 
r_{A}  
\left[
\begin{array}{ccc}
- s^2_{13}   & 0 & c_{13} s_{13} e^{ -i \delta}  \\
0 & 0 & 0 \\
c_{13} s_{13} e^{ i \delta}  & 0 & s^2_{13}  
\end{array}
\right] 
\right\} 
\nonumber \\ 
&+& \Delta r_{A}  
U_{13}^{\dagger}  
\left[
\begin{array}{ccc}
\tilde{\varepsilon}_{e e} & \tilde{\varepsilon}_{e \mu}  & \tilde{\varepsilon}_{e \tau}  \\
\tilde{\varepsilon}_{e \mu}^* & \tilde{\varepsilon}_{\mu \mu}  & \tilde{\varepsilon}_{\mu \tau}  \\
\tilde{\varepsilon}_{e \tau}^* & \tilde{\varepsilon}_{\mu \tau}^*  & \tilde{\varepsilon}_{\tau \tau} 
\end{array}
\right] 
U_{13} 
\label{H1-dtilde}
\end{eqnarray}
where 
$\Delta \equiv \frac{ \Delta m^2_{31} }{2E}$, 
$r_{\Delta} \equiv \frac{ \Delta m^2_{21} }{ \Delta m^2_{31} } $, 
$r_{A} \equiv \frac{ a }{ \Delta m^2_{31} } $. 
To simplify the expressions of the $S$ matrix elements 
we use the NSI elements in the tilde basis, 
$\tilde{\varepsilon}_{\alpha \beta} = 
(U_{23}^{\dagger})_{\alpha \gamma} \varepsilon_{\gamma \delta} (U_{23})_{\delta \beta} $, defined in (\ref{tilde-H1NSI}). 
Notice that $\tilde{\varepsilon}$'s are invariant under the extended 
transformation (\ref{transformation}).

The perturbative computation of the $S$ matrix elements can be done 
with the formulas similar to the ones in the tilde basis in Sec.~\ref{framework}. 
In this section we concentrate on the structural analysis of the NSI second-order 
oscillation probabilities e.g., for analysis of parameter determination. 
We collect all the resultant explicit expressions of $S$ matrix elements and 
the oscillation probabilities in Appendix~\ref{Smatrix-element} and 
\ref{2nd-order-formula}, respectively.  
The results of third-order calculation which are necessary to 
complete Table~\ref{order} are presented in Appendix~\ref{third-order-formula}.

\subsection{Electron neutrino sector}
\label{nu_e-sector}

The most distinctive feature of the NSI second-order oscillation probabilities 
in the $\nu_e$-related sector is that they have very similar forms as 
the SI second-order formulas \cite{golden} but with the generalized 
atmospheric and the solar variables: 
\begin{eqnarray} 
\Theta_{\pm} &\equiv& 
\st \frac{\mt}{a} + (s_{23} \varepsilon_{e \mu} + c_{23}  \varepsilon_{e \tau} ) e^{i \delta} = 
\pm \st \frac{ \delta m^2_{31}}{a}+  \vert \tilde{\varepsilon}_{e \tau} \vert  e^{i \hat{\phi}_{e \tau} }
\nonumber \\ 
\Xi &\equiv& 
\left( \ci \si \frac{\mn}{a}+ c_{23} \varepsilon_{e \mu} - s_{23} \varepsilon_{e \tau} \right) e^{i \delta} = 
\ci \si \frac{\mn}{a} e^{i \delta} + \vert \tilde{\varepsilon}_{e \mu} \vert e^{i \hat{\phi}_{e \mu} } 
\label{def-Theta-Xi}
\end{eqnarray}
and their antineutrino versions 
\begin{eqnarray} 
\bar{\Theta}_{\pm} &\equiv& 
- \st \frac{\mt}{a} + (s_{23} \varepsilon_{e \mu}^* + c_{23}  \varepsilon_{e \tau}^* ) e^{ - i \delta} = 
\mp \st \frac{ \delta m^2_{31}}{a}+  \vert \tilde{\varepsilon}_{e \tau} \vert  e^{ - i \hat{\phi}_{e \tau} }, 
\nonumber \\ 
\bar{\Xi} &\equiv& 
\left( - \ci \si \frac{\mn}{a}+ c_{23} \varepsilon_{e \mu}^* - s_{23} \varepsilon_{e \tau}^* \right) e^{ - i \delta} = 
- \ci \si \frac{\mn}{a} e^{ - i \delta} + \vert \tilde{\varepsilon}_{e \mu} \vert e^{ - i \hat{\phi}_{e \mu} }, 
\label{def-Theta-Xi-bar}
\end{eqnarray}
where $\hat{\phi}_{e \alpha} \equiv  \delta + \tilde{\phi}_{e \alpha} $ 
($\alpha = \mu, \tau$). 
The particular dependence on NSI elements in (\ref{def-Theta-Xi}) and 
(\ref{def-Theta-Xi-bar}) has root in the form of the perturbed Hamiltonian 
(\ref{tilde-H1NSI}) 
in the tilde-basis, from which it can be understood that 
$\tilde{\varepsilon}_{e \mu}$ and $\tilde{\varepsilon}_{e \tau}$ play 
the role of the mixing angles 
which govern 1-2 and 1-3 transitions, respectively.
At the second equality in the right-hand-side of these equations we have 
introduced a new notation $\Delta m^2_{31} = \pm \delta m^2_{31}$ 
where $\pm$ sign indicates the sign of $\Delta m^2_{31}$, 
the mass hierarchy, and $\delta m^2_{31} \equiv | \mt | > 0$. 
Note that $a \equiv 2\sqrt{2} G_F N_e E > 0$. 
For convenient notation we parametrize these quantities as 
\begin{eqnarray} 
\Theta_{\pm} = \vert \Theta_{\pm} \vert e^{ i \theta_{\pm} }, 
\hspace{1cm}
\Xi = \vert \Xi \vert e^{ i \xi}. 
\nonumber \\
\bar{\Theta}_{\pm} = \vert \bar{\Theta}_{\pm} \vert e^{ i \bar{\theta}_{\pm}}, 
\hspace{1cm}
\bar{\Xi} = \vert \bar{\Xi} \vert e^{ i \bar{\xi} }. 
\label{param} 
\end{eqnarray}
To represent the oscillation probability in a compact way we define 
\begin{eqnarray}
X_{\pm} &\equiv& \biggl (\frac{ a}{\delta m^2_{31} \mp a}\biggr )^2\sin ^2\frac{\delta m^2_{31} \mp a}{4E}L, 
\nonumber \\
Y_{\pm} &\equiv&
\biggl( \frac{ a }{\delta m^2_{31} \mp a} \biggr) \sin \frac{aL}{4E}\sin \frac{\delta m^2_{31} \mp a}{4E}L, 
\nonumber \\
Z &\equiv& \sin ^2\frac{aL}{4E}. 
\label{def-XYZ}
\end{eqnarray}
%
For anti-neutrinos we have flipped sign of $a$, and hence 
\begin{eqnarray}
\bar{X}_{\pm} &=& X_{\mp}, 
\nonumber \\
\bar{Y}_{\pm} &=& Y_{\mp}. 
\end{eqnarray}
$Z$ is obviously invariant under the sign change of $a$, $\bar{Z} = Z$. 

With these notations and by defining 
$\Delta_{31} \equiv \frac{\Delta m_{31}^2 L}{4 E}$
for simplicity of expressions, the oscillation probabilities 
$P(\nu _e \rightarrow \nu _e )$, 
$P(\nu _e \rightarrow \nu _\mu )$ and 
$P(\nu _e \rightarrow \nu _\tau )$ 
(together with the anti-neutrino counterparts of the latter two) 
can be written as
%
\begin{eqnarray} 
P(\nu _e \rightarrow \nu _e ) &=& 1 - 
4 X_{\pm} \vert \Theta_{\pm} \vert^2 - 
4 Z \vert \Xi \vert^2 
\label{Pee} \\
P(\nu _e \rightarrow \nu _\mu ) &=& 
4\sn^2 X_{\pm} \vert \Theta_{\pm} \vert^2 + 
4\cn^2 Z \vert \Xi \vert^2 + 
8\cn \sn Y_{\pm}
\vert \Xi \vert \vert \Theta_{\pm} \vert \cos(\xi - \theta_{\pm} - \vert \Delta_{31} \vert)
\label{Pemu} \\
P(\nu _e \rightarrow \nu _\tau ) &=& 
4 \cn^2 X_{\pm} \vert \Theta_{\pm} \vert^2 + 
4 \sn^2 Z \vert \Xi \vert^2 - 
8\cn \sn Y_{\pm}
\vert \Xi \vert \vert \Theta_{\pm} \vert \cos(\xi - \theta_{\pm} - \vert \Delta_{31} \vert)
\label{Petau} \\
P(\nu _\mu \rightarrow \nu _e ) &=& 
T[ P(\nu _e \rightarrow \nu _\mu ) ] 
\nonumber \\
&=&
4\sn^2 X_{\pm} \vert \Theta_{\pm} \vert^2 + 
4\cn^2 Z \vert \Xi \vert^2 + 
8\cn \sn Y_{\pm}
\vert \Xi \vert \vert \Theta_{\pm} \vert \cos(\xi - \theta_{\pm} + \vert \Delta_{31} \vert)
\label{Pmue} \\
P(\bar{\nu} _e \rightarrow \bar{\nu} _\mu ) &=& 
\text{CP} [ P(\nu _e \rightarrow \nu _\mu ) ] 
\nonumber \\
&=&
4\sn^2 X_{\mp} \vert \bar{\Theta}_{\pm} \vert^2 + 
4\cn^2 Z \vert \bar{ \Xi } \vert^2 + 
8\cn \sn Y_{\mp}
\vert \bar{\Xi} \vert \vert \bar{\Theta}_{\pm} \vert \cos(\bar{\xi} - \bar{\theta}_{\pm} - \vert \Delta_{31} \vert)
\label{Pemu-bar} \\
P(\bar{\nu} _e \rightarrow \bar{\nu} _\tau ) &=& 
\text{CP} [ P(\nu _e \rightarrow \nu _\tau ) ] 
\nonumber \\
&=&
4 \cn^2 X_{\mp} \vert \bar{\Theta}_{\pm} \vert^2 + 
4 \sn^2 Z \vert \bar{ \Xi } \vert^2 - 
8\cn \sn Y_{\mp}
\vert \bar{\Xi} \vert \vert \bar{\Theta}_{\pm} \vert \cos(\bar{\xi} - \bar{\theta}_{\pm} - \vert \Delta_{31} \vert) 
\label{Petau-bar} \\
P(\bar{\nu} _\mu \rightarrow \bar{\nu} _e ) &=& 
\text{T} [ P(\bar{\nu} _e \rightarrow \bar{\nu} _\mu ) ] = 
\text{TCP} [ P(\nu _e \rightarrow \nu _\mu ) ] 
\nonumber \\
&=&
4\sn^2 X_{\mp} \vert \bar{\Theta}_{\pm} \vert^2 + 
4\cn^2 Z \vert \bar{ \Xi } \vert^2 + 
8\cn \sn Y_{\mp}
\vert \bar{\Xi} \vert \vert \bar{\Theta}_{\pm} \vert \cos(\bar{\xi} - \bar{\theta}_{\pm} + \vert \Delta_{31} \vert) 
\label{Pmue-bar} 
%
\end{eqnarray}
%
The upper and the lower signs in the above equations 
are for the normal and the inverted hierarchies, respectively. 
The expression of $P(\nu _e\rightarrow \nu _e)$ is so simple because 
of the unitarity, 
$P(\nu _e\rightarrow \nu _e) = 1 - 
[ P(\nu _e \rightarrow \nu _\mu ) + P(\nu _e \rightarrow \nu _\tau ) ]$.

Notice that to second order in $\epsilon$, the oscillation probabilities in 
the $\nu_{e}$ related sector do not contain any NSI elements in 
the $\nu_{\mu} - \nu_{\tau}$ sector, $ \varepsilon_{\mu \tau}$ etc. 
It should not come as a surprise because in the $\nu_{\mu}$ 
and $\nu_{\tau}$ appearance channel from $\nu_e$ the leading term of 
the $S$ matrix is already of order $\epsilon$, and it can contain only the 
$\nu_{e}$ related NSI elements, 
$\varepsilon_{e\mu }$ and $\varepsilon_{e\tau }$.  
Therefore, to order $\epsilon^2$ there is no room for NSI elements in the 
$\nu_{\mu} - \nu_{\tau}$ sector in the appearance probabilities.  
We will see in the next subsection that this simple fact leads to a 
great simplification of the oscillation probabilities in the 
$\nu_{\mu} - \nu_{\tau}$ sector.

We note, in passing, that because of the relation 
$Y_{\pm} = \sqrt{ X_{\pm} Z }$ which is easily recognized by (\ref{def-XYZ}) 
it is evident that the oscillation probabilities 
can be written in a form of absolute square of addition of the atmospheric 
and the solar terms.\footnote{
We thank Stephen Parke for calling our attention to this point.
}
For example, 
$P(\nu _e \rightarrow \nu _\mu ) $ takes the form 
%
\begin{eqnarray} 
P(\nu _e \rightarrow \nu _\mu ) &=& 4 
\biggl | 
\sn \sqrt{ X_{\pm} } \vert \Theta_{\pm} \vert + 
\cn \sqrt{ Z } \vert \Xi \vert \text{exp} \left[ i \left( \xi - \theta_{\pm} - \vert \Delta_{31} \vert
\right) \right] \biggr |^2. 
\label{Pemu-squared} 
\end{eqnarray}
%
At the magic baseline, $\frac{ aL }{ 4\pi } = \pi$, the second term vanishes 
because $Z=0$, leaving a very simple expression of the oscillation probability, 
$ P(\nu _e \rightarrow \nu _\mu ) = 4 \sn^2 X_{\pm} \vert \Theta_{\pm} \vert^2$.

\subsection{ $\nu_{\mu} - \nu_{\tau}$ sector}
\label{mu-tau-sector}

As in the $\nu_e$-related sector there is a distinct characteristic feature of 
the oscillation probabilities with NSI in the $\nu_{\mu} - \nu_{\tau}$ sector. 
Namely, to second order in $\epsilon$ they can be decomposed into 
the three pieces with different dependences on NSI elements, 
the vacuum term, the ones with $\varepsilon_{\alpha \beta}$ in the 
$\nu_e$-related and the $\nu_{\mu} - \nu_{\tau}$ sectors, respectively:  
\begin{eqnarray} 
P(\nu _\alpha \rightarrow \nu _\beta; 
\varepsilon_{e \mu }, \varepsilon_{e \tau }, \varepsilon_{\mu \mu }, \varepsilon_{\mu \tau }, \varepsilon_{\tau \tau }) &=& 
P(\nu _\alpha \rightarrow \nu _\beta; \text{2 flavor in vacuum}) 
\nonumber \\
&+& 
P(\nu _\alpha \rightarrow \nu _\beta; 
\varepsilon_{e \mu }, \varepsilon_{e \tau }) 
\nonumber \\
&+& 
P(\nu _\alpha \rightarrow \nu _\beta; 
\varepsilon_{\mu \mu }, \varepsilon_{\mu \tau }, \varepsilon_{\tau \tau }) 
\label{Pmutau-decomposition}
\end{eqnarray}
where $\alpha$ and $\beta$ denote one of $\mu$ and $\tau$. 
The explicit expressions of these terms will be displayed in 
Appendix \ref{2nd-order-formula}.

The point is that the last term in (\ref{Pmutau-decomposition}) is universal, 
up to sign, among all the three channels, 
$P(\nu _\mu \rightarrow \nu _\mu)$, 
$P(\nu _\mu \rightarrow \nu _\tau) $, and 
$P(\nu _\tau \rightarrow \nu _\tau) $. 
Though it may look mysterious, it is in fact very simple to understand it. 
By unitarity it follows that 
\begin{eqnarray}
P(\nu _\mu \rightarrow \nu _\mu) + P(\nu _\mu \rightarrow \nu _\tau) &=& 
1 - P(\nu _\mu \rightarrow \nu _e), 
\nonumber \\
P(\nu _\tau \rightarrow \nu _\tau) + P(\nu _\tau \rightarrow \nu _\mu) &=& 
1 - P(\nu _\tau \rightarrow \nu _e). 
\label{unitarity}
\end{eqnarray}
We note that $P(\nu _\mu \rightarrow \nu _e)$ and 
$P(\nu _\tau \rightarrow \nu _e)$ do not contain 
$\varepsilon_{\mu \mu }$, $\varepsilon_{\tau \tau }$, and 
$\varepsilon_{\mu \tau }$ to second order in $\epsilon$. 
Then, it follows from the first equation in (\ref{unitarity}) that 
$P(\nu _\mu \rightarrow \nu _\tau; \varepsilon_{\mu \mu }, \varepsilon_{\mu \tau }, \varepsilon_{\tau \tau }) = - 
P(\nu _\mu \rightarrow \nu _\mu; \varepsilon_{\mu \mu }, \varepsilon_{\mu \tau }, \varepsilon_{\tau \tau })$. 
Noticing that the terms related to 
$\varepsilon$'s in the $\nu_{\mu} - \nu_{\tau}$ sector are T-invariant, 
the relations 
$P(\nu _\tau \rightarrow \nu _\tau; \varepsilon_{\mu \mu }, \varepsilon_{\mu \tau }, \varepsilon_{\tau \tau }) = - 
P(\nu _\mu \rightarrow \nu _\tau; \varepsilon_{\mu \mu }, \varepsilon_{\mu \tau }, \varepsilon_{\tau \tau }) $ must also hold. 
Therefore, the $\varepsilon_{\alpha \beta}$ ($\alpha, \beta = \mu, \tau$)
dependent term in the three channels are all equal up to sign. 
The equality 
$P(\nu _\mu \rightarrow \nu _\mu; \varepsilon_{\mu \mu }, \varepsilon_{\mu \tau }, \varepsilon_{\tau \tau }) = 
P(\nu _\tau \rightarrow \nu _\tau; \varepsilon_{\mu \mu }, \varepsilon_{\mu \tau }, \varepsilon_{\tau \tau }) $ 
also follows from the relationship between $S$ 
matrix elements due to the extended transformation (\ref{transformation}).


\section{Parameter determination in neutrino oscillation with NSI}
\label{determination}

Thanks to the NSI second-order probability formulas derived in 
the previous section, we can now address the question of how 
simultaneous measurement of the SI and the NSI 
parameters can be carried out. 
However, we must first warn the readers that our discussions in this section 
are based solely on the NSI second-order formulas, 
and hence its validity may be limited. 
Nonetheless, we believe that ignoring the $\epsilon^3$ effects 
is quite safe because we anticipate $\epsilon \sim 10^{-2}$ 
in our perturbative framework.

\subsection{SI-NSI confusion} 
\label{confusion}

One of the most distinctive features of the oscillation probability formulas 
in Sec.~\ref{formula} is that the NSI parameters $\varepsilon_{e \alpha}$ 
($\alpha = \mu, \tau)$ appears in the particular combination with 
the SI parameters as in (\ref{def-Theta-Xi}) and (\ref{def-Theta-Xi-bar}). 
What that means in the context of parameter determination? 
It means that, in general, determination of SI mixing parameters, 
$\theta_{13}$ and $\delta$, has severe confusion with determination 
of NSI parameters $\varepsilon_{\alpha \beta}$, and vice versa. 
However, it should be noticed that it does {\em not} mean something like 
``No Go'' theorem. 
Namely, there is a way to circumvent this problem. 
It is a complete determination of the SI and the NSI parameters, 
the possibility we address later in this section.

Nonetheless, we should note the following: 
If such complete determination is somehow not feasible 
experimentally, our result may be interpreted as an analytic proof of 
the ``NSI-SI confusion theorem''.\footnote{
We note that a different type of the confusion theorem was derived 
in \cite{confusion2} which involves $\theta_{13}$ and NSI parameters 
in production and in propagation processes that obey a special relationship. 
}
It is a powerful statement because it not only reveals the existence of 
confusion but also illuminates which SI parameters are confused with 
which NSI parameters via which manner.

In fact, the characteristic feature in (\ref{def-Theta-Xi}), namely, 
$\tilde{\varepsilon}_{e \mu}$ only couples to the solar scale oscillation 
and $\tilde{\varepsilon}_{e \tau}$ the atmospheric one, would affect 
the resolution of the $\theta_{13}$-NSI and the two-phase confusions. 
Coupling between the solar and the atmospheric degrees of 
freedom bridged by a NSI element 
is the key to the resolution of the $\theta_{13}$-NSI confusion by the 
two-detector method \cite{NSI-nufact}. 
Therefore, the resolution mechanism might be affected by the simultaneous 
presence of two $\varepsilon$'s, which ``decouples'' the solar and 
the atmospheric degrees of freedom. 
This point deserves a careful investigation.

\subsection{Strategy for parameter determination}
\label{strategy}

To gain a hint of how we can proceed let us look at Table.~\ref{order}. 
We first note that it is not possible to detect the effects of 
$\varepsilon_{ee}$ because it is of third order in all channels, 
and hence we have to omit it from our subsequent discussions.\footnote{
If we take the setting with only $\varepsilon_{ee}$ as NSI,    
it can be regarded as uncertainty in the matter density and it is known that 
neutrino factory has a great sensitivity to it \cite{mina-uchi,gandhi-winter}. 
However, in our current setting the issue of matter density uncertainty is 
much more severe and universal;  It produces uncertainties in determining 
all the NSI elements.    
Clearly, the discussion of this point is beyond the scope of the present paper. 
}
%
It is also well known and is obvious from the probability formulas 
in Appendix~\ref{2nd-order-formula} that 
$\varepsilon_{\mu \mu}$ and 
$\varepsilon_{\tau \tau}$ come in through the form 
$\varepsilon_{\mu \mu} - \varepsilon_{\tau \tau}$ 
and therefore only their difference is measurable.

Next, we observe that in $\nu_{\mu}$ and $\nu_{\tau}$ appearance 
channels from $\nu_e$, only the $\nu_{e}$ related NSI, 
$\varepsilon_{e \mu}$ and $\varepsilon_{e \tau}$ appear to second 
order in $\epsilon$. 
Therefore, the obvious strategy is to use these channels for complete 
determination of them simultaneously with $\theta_{13}$ and $\delta$. 
Then, we may be able to determine the rest of the NSI parameters in the 
$\nu_{\mu} - \nu_{\tau}$ sector by disappearance and appearance 
measurement in that sector.

The important point is therefore that one can explore the 
effects of $\varepsilon_{e \mu}$ and $\varepsilon_{e \tau}$ in 
$\nu_{e}$ related channels while ignoring 
$\varepsilon_{\mu \tau}$, $\varepsilon_{\mu \mu}$, and 
$\varepsilon_{\tau \tau}$. 
It is a good news because the appearance channels, assuming 
excellent detection capability of $\nu_{\mu}$ and $\nu_{\tau}$, 
have great potential of detecting the effects of NSI \cite{NSI-nufact}. 
Once $\varepsilon_{e \mu}$ and $\varepsilon_{e \tau}$ are measured 
one can proceed to determine the rest of the NSI elements 
$\varepsilon_{\mu \tau}$, $\varepsilon_{\mu \mu}$, and 
$\varepsilon_{\tau \tau}$ using the oscillation probabilities 
in the $\nu_{\mu} - \nu_{\tau}$ sector.\footnote{
If $\theta_{23}$ is deviated significantly from the maximal so that 
$\cos 2 \theta_{23} \gg \epsilon$, then the terms with 
$\varepsilon_{\mu \mu} - \varepsilon_{\tau \tau}$ can have sizes of 
order $\epsilon$. 
In this case, it may be possible to detect the effects of 
$\varepsilon_{\mu \mu} - \varepsilon_{\tau \tau}$ and measure 
(or constrain) it even without having a priori knowledges of 
$\varepsilon_{e \mu}$ and $\varepsilon_{e \tau}$. 
}

\subsection{Complete measurement of the SI and the NSI parameters; 
$\theta_{13}$, $\delta$, $\varepsilon_{e \mu}$ and $\varepsilon_{e \tau}$ }
\label{complete}

Now, we start to formulate a recipe for complete determination of the 
SI and the NSI parameters. 
Based on consideration in the previous subsection, we concentrate on 
$P(\nu _e \rightarrow \nu _\mu )$ and $P(\nu _e \rightarrow \nu _\tau )$ 
and their CP and T conjugates. 
By looking into the expressions of oscillation probabilities in 
(\ref{Pemu}) and (\ref{Petau}) 
(and other related ones which will be given below) 
one notices that the observable quantities are of the forms 
\begin{eqnarray} 
&& \vert \Theta_{\pm} \vert^2, \vert \Xi \vert^2, 
\xi - \theta_{\pm}, \text{ in neutrino sector, and}
\nonumber \\ 
&& \vert \bar{\Theta}_{\pm} \vert^2, \vert \bar{\Xi} \vert^2, 
\bar{\xi} - \bar{\theta}_{\pm}, \text{ in antineutrino sector.}
\label{observable}
\end{eqnarray} 
where the phase $\xi$, $\theta_{\pm}$, etc are defined in (\ref{param}). 
There are altogether six quantities.

Suppose now that somehow we were able to determine all these quantities. 
We discuss in the following subsections how it can be done. 
Here, we show how they determine the SI and the NSI parameters, 
$s_{13}$, $\delta$, $\vert \tilde{\varepsilon}_{e \mu} \vert$, 
$\vert \tilde{\varepsilon}_{e \tau} \vert$, $\phi_{e \mu}$, and $\phi_{e \tau}$.
It may be sufficient, assuming that the inversion is possible, to express 
the observable in terms of the physical parameters. 
We start with the neutrino sector: 
\begin{eqnarray} 
\vert \Theta_{\pm} \vert^2 &=& 
s^2_{13} \left( \frac{\delta m^2_{31}}{a} \right)^2 + 
\vert \tilde{\varepsilon}_{e \tau} \vert^2 \pm 
2 s_{13} \vert \tilde{\varepsilon}_{e \tau} \vert 
\left( \frac{\delta m^2_{31}}{a} \right) 
\cos \hat{\phi}_{e \tau}, 
\nonumber \\ 
\vert \Xi \vert^2 &=& 
\left( c_{12} s_{12} \frac{\Delta m^2_{21}}{a} \right)^2 + 
\vert \tilde{\varepsilon}_{e \mu} \vert^2 + 
2 c_{12} s_{12} \vert \tilde{\varepsilon}_{e \mu} \vert 
\frac{\Delta m^2_{21}}{a} 
\cos (\delta - \hat{\phi}_{e \mu} ). 
\label{Theta-Xi} 
\end{eqnarray}
For phase difference we obtain 
%
\begin{eqnarray} 
&& \frac{ \Theta_{\pm}^* \Xi }{ \Theta_{\pm} \Xi^* } = 
e^{ 2i ( \xi - \theta_{\pm} ) } = 
\frac{ 1 } { \vert \Theta_{\pm} \vert^2 \vert \Xi \vert^2 } 
\nonumber \\ 
&\times&
\left[ 
s^2_{13} \left( \frac{\delta m^2_{31}}{a} \right)^2 
\Bigl\{
\left( c_{12} s_{12} \frac{\Delta m^2_{21}}{a} \right)^2 e^{2i \delta} + 
\vert \tilde{\varepsilon}_{e \mu} \vert^2 e^{ 2i \hat{\phi}_{e \mu} } + 
2 c_{12} s_{12} \vert \tilde{\varepsilon}_{e \mu} \vert 
\frac{\Delta m^2_{21}}{a} 
e^{ i (\delta + \hat{\phi}_{e \mu}) } 
\Bigr\}
\right.
\nonumber \\
&&\hspace*{-4mm} {} +
\left.
\vert \tilde{\varepsilon}_{e \tau} \vert^2 
\Bigl\{
\left( c_{12} s_{12} \frac{\Delta m^2_{21}}{a} \right)^2 e^{2i (\delta - \hat{\phi}_{e \tau} ) } + 
\vert \tilde{\varepsilon}_{e \mu} \vert^2 e^{ 2i (\hat{\phi}_{e \mu} - \hat{\phi}_{e \tau} ) } + 
2 c_{12} s_{12} \vert \tilde{\varepsilon}_{e \mu} \vert 
\frac{\Delta m^2_{21}}{a} 
e^{ i (\delta + \hat{\phi}_{e \mu} - 2\hat{\phi}_{e \tau} ) } 
\Bigr\} 
\right.
\nonumber \\
&&\hspace*{-12mm} {} \pm
\left.
2 s_{13} \vert \tilde{\varepsilon}_{e \tau} \vert 
\left( \frac{\delta m^2_{31}}{a} \right) 
\Bigl\{
\left( c_{12} s_{12} \frac{\Delta m^2_{21}}{a} \right)^2 e^{ i (2 \delta - \hat{\phi}_{e \tau} ) } + 
\vert \tilde{\varepsilon}_{e \mu} \vert^2 e^{ i (2 \hat{\phi}_{e \mu} - \hat{\phi}_{e \tau} ) } + 
2 c_{12} s_{12} \vert \tilde{\varepsilon}_{e \mu} \vert 
\frac{\Delta m^2_{21}}{a} 
e^{ i (\delta + \hat{\phi}_{e \mu} - \hat{\phi}_{e \tau} ) } 
\Bigr\} 
\right]. 
\nonumber \\
\label{phase-diff} 
\end{eqnarray}
%
By taking the real and the imaginary parts of (\ref{phase-diff}) one can 
obtain $ \cos 2( \xi - \theta_{\pm} )$ and $\sin 2( \xi - \theta_{\pm} )$, 
respectively. 
%
For antineutrinos we obtain 
\begin{eqnarray} 
\vert \bar{\Theta}_{\pm} \vert^2 &=& 
s^2_{13} \left( \frac{\delta m^2_{31}}{a} \right)^2 + 
\vert \tilde{\varepsilon}_{e \tau} \vert^2 \mp 
2 s_{13} \vert \tilde{\varepsilon}_{e \tau} \vert 
\left( \frac{\delta m^2_{31}}{a} \right) 
\cos \hat{\phi}_{e \tau}, 
\nonumber \\ 
\vert \bar{\Xi} \vert^2 &=& 
\left( c_{12} s_{12} \frac{\Delta m^2_{21}}{a} \right)^2 + 
\vert \tilde{\varepsilon}_{e \mu} \vert^2 - 
2 c_{12} s_{12} \vert \tilde{\varepsilon}_{e \mu} \vert 
\frac{\Delta m^2_{21}}{a} 
\cos (\delta - \hat{\phi}_{e \mu} ). 
\label{Theta-Xi-bar} 
\end{eqnarray}
Similarly, the equation for the phase difference $\bar{\theta}_{\pm} - \bar{\xi}$ 
similar to (\ref{phase-diff}) can be obtained by 
making the transformation 
$a \rightarrow -a$, 
$\delta  \rightarrow -\delta$, 
$\hat{\phi}_{e \mu}   \rightarrow -\hat{\phi}_{e \mu} $, 
$\hat{\phi}_{e \tau}   \rightarrow -\hat{\phi}_{e \tau} $ 
in (\ref{phase-diff}).

Having the six equations altogether with given six observable, 
$\vert \Theta_{\pm} \vert$, 
$\vert \bar{\Theta}_{\pm} \vert$, 
$\vert \Xi \vert$, $\vert \bar{\Xi} \vert$, 
$ \bar{\xi} - \bar{\theta}_{\pm} $, and 
$ \xi - \theta_{\pm} $, 
they can be solved for the six unknowns, 
$s_{13}$, $\delta$, 
two complex numbers $\tilde{\varepsilon}_{e \mu}$, and 
$\tilde{\varepsilon}_{e \tau}$. 
Given the latter two numbers one can determine the original 
$\varepsilon_{e \mu}$ and $\varepsilon_{e \tau}$. 
Therefore, the rest of the problem in simultaneous determination of 
the SI and the NSI parameters is how to measure the above six observable.

\subsection{Measurement with a monochromatic neutrino beam; $\nu_e$ sector} 
\label{mono-energetic}

In this subsection, we discuss a way of determining the SI-NSI 
combined parameters in (\ref{observable}) by assuming a set of 
measurement at an energy $E$, aiming at their complete determination.\footnote{
It was proposed that such a monochromatic neutrino beam can be 
prepared for $\nu_{e}$ and $\bar{\nu}_{e}$ beams \cite{sato,bernabeu}. 
}
%
Though it might not be a practical way, by describing a concrete 
method we try to illuminate characteristic 
features of the problem of complete determination. 
With the six unknowns we have to prepare neutrino oscillation 
measurement of six different channels.

Suppose that one measures the following six probabilities 
at a neutrino energy $E$, 
$P(\nu _e \rightarrow \nu _\mu ) $, 
$P(\nu _e \rightarrow \nu _\tau ) $, 
$P(\nu _\mu \rightarrow \nu _e ) = \text{T} [ P(\nu _e \rightarrow \nu _\mu ) ]$, 
$P(\bar{\nu} _e \rightarrow \bar{\nu} _\mu ) = 
\text{CP} [ P(\nu _e \rightarrow \nu _\mu ) ] $, 
$P(\bar{\nu} _e \rightarrow \bar{\nu} _\tau ) = 
\text{CP} [ P(\nu _e \rightarrow \nu _\tau ) ] $, 
$P(\bar{\nu} _\mu \rightarrow \bar{\nu} _e ) = 
\text{T} [ P(\bar{\nu} _e \rightarrow \bar{\nu} _\mu ) ] $. 
Notice that we have intensionally avoided to use the channels 
which require $\nu_{\tau}$ beam which, if not impossible, 
would be very difficult to prepare. 
From (\ref{Pemu}), (\ref{Petau}), and (\ref{Pmue}), it is easy to obtain 
\begin{eqnarray} 
P(\nu _e \rightarrow \nu _\mu ) + P(\nu _\mu \rightarrow \nu_e ) 
&=&
8 \sn^2 X_{\pm} \vert \Theta_{\pm} \vert^2 + 
8 \cn^2 Z \vert \Xi \vert^2
\nonumber \\
&+& 
16 \cn \sn Y_{\pm}
\vert \Xi \vert \vert \Theta_{\pm} \vert \cos(\xi - \theta_{\pm}) \cos\vert \Delta_{31} \vert
\label{emu+} \\
P(\nu _e \rightarrow \nu _\mu ) - P(\nu _\mu \rightarrow \nu_e ) 
&=&
16 c_{23} s_{23} Y_{\pm} 
\vert \Xi \vert \vert \Theta_{\pm} \vert \sin(\xi - \theta_{\pm}) \sin\vert \Delta_{31} \vert
\label{emu-} \\
P(\nu _e \rightarrow \nu _\mu ) + P(\nu _e \rightarrow \nu_\tau ) 
&=&
4 X_{\pm} \vert \Theta_{\pm} \vert^2 + 
4 Z \vert \Xi \vert^2 
\label{etau+}
\end{eqnarray}
Similarly, for the antineutrino channels, we obtain from (\ref{Pemu-bar}), (\ref{Petau-bar}), and (\ref{Pmue-bar}), 
\begin{eqnarray} 
P(\bar{\nu} _e \rightarrow \bar{\nu} _\mu ) + P(\bar{\nu} _\mu \rightarrow \bar{\nu}_e ) 
&=&
8 \sn^2 X_{\mp} \vert \bar{\Theta}_{\pm} \vert^2 +
8 \cn^2 Z \vert \bar{\Xi} \vert^2
\nonumber \\
&+& 
16 \cn \sn Y_{\mp}
\vert \bar{\Xi} \vert \vert \bar{\Theta}_{\pm} \vert \cos(\bar{\xi} - \bar{\theta}_{\pm}) \cos\vert \Delta_{31} \vert
\label{emubar+} \\
P(\bar{\nu} _e \rightarrow \bar{\nu} _\mu ) - P(\bar{\nu} _\mu \rightarrow \bar{\nu}_e ) 
&=&
16 c_{23} s_{23} Y_{\mp}
\vert \bar{\Xi} \vert \vert \bar{\Theta}_{\pm} \vert \sin(\bar{\xi} - \bar{\theta}_{\pm}) \sin\vert \Delta_{31} \vert
\label{emubar-} \\
P(\bar{\nu} _e \rightarrow \bar{\nu} _\mu ) + P(\bar{\nu} _e \rightarrow \bar{\nu}_\tau ) 
&=&
4 X_{\mp} \vert \bar{\Theta}_{\pm} \vert^2 + 
4 Z \vert \bar{\Xi} \vert^2 
\label{etaubar+} 
\end{eqnarray}
It is easy to solve these equations to obtain 
$\vert \Theta_{\pm} \vert$, 
$\vert \Xi \vert$, and
$(\xi - \theta_{\pm})$
(for neutrinos), and 
$\vert \bar{\Theta}_{\pm} \vert$, 
$\vert \bar{\Xi} \vert$, and 
$(\bar{\xi} -\bar{\theta}_{\pm})$ (for antineutrinos). 

It may be obvious that the above analysis can be converted to the 
rate only analysis by replacing the probabilities 
$P(\nu _\alpha \rightarrow \nu _\beta )$ by energy integrated number 
of events with fluxes and cross sections 
$\int dE F_{\alpha} \sigma_{\nu N} P(\nu _\alpha \rightarrow \nu _\beta )$, 
and the similar integrated quantities of $X_{\pm}$ etc.

\subsection{Determining the NSI parameters in the $\nu_\mu - \nu_\tau$ sector}
\label{NSI-mutau}

After measurement of $\theta_{13}$, $\delta$, and $\varepsilon_{e \mu}$ 
and $\varepsilon_{e \tau}$ as described in the previous subsection, 
one can proceed to determination of the NSI parameters in the 
$\nu_{\mu} - \nu_{\tau}$ sector with (for concreteness) mono-energetic beam. 
As we saw in Sec.~\ref{mu-tau-sector} the $\varepsilon_{\mu \mu}$ and $\varepsilon_{\mu \tau}$ dependent term in the oscillation probabilities 
is universal in 
$P(\nu_{\mu} \rightarrow \nu_{\mu})$, 
$P(\nu_{\mu} \rightarrow \nu_{\tau})$, and 
$P(\nu_{\tau} \rightarrow \nu_{\tau})$. 
Therefore, one can simply use one of the above three channels, 
which means that $\tau$ neutrino beam, even if it were prepared, 
does not help.

The oscillation probability $P(\nu_{\mu} \rightarrow \nu_{\mu})$ 
derived in Sec.~\ref{formula} can be written as 
%
\begin{eqnarray}
&& \hspace{-20mm}
P(\nu _\mu \rightarrow \nu _\mu; 
\varepsilon_{e \mu }, \varepsilon_{e \tau }, \varepsilon_{\mu \mu }, \varepsilon_{\mu \tau }, \varepsilon_{\tau \tau }) = 
P(\nu _\mu \rightarrow \nu _\mu; 
\varepsilon_{e \mu }, \varepsilon_{e \tau }) 
\nonumber \\
&+& 
\mathcal{D}^{(0)}_{\pm} (\varepsilon_{\mu \mu }-\varepsilon_{\tau \tau } ) + 
\mathcal{R}^{(0)}_{\pm} \text{Re} (\varepsilon_{\mu \tau }) + 
\mathcal{D}^{(1)} (\varepsilon_{\mu \mu} - \varepsilon_{\tau \tau }) +
\mathcal{R}^{(1)} \text{Re} (\varepsilon_{\mu \tau })
\nonumber \\
&+& 
\mathcal{S}^{(0)} (\varepsilon_{\mu \mu }-\varepsilon_{\tau \tau } )^2 + 
\mathcal{W}^{(0)} (\varepsilon_{\mu \mu }-\varepsilon_{\tau \tau } )\text{Re} (\varepsilon_{\mu \tau }) + 
\mathcal{Q}^{(0)} \text{Re} (\varepsilon_{\mu \tau })^2 + 
\mathcal{I}^{(0)} \text{Im} (\varepsilon_{\mu \tau })^2 
\label{Pmumu2}
\end{eqnarray}
%
where the explicit form of the coefficients can be easily read off from 
the expressions in Appendix~\ref{2nd-order-formula}
and we have the similar expression for antineutrinos. 
We have obtained two equations for the three unknowns, 
$\varepsilon_{\mu \mu }-\varepsilon_{\tau \tau }$, 
$\text{Re} (\varepsilon_{\mu \tau })$, and 
$\text{Im} (\varepsilon_{\mu \tau })$. 
Clearly we need one more equation to determine the three unknowns, 
which is unavailable under the current setting. 
Thus, we have to conclude that a complete determination of the 
NSI elements in the $\nu_{\mu} - \nu_{\tau}$ sector is not possible by 
measurement at a monochromatic beam or the rate only analysis.

\subsection{Necessity of spectrum analysis} 
\label{spectrum}

Doing measurement at six different channels is {\em not} the unique way 
of carrying out complete determination of the six parameters. 
Even in the case where only the ``golden channel'', 
$ P(\nu _e \rightarrow \nu _\mu ) $ and 
$P(\bar{\nu} _e \rightarrow \bar{\nu} _\mu )$, 
is available, one can in principle determine 
$\vert \Theta_{\pm} \vert^2$, 
$\vert \Xi \vert^2$, $\xi - \theta_{\pm}$, 
and their antineutrino counterparts by spectrum analysis. 
It is because the energy and baseline dependences of the coefficients of 
these quantities in the oscillation probabilities in (\ref{Pemu}) and 
(\ref{Pemu-bar}) are different with each other. 
In the $\nu_{\mu} - \nu_{\tau}$ sector all the NSI elements cannot be 
determined by the rate only analysis, and need for the spectrum information 
is mandatory in this sector.

It appears that one of the most promising ways to carry this out 
is the two-detector method \cite{MNplb97}. 
It has been applied to the Tokai-to-Kamioka-Korea (T2KK) 
two-detector complex which receives neutrino beam from 
J-PARC \cite{T2KK1st,T2KK2nd,T2KK3rdWS}.\footnote{
Other possibility would be the one called the ``on axis wide-band beam 
approach'' which was proposed in a concrete form in the project 
description for Brookhaven National Laboratory \cite{BNL}. 
Precise estimation of the potential in doing spectrum analysis, however, 
depends upon which kind of detector is chosen and the actual 
performance of the detector. 
}
%
In the context of neutrino parameter determination in neutrino factory 
with NSI as well as SI, 
this method was examined in detail in \cite{NSI-nufact}.\footnote{
See \cite{kopp3} for effects of the systematic errors and optimization of 
the similar two-detector setting in parameter determination in neutrino factory. 
}

\subsection{Parameter degeneracy; Old and new}
\label{degeneracy}

\subsubsection{NSI-enriched conventional type degeneracy}

The parameter degeneracy is the problem of multiple solutions in 
determination of lepton mixing parameters \cite{intrinsicD,MNjhep01,octant}. 
It is known to be a notorious problem for their precision measurement. 
See \cite{BMW02,MNP2} for a global overview of the degeneracy, and 
\cite{MNnpps02,MNP2} for pictorial representation. 

We give evidences that the phenomenon has an extension to the system 
with NSI.\footnote{
Notice that introduction of NSI parameters leads to a new solution 
of the solar neutrino problem \cite{NSI-solar}. 
}
%
Our discussion based on the matter perturbation theory in 
Sec.~\ref{matter-perturbation} indicates that the parameter degeneracy 
prevails in system with NSI but with new form which involve NSI parameters. 
Set of equations for observable we have derived in Secs.~\ref{complete} 
and \ref{mono-energetic} shows that the sign-$\Delta m^2_{31}$ and 
the $\theta_{23}$ octant degeneracies exist because the equations take 
different form for different mass hierarchies and octant for a given set of 
observable. 
It is also very likely that the intrinsic-type degeneracy survives with a 
NSI-enriched form, as one can see in the bi-probability diagram 
\cite{MNjhep01} given in Fig.~2 of \cite{NSI-nufact}.

\subsubsection{New type of degeneracy}

Here, we present a completely new type of parameter degeneracy 
which may be called as the ``atmospheric-solar variable exchange'' degeneracy. 
We work with the setting of measurement of six channels at a 
monochromatic energy. 
First of all, one notices that determination of the neutrino (un-barred) 
and the antineutrino (over-barred) variables decouples with each other. 
We discuss only the neutrino variables below because the antineutrino 
ones is so similar. 
To simplify the expressions we restrict ourselves to the case of maximal 
$\theta_{23}$.
By combining (\ref{emu+}), (\ref{emu-}), and (\ref{etau+}) it is easy to show 
that the phase variable can be determined as 
\begin{eqnarray} 
\tan(\xi - \theta_{\pm}) = 
\cot \vert \Delta_{31} \vert 
\frac{ P_{e \mu}  - P_{\mu e}  }{ P_{\mu e}  - P_{e \tau}  }. 
\label{xi-theta} 
\end{eqnarray}
where we have used a simplified notation 
$P_{\alpha \beta} \equiv P(\nu_\alpha \rightarrow \nu_\beta )$.

It is easy to show that if the mass hierarchy is known the solution of 
this equation is unique in the physical region $-\pi \leq  \xi \leq \pi$ and 
$-\pi \leq \theta_{\pm} \leq \pi$. 
Then, the solutions for $\vert \Theta_{\pm} \vert$ and $\vert \Xi \vert$ are given by 
\begin{eqnarray} 
\vert \Theta_{\pm} \vert^2 &=& 
\frac{ P_{e \mu}  + P_{e \tau}  }{ 8 X_{\pm} } 
\left[ 
1 \pm \sqrt{ 1 - \frac{ 1 }{ \sin^2 (\xi - \theta_{\pm}) \sin^2 \Delta_{31} } 
\left(  \frac{ P_{e \mu}  - P_{\mu e}  }{ P_{e \mu}  + P_{e \tau}  }  \right)^2 }
\right], 
\nonumber \\
\vert \Xi \vert^2 &=& 
\frac{ P_{e \mu}  + P_{e \tau}  }{ 8 Z } 
\left[ 
1 \mp \sqrt{ 1 - \frac{ 1 }{ \sin^2 (\xi - \theta_{\pm}) \sin^2 \Delta_{31} } 
\left(  \frac{ P_{e \mu}  - P_{\mu e}  }{ P_{e \mu}  + P_{e \tau}  }  \right)^2 }
\right]. 
\label{exchange-sol} 
\end{eqnarray}
Notice that the degeneracy is quite new; 
It is the solar-atmospheric variable exchange degeneracy. 
That is, if there is a solution 
$\vert \Theta_{\pm}^{(1)} \vert$ and $\vert \Xi^{(1)} \vert$, then 
the second solution 
$\vert \Theta_{\pm}^{(2)} \vert = \sqrt{  \frac{ Z }{ X_{\pm} } } \vert \Xi^{(1)} \vert$ and 
$\vert \Xi^{(2)} \vert =  \sqrt{  \frac{ X_{\pm} }{ Z } }  \vert \Theta_{\pm}^{(1)} \vert$  
exists. 
Notice that the new degeneracy does not survive when NSI is switched off 
where $\xi = \delta$ and $\theta_{\pm}=0$. 
Namely, there is no phase degree of freedom in the atmospheric variable 
in the limit, while only phase degree of freedom exists in the solar variable.

Now we turn to the sign-$\Delta m^2$ degeneracy. 
At first sight there is no sign-$\Delta m^2$ degeneracy because 
the sign-$\Delta m^2$ flipped solution of $\xi - \theta_{\pm}$ has to 
satisfy the same equation (\ref{xi-theta}) which has no explicit 
dependence on the sign. 
Nevertheless, there is indeed a sign-$\Delta m^2$ flipped solution. 
If $\xi^{(1)}$ and $\theta_{+}^{(1)}$ are the solution to (\ref{xi-theta}) 
then there are another solutions 
$\xi^{(2)} = \xi^{(1)} \pm \pi$ and 
$\theta_{-}^{(2)} = \theta_{+}^{(1)} \mp \pi$. 
It means the existence of the sign-flipped solution of $\Theta$ and $\Xi$, 
which can be another  solutions if accompanied by 
$(\Delta m^2_{31})^{(2)} = - (\Delta m^2_{31})^{(1)} $.  
With these solutions of the phase equation there exist the similar 
degenerate solutions as in (\ref{exchange-sol}). 
Again the sign-$\Delta m^2$ degeneracy does not survive in the no 
NSI limit because of no degrees of freedom of $\theta_{\pm}$ in the limit. 
In conclusion we have uncovered new degeneracies of the intrinsic 
and the sign-$\Delta m^2$ flipped type which exist as a consequence 
of the presence of NSI.

\section{Matter perturbation theory with NSI}
\label{matter-perturbation}

As a first step toward understanding the degeneracy we examine 
neutrino oscillation with NSI by matter perturbation theory following 
the treatment in \cite{T2KK2nd}. 
It is known \cite{MNjhep01} that structure of parameter 
degeneracy is particularly transparent in the region where the 
matter effect can be treated as a perturbation, as explicitly 
verified in the analyses in \cite{T2KK1st,T2KK2nd}. 
See \cite{HQL06} for further explanation of this point.

For simplicity, we restrict our discussion to $\nu_{e}$ related 
appearance measurement in this section. 
In concordance to these works we consider $\nu_{e}$ and 
$\bar{\nu}_{e}$ appearance measurement 
with conventional muon neutrino beam and its antiparticles.

\subsection{Structure of the oscillation probability with NSI 
in matter perturbation theory}
\label{probability-mattP}

If we restrict ourselves into the first order in $a$, the matter effect coefficient, 
the only terms that survive are the ones up to first order in 
$\varepsilon_{e \mu} $ or $\varepsilon_{e \tau} $. 
%
The oscillation probability in $\nu_{\mu} \to \nu_{e}$ channel is given 
to first order in matter perturbation theory as 
\begin{eqnarray}
P( \nu_{\mu} \to \nu_{e}; \varepsilon_{e \tau}, \varepsilon_{e \mu} ) &=& 
P( \nu_{\mu} \to \nu_{e}; \varepsilon=0)_{AKS} 
\nonumber \\
&+& 
P( \nu_{\mu} \to \nu_{e}; \varepsilon_{e \tau} )_{NSI} + 
P( \nu_{\mu} \to \nu_{e}; \varepsilon_{e \mu} )_{NSI}, 
\label{Pmue-mattP}
\end{eqnarray}
where the leading term is the Arafune-Koike-Sato (AKS) formula without 
NSI \cite{AKS}\footnote{ 
We got rid of a higher order $\epsilon^3$ term which was kept 
in our previous references, e.g., \cite{MNjhep01,T2KK1st,T2KK2nd}. 
}
%
\begin{eqnarray}
P( \nu_{\mu} \rightarrow \nu_{\rm e}; \varepsilon = 0 )_{AKS} &=& 
\sin^2{2\theta_{13}} s^2_{23} \sin^2 \Delta_{31} + 
c^2_{23} \sin^2{2\theta_{12}} 
\left( \frac{ \Delta m^2_{21} }{ \Delta m^2_{31} } \right)^2 
\Delta_{31}^2 
\nonumber \\
&+& 
4 J_{r} \left( \frac{ \Delta m^2_{21} }{ \Delta m^2_{31} } \right)
\Delta_{31}
\left[
\cos{\delta}
\sin 2 \Delta_{31} - 
2 \sin{\delta}
\sin^2 \Delta_{31} 
\right] 
\nonumber \\
&+& 
2 \sin^2{2\theta_{13}} s^2_{23} 
\left( \frac{aL}{4 E} \right)
\left[
\frac{1}{ \Delta_{31} }
\sin^2 \Delta_{31} 
- \frac{1}{2} \sin 2 \Delta_{31} 
\right]. 
\label{AKS}
\end{eqnarray}
In (\ref{AKS}), 
$\Delta_{31} \equiv \frac{ \Delta m^2_{31} L} {4 E} $
$a\equiv 2 \sqrt{2} G_{F} N_{e} E $ as before. 
$J_r$ $(\equiv c_{12} s_{12} c_{13}^2 s_{13} c_{23} s_{23} )$ 
denotes the reduced Jarlskog factor.

The first order matter corrections which include the first order NSI effects 
in $\varepsilon$'s can be obtained by taking the first order term in $a$ as 
%
\begin{eqnarray}
&&P( \nu_{\mu} \to \nu_{e}; \varepsilon_{e \tau} )_{NSI} = 8 \left( \frac{aL}{4E} \right) 
\nonumber \\
&\times& 
%
\left[
c_{23} s^2_{23} s_{13} 
\left\{ 
\vert \varepsilon_{e \tau} \vert \cos (\delta + \phi_{e \tau} )
\left( 
\frac{ \sin^2 \Delta_{31} }{ \Delta_{31} } - \frac{1}{2} \sin 2\Delta_{31}
\right) 
+ \vert \varepsilon_{e \tau} \vert \sin (\delta + \phi_{e \tau} )
\sin^2 \Delta_{31} 
\right\}
\right.
\nonumber \\
&&\hspace*{0mm} {} -
\left.
c_{12} s_{12} c^2_{23} s_{23} 
\frac{ \Delta m^2_{21} }{ \Delta m^2_{31} } 
\left\{
\vert \varepsilon_{e \tau} \vert \cos \phi_{e \tau} 
\left( \Delta_{31} - \frac{1}{2} \sin 2\Delta_{31} \right) 
- 
\vert \varepsilon_{e \tau} \vert \sin \phi_{e \tau} 
\sin^2 \Delta_{31} 
\right\} 
\right], 
\label{P-NSI-etau}
\end{eqnarray}
%
%
%
\begin{eqnarray}
&&P( \nu_{\mu} \to \nu_{e}; \varepsilon_{e \mu} )_{NSI} = - 8 \left( \frac{aL}{4E} \right) 
\nonumber \\
&\times& 
\left[
s_{23} s_{13} 
\left\{ 
\vert \varepsilon_{e \mu} \vert \cos (\delta + \phi_{e \mu} )
\left( 
s^2_{23} \frac{ \sin^2 \Delta_{31} }{ \Delta_{31} } - 
\frac{c^2_{23} }{2} \sin 2\Delta_{31}
\right) 
+ c^2_{23} \vert \varepsilon_{e \mu} \vert \sin (\delta + \phi_{e \mu} )
\sin^2 \Delta_{31} 
\right\}
\right.
\nonumber \\
&&\hspace*{-6mm} {} -
\left.
c_{12} s_{12} c_{23} 
\frac{ \Delta m^2_{21} }{ \Delta m^2_{31} } 
\left\{
\vert \varepsilon_{e \mu} \vert \cos \phi_{e \mu} 
\left( c^2_{23} \Delta_{31} + \frac{s^2_{23} }{2} \sin 2\Delta_{31} \right) 
+ 
s^2_{23} 
\vert \varepsilon_{e \mu} \vert \sin \phi_{e \mu} \sin^2 \Delta_{31} 
\right\} 
\right]. 
\label{P-NSI-emu}
\end{eqnarray}
%
The antineutrino probability 
$ P( \bar{\nu}_{\mu} \rightarrow \bar{\nu}_{\rm e}; \varepsilon_{e \tau}, \varepsilon_{e \mu} )$
can be obtained by making the replacement in (\ref{Pmue-mattP}); 
$a \rightarrow - a$, $\delta \rightarrow 2\pi - \delta$. 
$\varepsilon_{\alpha \beta} \rightarrow \varepsilon_{\alpha \beta}^*$. 
Notice that both of the CP violating leptonic KM phase $\delta$ 
and $\phi_{\alpha \beta}$ due to NSI elements changes sign 
when we discuss the time reversal process $\nu_{\mu} \to \nu_{e} $, 
as opposed to $\nu_{e} \to \nu_{\mu} $ in the previous sections.

For the oscillation probabilities in the $\nu_{\mu} - \nu_{\tau} $ sector 
we only deal with the one in $\nu_\mu$ disappearance channel 
(which may be easiest to measure) to first order in $\epsilon$:
%
\begin{eqnarray}
&& \hspace{-6mm}
P(\nu _\mu \rightarrow \nu _\mu; \text{1st order in}~\epsilon ) 
\nonumber \\
&=& 1- 4\cn ^2\sn ^2\sin ^2 \Delta_{31} 
+ 4 \ci ^2 \cn ^2\sn ^2 \biggl ( \frac{ \Delta m^2_{21} }{ \Delta m^2_{31} } \biggr ) \Delta_{31} \sin 2 \Delta_{31} 
\nonumber \\
&+& 2\cn ^2\sn ^2 \biggl [ (\cn ^2-\sn ^2)(\varepsilon_{\mu \mu }-\varepsilon_{\tau \tau } ) - 4 \cn \sn \text{Re} ( \varepsilon_{\mu \tau } ) \biggr] \frac{aL}{2E} \sin 2 \Delta_{31} 
\nonumber \\
&-& 8\cn \sn (\cn ^2-\sn ^2) \biggl [\cn \sn (\varepsilon_{\mu \mu }-\varepsilon_{\tau \tau })+ (\cn ^2 - \sn ^2 ) \text{Re} ( \varepsilon_{\mu \tau } ) \biggr ] \frac{a}{\mt}\sin ^2 \Delta_{31}. 
\label{Pmumu-1st}
\end{eqnarray}
%
Notice that (\ref{Pmumu-1st}) is already in the form of first-order formula 
in matter perturbation theory.

\subsection{Sign-$\Delta m^2$ and $\theta_{23}$ octant degeneracies prevail in the presence of NSI}

In this subsection, we discuss the fate of the sign-$\Delta m^2$ and 
the $\theta_{23}$ octant degeneracies in the presence of NSI. 
In the conventional cases without NSI, 
they are known as notorious ones among the three types of degeneracies 
because they are hard to resolve and the former can confuse CP violation 
with CP conservation. 
The sign-$\Delta m^2$ degeneracy was uncovered in systems without 
NSI by noticing that the oscillation probability $P( \nu_{\mu} \to \nu_{e})$ 
in vacuum is invariant under the transformation 
$\Delta m^2_{31} \rightarrow - \Delta m^2_{31}$, 
$\delta \rightarrow \pi - \delta$ without changing $\theta_{13}$ 
\cite{MNjhep01}.
It maps a positive $\Delta m^2_{31}$ solution to the negative one, 
and vice versa. 
The presence of the symmetry as well as the fact that it is broken 
by the first order matter terms can be seen in (\ref{AKS}).

Now, we observe that the sign-$\Delta m^2$ degeneracy prevails in 
the presence of NSI. 
That is, the NSI induced terms in the probability
(\ref{P-NSI-etau}) and (\ref{P-NSI-emu}), though they are ``matter terms'', 
are invariant under the extended transformation 
\begin{eqnarray}
\Delta m^2_{31} &\rightarrow& - \Delta m^2_{31}, 
\nonumber \\ 
\delta &\rightarrow& \pi - \delta, 
\nonumber \\ 
\phi_{e \alpha} &\rightarrow& 2\pi - \phi_{e \alpha}. 
\label{extended-symmetry}
\end{eqnarray}
while keeping $\theta_{13}$ and $\vert \varepsilon_{e \alpha} \vert$ 
fixed, where $\alpha = \mu, \tau$.\footnote{
Under the transformation (\ref {extended-symmetry}), 
the trigonometric factors in (\ref{P-NSI-emu}) and (\ref{P-NSI-etau}) 
transform as follows: 
$\cos (\delta + \phi_{e \alpha}) \rightarrow - \cos (\delta + \phi_{e \alpha}) $, 
$\sin (\delta + \phi_{e \alpha}) \rightarrow + \sin (\delta + \phi_{e \alpha}) $, 
$\cos \phi_{e \alpha} \rightarrow + \cos \phi_{e \alpha} $, and 
$\sin \phi_{e \alpha} \rightarrow - \sin \phi_{e \alpha} $. 
}
%
The symmetry is broken only by the matter term in (\ref{AKS}) 
which is independent of NSI; 
The symmetry is broken by the matter effect which has exactly the 
same magnitude in systems with and without NSI.
Therefore, to first order in matter perturbation theory, the sign-$\Delta m^2$ 
degeneracy exists in systems with NSI to the same extent as it 
does in the system without NSI. 
Given the robustness of the sign-$\Delta m^2$ degeneracy in the 
conventional case we suspect that the degeneracy in systems 
with NSI has the similar robustness.

Similarly, one can easily show that the $\theta_{23}$ octant degeneracy 
survives the presence of NSI. 
It can be readily observed that 
$P(\nu _\mu \rightarrow \nu _\mu; \text{1st order in}~\epsilon ) $ 
in (\ref{Pmumu-1st}) is invariant under the transformation 
\begin{eqnarray}
c_{23} &\rightarrow& s_{23}, 
\nonumber \\ 
s_{23} &\rightarrow& c_{23}, 
\nonumber \\ 
(\varepsilon_{\mu \mu }-\varepsilon_{\tau \tau }) &\rightarrow& 
- (\varepsilon_{\mu \mu }-\varepsilon_{\tau \tau }). 
\label{extended-symmetry2} 
\end{eqnarray}
It means that the $\theta_{23}$ octant degeneracy prevails in the 
presence of NSI, and actually in an extended form which involves 
NSI parameter $\varepsilon_{\mu \mu }-\varepsilon_{\tau \tau }$. 
Since this NSI parameter decouples from $P( \nu_{\mu} \to \nu_{e} )$ 
to second-order in $\epsilon$, the presence of the $\theta_{23}$ 
octant degeneracy remains intact when the NSI is included though values 
of the degenerate solutions themselves are affected by the presence of 
$\varepsilon_{e \alpha }$.

It is interesting to note that both of the two degeneracies discussed 
in this subsection have common features. 
Their presence can be discussed based on (approximate) invariance 
under some discrete transformations, and with NSI the transformations 
are extended to the ones which involve NSI parameters. 
Most probably, our treatment here is the first one to signal the existence 
of the degenerate solutions which involves both the SI 
($\theta_{13}$ and $\delta$) and the NSI parameters.

\subsection{Decoupling between the degeneracies in the presence of NSI}
\label{decoupling}

In \cite{T2KK2nd} the property called ``decoupling between degeneracies'' 
are shown to exist for experimental settings with baseline shorter 
than $\sim1000$ km which may allow treatment based on matter 
perturbation theory. 
See also \cite{resolve23} and \cite{SNOW-mina} for preliminary 
discussions. 
The property of decoupling between degeneracies A and B 
guarantees that when one tries to resolve the degeneracy A 
one can forget about the presence of the degeneracy B, 
and vice versa. 
Existence of NSI terms, in general, influences the discussion of decoupling. 
It is the purpose of this and the next subsections to fully discuss the fate 
of the decoupling in the presence of NSI.
Since it is one of the most significant characteristic features of the 
degeneracies in matter perturbative regime, we believe it worth to present 
a complete treatment.

\subsubsection{Definition of decoupling between degeneracies}

To define the concept of decoupling between degeneracies A and B, 
we introduce, following \cite{T2KK2nd}, the probability difference 
\begin{eqnarray}
\Delta P^{ab}(\nu_{\alpha} \rightarrow \nu_{\beta}) 
&\equiv& 
P \left( \nu_{\alpha} \rightarrow \nu_{\beta}; (\Delta m^2_{31})^{(a)}, \theta_{23}^{(a)}, \theta_{13}^{(a)}, \delta^{(a)}, 
\varepsilon_{\alpha \beta}^{(a)} \right) 
\nonumber \\
&-& 
P \left( \nu_{\alpha} \rightarrow \nu_{\beta}; (\Delta m^2_{31})^{(b)}, \theta_{23}^{(b)}, \theta_{13}^{(b)}, \delta^{(b)}, 
\varepsilon_{\alpha \beta}^{(b)} \right), 
\label{DeltaPdef}
\end{eqnarray}
where the superscripts $a$ and $b$ label the degenerate solutions. 
Suppose that we are discussing the degeneracy A. 
The decoupling between the degeneracies A and B 
holds if $\Delta P^{ab}$ defined in (\ref{DeltaPdef}) for the degeneracy A 
is invariant under the replacement of the mixing parameters 
corresponding to the degeneracy B, and vice versa.

\subsubsection{Matter-perturbative treatment of the degenerate solutions}
\label{perturbative-degene}

We follow \cite{T2KK2nd} to define the degenerate solutions in 
a perturbative manner.\footnote{
More precise meaning of the term ``perturbative'' is as follows: 
Since the disappearance probability by which $\theta_{23}$ is 
determined is of order unity we disregard quantities of order 
$\epsilon$ or higher. 
They include the matter effect, $\theta_{13}$, and NSI. 
Similarly, $\nu_{e}$ appearance probability is of order 
$\epsilon^2$ the relationship between the two degenerate solution 
inevitably contains a small quantity, which is $\theta_{13}$ in this case.
But, all the quantities of higher order are neglected. 
If the near-far two detectors are involved, like in the case of T2KK 
\cite{T2KK1st,T2KK2nd}, the degenerate solutions are essentially 
defined by the near detector. In this case, the second detector is meant 
to give raise to perturbation effect to lift the degeneracy. 
For more concrete example of this feature, see \cite{T2KK2nd}.
}
Throughout the discussion in this section we assume that 
deviation of $\theta_{23}$ from the maximal angle $\pi/4$ is small. 
A disappearance measurement, $\nu_{\mu} \rightarrow \nu_{\mu}$, 
determines $s^2_{23}$ to first order in $s^2_{13}$ as 
$(s^2_{23})^{(1)}= (s^2_{23})^{(0)} (1+ s^2_{13})$, 
where 
$(s^2_{23})^{(0)}$ is the solution obtained by ignoring $s^2_{13}$. 
It is given by 
$(s^2_{23})^{(0)}= \frac{1}{2} \left[ 1 \pm \sqrt{1- \sin^2{2\theta_{23}}} \right]$. 
In leading order the relationship between the first and the second octant solutions of $\theta_{23}$ is given by 
$s_{23}^{{\text 1st}} = c_{23}^{{\text 2nd} } $. 

A $\nu_{e}$ appearance measurement determines the 
combination $s^2_{23} \sin^2 2\theta_{13}$. 
The first and the second octant solutions of $\theta_{23}$ are also 
related to leading order by 
$s_{23}^{{\text 1st}} s_{13}^{{\text 1st}} = 
s_{23}^{{\text 2nd}} s_{13}^{{\text 2nd}} $. 
In an environment where the vacuum oscillation approximation applies the 
solutions corresponding to the intrinsic degeneracy are given in 
Appendix~\ref{intrinsic-vac} as
\begin{eqnarray} 
\theta_{13}^{(2)} &=& \sqrt { (\theta_{13}^{(1)})^2 + 
2 \left( \frac{Y_{c}}{X} \right) 
\theta_{13}^{(1)} \cos \delta_{1} + 
\left( \frac{Y_{c}}{X} \right)^2 } 
\nonumber \\
\sin \delta_{2} &=& \frac{ \theta_{13}^{(1)} }{ \theta_{13}^{(2)} } \sin \delta_{1} 
\nonumber \\
%
\cos \delta_{2} &=& \mp \frac{ 1 }{ \theta_{13}^{(2)} } 
\left(
\theta_{13}^{(1)} \cos \delta_{1} + \frac{Y_{c}}{X} 
\right)
\label{intrinsic-solution}
\end{eqnarray}
where 
\begin{eqnarray} 
\frac{Y_{c}}{X} \equiv 
\sin 2\theta_{12} \cot \theta_{23} \Delta_{21} \cot \Delta_{31}. 
\label{Y/X-def}
\end{eqnarray}
and the superscripts (1) and (2) label the solutions due to the 
intrinsic degeneracy. 
The sign $\mp$ for $\cos \delta_{2}$ are for $Y_{c} = \pm |Y_{c}|$, 
and $\theta_{13}^{(2)} $ in the solution of $\delta$ is meant to be the 
$\theta_{13}^{(2)} $ solution given in the first line in (\ref{intrinsic-solution}).

As we saw in the previous section, an extended form of the 
sign-$\Delta m^2$ degeneracy is given under the same 
approximation (mod. $2\pi$) as 
\begin{eqnarray}
\theta_{13}^{\,\text{norm}} = \theta_{13}^{\,\text{inv}}, 
\hspace{0.3cm}
(\Delta m^2_{31})^{\,\text{norm}} = - (\Delta m^2_{31})^{\,\text{inv}}, 
\hspace{0.3cm}
\delta^{\,\text{norm}} = \pi - \delta^{\,\text{inv}}, 
\hspace{0.3cm}
(\phi_{\alpha \beta})^{\,\text{norm}} = - (\phi_{\alpha \beta})^{\,\text{inv}}, 
\label{signdm2-dgen-sol}
\end{eqnarray}
where the superscripts ``norm'' and ``inv'' label the solutions with 
the positive and the negative sign of $\Delta m^2_{31}$, and 
$\phi_{\alpha \beta}$ denotes the phase of 
$\varepsilon_{\alpha \beta}$. 
The validity of these approximate relationships in the actual experimental 
setup in the T2K II measurement is explicitly verified in \cite{T2KK1st,T2KK2nd}. 
It should be noticed that even if sizable matter effect is present 
the relation (\ref{signdm2-dgen-sol}) holds in a good approximation 
if the energy is tuned to the one corresponding to the vacuum oscillation 
maximum, or more precisely, the shrunk ellipse limit \cite{KMN02}.

\subsection{Decoupling between the sign-$\Delta m^2$ and 
the $\theta_{23}$ octant degeneracies}
\label{sign-octant}

Let us start by treating the sign-$\Delta m^2$ degeneracy. 
For this purpose, we calculate 
$\Delta P^{\,\text{norm~inv}} (\nu_{\mu} \rightarrow \nu_{e}) $
as defined in (\ref{DeltaPdef}). 
Thanks to the extended symmetry (\ref{extended-symmetry}) 
obeyed by the appearance probability, 
it is given by the same result obtained without NSI in \cite{T2KK2nd}: 
\begin{eqnarray}
&&\Delta P^{\,\text{norm~inv}} (\nu_{\mu} \rightarrow \nu_{e}) 
\nonumber \\
&=&
\sin^2{2\theta_{13}^{\,\text{norm}}} (s_{23}^{\,\text{norm}})^2
\left( \frac{aL}{E} \right)
\left[ 
\frac{1}{ (\Delta_{31})^{\,\text{norm}} }
\sin^2 (\Delta_{31})^{\,\text{norm}} - 
\frac{1}{2} \sin 2(\Delta_{31})^{\,\text{norm}} 
\right] 
\label{DeltaP_sign}
\end{eqnarray}
where the superscripts ``norm'' and ``inv'' can be exchanged 
if one want to start from the inverted hierarchy. 
Therefore, breaking the sign-$\Delta m^2$ degeneracy 
requires the matter effect but not more than that required 
in resolving it in systems without NSI; 
NSI does not contribute resolution of the sign-$\Delta m^2$ 
degeneracy but it does not add more difficulties. 

By following the same discussion as in \cite{T2KK2nd}, 
we observe that $\Delta P^{\,\,\text{norm~inv}}$
is invariant under the transformation 
$\theta_{23}^{\,\text{1st}} \leftrightarrow \theta_{23}^{\,\text{2nd}}$ and 
$\theta_{13}^{\,\text{1st}} \leftrightarrow \theta_{13}^{\,\text{2nd}}$, 
because $\Delta P^{\,\,\text{norm~inv}}$ depends upon 
$\theta_{13}$ and $\theta_{23}$ only through the combination 
$\sin^2{2\theta_{13}} s_{23}^{2}$ within our approximation. 
Therefore, resolution of the sign-$\Delta m^2_{31}$ can be done 
in the presence of the $\theta_{23}$ octant degeneracy.

What is the influence of the $\nu _\mu$ disappearance channel in 
the discussion of decoupling? 
Using the first-order formula in (\ref{Pmumu-1st}), 
$\Delta P^{\,\text{norm~inv}} (\nu _\mu \rightarrow \nu _\mu) $ can be 
computed as
%
\begin{eqnarray}
&& \hspace{-8mm} 
\Delta P^{\,\text{norm~inv}} (\nu _\mu \rightarrow \nu _\mu) = 
8 \ci ^2 \cn ^2\sn ^2 \biggl ( \frac{ \Delta m^2_{21} L}{4 E} \biggr ) \sin \frac{\mt L}{2E} 
\nonumber \\
&+& 4\cn ^2\sn ^2 \biggl [ (\cn ^2-\sn ^2)(\varepsilon_{\mu \mu }-\varepsilon_{\tau \tau } ) - 4 \cn \sn \text{Re} ( \varepsilon_{\mu \tau } ) \biggr] \frac{aL}{2E}\sin \frac{\mt L}{2E}. 
\label{DeltaP-Pmumu-1st}
\end{eqnarray}
%
It is manifestly invariant under that the transformation in 
(\ref{extended-symmetry2}), and hence the sign-$\Delta m^2_{31}$ degeneracy 
decouples from the $\theta_{23}$ octant degeneracy. 
Presence of the $\Delta P^{\,\text{norm~inv}} (\nu _\mu \rightarrow \nu _\mu)$ 
in first order in $\epsilon$ indicates that 
the $\nu_{\mu}$ disappearance channel would play a role in lifting 
the sign-$\Delta m^2_{31}$ degeneracy 
if the measurement is done off the vacuum oscillation maximum. 

Now, we discuss the inverse problem, namely, 
whether the $\theta_{23}$ octant degeneracy can be resolved 
in the presence of the sign-$\Delta m^2_{31}$ degeneracy. 
By noting that $J_r^{\,\text{1st}} - J_r^{\,\text{2nd}} 
= \cos 2\theta_{23}^{\,\text{1st}} J_r^{\,\text{1st}}$ 
in leading order in $\cos{2\theta_{23}}$, 
the difference between probabilities with the first and the second octant 
solutions can be given by 
\begin{eqnarray}
&&\Delta P^{\,\text{1st 2nd}}(\nu_{\mu} \rightarrow \nu_{e}) 
\nonumber \\
&=&
\cos{2\theta_{23}^{\,\text{1st}}} 
\Delta_{21}
\left[ 
\sin^2{2\theta_{12}} 
\Delta_{21} +
4J_{r}^{\,\text{1st}} 
\left( 
\cos{\delta}
\sin 2 \Delta_{31} - 
2 \sin{\delta}
\sin^2 \Delta_{31} 
\right)
\right] 
\nonumber \\
&+& 
\Delta P^{\,\text{1st 2nd}}(\nu_{\mu} \rightarrow \nu_{e}; 
\varepsilon_{e \tau})_{NSI} + 
\Delta P^{\,\text{1st 2nd}}(\nu_{\mu} \rightarrow \nu_{e}; 
\varepsilon_{e \mu})_{NSI}, 
\label{DeltaP_23}
\end{eqnarray}
where
%
\begin{eqnarray}
\Delta P^{\,\text{1st 2nd}}(\nu_{\mu} &\rightarrow& \nu_{e}; 
\varepsilon_{e \tau})_{NSI} = 
- 2 \sqrt{2} c_{12} s_{12}
\cos{2\theta_{23}^{\,\text{1st}} } \sin{2\theta_{23}^{\,\text{1st}} } 
\left( \frac{aL}{4E} \right)
\left( \frac{ \Delta m^2_{21} }{ \Delta m^2_{31} } \right) 
\nonumber \\
&\times&
\left[
\vert \varepsilon_{e \tau} \vert \cos \phi_{e \tau} 
\left( \Delta_{31} - \frac{1}{2} \sin 2\Delta_{31}
\right) 
- \vert \varepsilon_{e \tau} \vert \sin \phi_{e \tau} 
\sin^2 \Delta_{31} 
\right], 
\label{DeltaP_23-etau}
\end{eqnarray}
%
\begin{eqnarray}
&&\Delta P^{\,\text{1st 2nd}}(\nu_{\mu} \rightarrow \nu_{e}; 
\varepsilon_{e \mu})_{NSI} = 
8 \cos{2\theta_{23}^{\,\text{1st}} } 
\left( \frac{aL}{4E} \right) 
\nonumber \\
&\times& 
\left[ 
s_{23}^{\,\text{1st}} s_{13} 
\left\{ 
\vert \varepsilon_{e \mu} \vert \cos (\delta + \phi_{e \mu} )
\left( 
\frac{ \sin^2 \Delta_{31} }{ \Delta_{31} } + \frac{1}{2} \sin 2\Delta_{31}
\right) 
- \vert \varepsilon_{e \mu} \vert \sin (\delta + \phi_{e \mu} )
\sin^2 \Delta_{31} 
\right\}
\right.
\nonumber \\
&&\hspace*{-12mm} {} + 
\left.
\frac{ c_{12} s_{12} }{ 2 \sqrt{2} }
\left( \frac{ \Delta m^2_{21} }{ \Delta m^2_{31} } \right)
\left\{ 
\vert \varepsilon_{e \mu} \vert \cos \phi_{e \mu} 
\left( 3 \Delta_{31} - \frac{1}{2} \sin 2\theta_{23}^{\,\text{1st}} \sin 2\Delta_{31}
\right) 
- \vert \varepsilon_{e \mu} \vert \sin \phi_{e \mu} 
\sin 2\theta_{23}^{\,\text{1st}} \sin^2 \Delta_{31} 
\right\}
\right]. 
\nonumber \\
\label{DeltaP_23-emu}
\end{eqnarray}

The first term of 
$\Delta P^{\,\text{1st~2nd}}$ in (\ref{DeltaP_23}), being composed only of 
the vacuum oscillation terms, 
is obviously invariant under the replacement 
$normal \leftrightarrow inverted$ solutions.
The remarkable feature of (\ref{DeltaP_23-etau}) and (\ref{DeltaP_23-emu})
is that they are also invariant under the replacement relation 
between different hierarchy solutions given in 
(\ref{extended-symmetry}) 
which is extended to include NSI phases. 
%
The disappearance channel does not play a role in the present discussion 
under the approximation taken in deriving (\ref{Pmumu-1st}), 
because then $\Delta P^{\,\text{1st~2nd}} (\nu_\mu \rightarrow \nu_\mu )$ vanishes. 
Therefore, even in the presence of NSI, the resolution of the 
$\theta_{23}$ octant degeneracy can be carried out without 
worrying about the presence of the sign-$\Delta m^2_{31}$ degeneracy. 
The sign-$\Delta m^2$ and the $\theta_{23}$ octant 
degeneracies decouple with each other even in the presence of NSI 
in matter perturbative regime.

\subsection{Non-Decoupling of Intrinsic degeneracy}
\label{intrinsic}

Now we discuss the intrinsic degeneracy for which the situation 
is somewhat different.
First of all, this is the degeneracy which is somewhat different in nature. 
Unlike the case of the sign-$\Delta m^2_{31}$ degeneracy, 
this degeneracy is known to be fragile to the spectrum analysis; 
In many cases it can be resolved by including informations of energy 
dependence in the reconstructed events. 
An example for this is the T2KK setting which 
receives an intense neutrino beam from J-PARC \cite{T2KK1st,T2KK2nd}. 
It means that in this case there is no intrinsic degeneracy from the beginning. 
Nonetheless, anticipating possible circumstances in which spectrum 
informations are not available, and for completeness, 
we discuss below if resolving the intrinsic degeneracy decouple 
to lifting the other two degeneracies. 
We disregard the $\nu_\mu$ disappearance channel in this subsection 
because it does not appear to play a major role in resolving the intrinsic degeneracy. 
The discussions in this subsection are also meant to partly correct 
and append the ones given in Sec.~III in \cite{T2KK2nd}.

\subsubsection{Non-Decoupling of Intrinsic degeneracy without NSI}
\label{non-decoupling1}

Let us first discuss the problem of decoupling with intrinsic degeneracy 
without NSI. In our perturbative approach 
$\Delta P^{12}(\nu_{\mu} \rightarrow \nu_{e}) $ arises only from 
the first order matter term in (\ref{AKS}) because 
the degenerate solutions in vacuum, by definition, 
gives the same vacuum oscillation probabilities. 
It reads 
\begin{eqnarray}
&& \Delta P^{12}(\nu_{\mu} \rightarrow \nu_{e}) = - 4 
\left( \frac{ aL }{ 4E} \right) \Delta \theta^2 
\left(
\frac{1}{ \Delta_{31} }
\sin^2 \Delta_{31} 
- \frac{1}{2} \sin 2 \Delta_{31} 
\right). 
\label{DeltaP-intrinsic}
\end{eqnarray}
where 
\begin{eqnarray}
\Delta \theta^2 &\equiv& 
(\theta^{(1)}_{13})^2 - (\theta^{(2)}_{13})^2 
\nonumber \\ 
&=& - 
\sin 2\theta_{12} \cot \theta_{23} \Delta_{21} \cot \Delta_{31} 
\left(
2 \theta_{13}^{(1)} \cos \delta^{(1)} + 
\sin 2\theta_{12} \cot \theta_{23} \Delta_{21} \cot \Delta_{31} 
\right)
\label{Del-theta2-def}
\end{eqnarray}
Based on the result of $\Delta P^{12}$ in (\ref{DeltaP-intrinsic}) 
we discuss possible decoupling of the sign-$\Delta m^2$ and 
the octant $\theta_{23}$ degeneracies from the intrinsic one.

We start from the sign-$\Delta m^2$ degeneracy. 
It can be readily seen that $\Delta P^{12}$ is odd under interchange 
of the normal and the inverted hierarchy solutions as dictated 
in (\ref{signdm2-dgen-sol}). 
It means that 
$\Delta P^{12} (normal) - \Delta P^{12} (inverted) = 2 \Delta P^{12}$ .
Clearly, the sign-$\Delta m^2$ degeneracy do not decouple from 
the intrinsic one.

Now we turn to the octant $\theta_{23}$ degeneracy. 
From (\ref{DeltaP-intrinsic}), 
$\Delta P^{12} (\text{1st}) - \Delta P^{12} (\text{2nd)}$ reads 
\begin{eqnarray}
\Delta P^{12} (\text{1st}) &-& \Delta P^{12} (\text{2nd)}= 8 
\left( \frac{ aL }{ 4E} \right) 
\sin 2\theta_{12} \Delta_{21} \cot \Delta_{31} 
\left(
\frac{1}{ \Delta_{31} }
\sin^2 \Delta_{31} 
- \frac{1}{2} \sin 2 \Delta_{31} 
\right) 
\nonumber \\
& \times&
\cos 2\theta_{23} 
\left[ 
\theta_{13}^{({1})} \cos \delta^{({1})} 
\frac{ 1 + c_{23} s_{23} }{ c^2_{23} s_{23} ( c_{23} + s_{23}) } + 
\frac { 2 }{ \sin^2 2\theta_{23} } \sin 2\theta_{12} \Delta_{21} \cot \Delta_{31} 
\right] 
\label{DeltaP-diff-octant}
\end{eqnarray}
where $\theta_{13}$ and $s_{23}$ etc. in (\ref{DeltaP-diff-octant}) are 
meant to be the ones in the first octant. 
It is small in the sense that it is proportional to $\cos 2\theta_{23} $ 
which vanishes in the limit of maximal $\theta_{23}$. 
But, this is the factor of kinematical origin which inevitably exists 
because the measure for breaking of the octant degeneracy has 
to vanish at $\theta_{23} = \pi/4$. 
Therefore, we conclude that there is no dynamical decoupling 
of the $\theta_{23}$ octant degeneracy from the intrinsic one. 

Now, we discuss the inverse problem, namely, 
whether the sign-$\Delta m^2$ and the $\theta_{23}$ 
octant degeneracies can be resolved independently of the 
intrinsic degeneracy. 
The measure for resolving the sign-$\Delta m^2$ degeneracy 
is given in (\ref{DeltaP_sign}) 
\begin{eqnarray}
&&\Delta P^{\,\text{norm~inv}} (1) - \Delta P^{\,\text{norm~inv}} (2)
\nonumber \\
&=&
4 \Delta \theta^2 s^2_{23} 
\left( \frac{aL}{E} \right)
\left[ 
\frac{1}{ \Delta_{31} }
\sin^2 \Delta_{31} - 
\frac{1}{2} \sin 2\Delta_{31} 
\right] 
\biggr |_{ \text{norm} }^{(1)}
\label{DeltaP-sign-12}
\end{eqnarray}
where all the quantities in (\ref{DeltaP-sign-12}) is to be evaluated 
by using the normal hierarchy and intrinsic first solution. 
Clearly, the intrinsic degeneracy does not decouple from 
the sign-$\Delta m^2$ one.

How about the $\theta_{23}$ octant degeneracy? 
The appropriate measure for the question is given by 
\begin{eqnarray}
&&\Delta P^{\,\text{1st 2nd}}(1) - \Delta P^{\,\text{1st 2nd}}(2)
= \cos{2\theta_{23}^{\,\text{1st}}} 
\Delta_{21}
\nonumber \\
&\times&
\frac{ 1 }{ \theta_{13}^{(2)} } 
\left[ 
\Bigl\{ 4J_{r}^{\,\text{1st}} 
\left( \theta_{13}^{(1)} + \theta_{13}^{(2)} \right) 
\cos{\delta}^{(1)} + \frac{Y_{c}}{X} \Bigr\}
\sin 2 \Delta_{31} - 
2 \frac{ \sin{\delta}^{(1)} } { \theta_{13}^{(1)} + \theta_{13}^{(2)} } 
\Delta \theta^2 \sin^2 \Delta_{31} 
\right] 
\label{DeltaP-octant-12}
\end{eqnarray}
where $\theta_{13}^{(2)}$ implies to insert the expression in 
(\ref{intrinsic-solution}). 
Again there is no sign of the decoupling.

Nonetheless, there are some cases in which the decoupling 
with the intrinsic degeneracy still holds in a good approximation. 
For example, 
$\Delta P^{12} (\text{1st}) - \Delta P^{12} (\text{2nd)}$ 
in (\ref{DeltaP-diff-octant}) and 
$\Delta P^{\,\text{norm~inv}} (1) - \Delta P^{\,\text{norm~inv}}$ in 
(\ref{DeltaP-sign-12}) may be small numerically. 
It is the case at relatively short baseline $L \lsim 1000$ km 
where it is further suppressed by $\frac{ aL }{ 4E} $. 
The $\Delta P$ differences between the two $\theta_{23}$ octant 
solutions are always suppressed by $ \cos 2\theta_{23}$, and 
hence they may be small at $\theta_{23}$ very close to the maximal.

It is significant to observe that at the vacuum oscillation maxima, 
$\Delta_{31} = (2n+1) \frac{\pi}{2}$, 
the decoupling is realized in all pairs of degeneracies. 
Therefore, if the experimental set up is near the vacuum oscillation 
maxima the decoupling with the intrinsic degeneracy perfectly holds. 
The identical two detector setting in T2KK \cite{T2KK1st,T2KK2nd}, 
whose intermediate (far) detector is near the first (second) oscillation 
maximum provides a good example for such ``accidental decoupling''.

\subsubsection{Decoupling and non-decoupling of Intrinsic degeneracy with NSI}
\label{non-decoupling2}

We concisely describe what happens in the decoupling 
between the intrinsic and the other two degeneracies when 
NSI is introduced. 
We explicitly discuss below the case with $\varepsilon_{e \tau}$ 
because the equations are slightly simpler, 
but we have verified that the same conclusion holds for the case 
with $\varepsilon_{e \mu}$, and hence in the full system. 

$\varepsilon_{e \tau}$ type NSI gives rise to contribution to the 
difference of the probabilities with the first and the second solutions 
of intrinsic degeneracy of the following form 
%
\begin{eqnarray}
&&\Delta P^{12}( \nu_{\mu} \to \nu_{e}; \varepsilon_{e \tau} ) = 
8 \left( \frac{aL}{4E} \right) | \varepsilon_{e \tau}|
c_{23} s^2_{23} 
\nonumber \\
&\times& 
\left(
2 \theta_{13}^{(1)} \cos \delta^{(1)} + \frac{Y_{c}}{X} 
\right)
\left[ \cos \phi_{e \tau}
\left( 
\frac{ \sin^2 \Delta_{31} }{ \Delta_{31} } - \frac{1}{2} \sin 2\Delta_{31}
\right) 
+ \sin \phi_{e \tau} \sin^2 \Delta_{31} 
\right], 
\label{DeltaP-NSI-etau}
\end{eqnarray}
%
%
where use has been made of the relation (\ref{reduct2b}).
Notice that the terms proportional to the solar $\Delta m^2_{21}$ 
do not contribute, and $\sin \delta$ terms cancel out owing to the 
relation (\ref{reduct2a}).

We observe that 
$\Delta P^{12}(\varepsilon_{e \tau} )$ 
are invariant under interchange between the normal and 
the inverted hierarchies, (\ref{signdm2-dgen-sol}). 
Therefore, NSI induced oscillation probability, by itself, fulfills 
the decoupling condition with the sign-$\Delta m^2$ degeneracy.

The situation is different in relationship with the $\theta_{23}$ 
octant degeneracy.
%
With $\varepsilon_{e \tau}$ one can derive the similar expression as 
(\ref{DeltaP-diff-octant}): 
\begin{eqnarray}
\Delta P^{12} (\varepsilon_{e \tau}; \text{1st}) &-& \Delta P^{12} (\varepsilon_{e \tau}; \text{2nd)} = 4 \sqrt{ 2 } \cos 2\theta_{23}
\left( \frac{ aL }{ 4E} \right) | \varepsilon_{e \tau}| 
\sin 2\theta_{12} \Delta_{21} \cot \Delta_{31} 
\nonumber \\
& \times&
\left[ \cos \phi_{e \tau}
\left( 
\frac{ \sin^2 \Delta_{31} }{ \Delta_{31} } - \frac{1}{2} \sin 2\Delta_{31}
\right) 
+ \sin \phi_{e \tau} \sin^2 \Delta_{31} 
\right]. 
\label{DeltaP-diff-octant-etau}
\end{eqnarray}
Though the intrinsic degeneracy does not decouple with the $\theta_{23}$ 
octant degeneracy, the suppression factor 
$\cos 2\theta_{23}
\left( \frac{ aL }{ 4E} \right) | \varepsilon_{e \tau}| $
may be very small if baseline is relatively short and 
$\theta_{23}$ is near maximal, assuming the likely possibility that 
$| \varepsilon_{e \tau}| $ is small. 
Again, the decoupling holds at the vacuum oscillation maxima. 
%

General conclusion in the last two subsections is that although 
the decoupling between the sign-$\Delta m^2_{31}$ and the 
$\theta_{23}$ octant degeneracies holds, there is no decoupling 
between the intrinsic degeneracy and the other two types of degeneracies. 
The conclusion applies to the cases with and without NSI.

\section{Concluding remarks}

In this paper, we have discussed various aspects of neutrino oscillation 
with NSI, the exactly hold properties as well as the properties best 
illuminated by a perturbative method. 
The former category includes the relation between the $S$ matrix 
elements and the probabilities that arises due to an invariance of the 
Hamiltonian under the transformation (\ref{transformation}) 
which involves $\theta_{23}$ and the NSI elements 
$\varepsilon_{\alpha \beta}$ ($\alpha, \beta = e, \mu, \tau$). 
It allows us to connect the probabilities of various flavor conversion channels, 
which is powerful enough to strongly constrain the way how various 
NSI elements $\varepsilon_{\alpha \beta}$ 
enter into the oscillation probabilities. 
This category also includes the phase reduction theorem 
which guarantees reduction of number of CP violating phases 
when the solar $\Delta m^2_{21}$ is switched off.

By taking the following three quantities, 
$\frac{ \Delta m^2_{21} }{ \Delta m^2_{31} }$, 
$s_{13}$, and the NSI elements $\varepsilon_{\alpha \beta}$, 
as small expansion parameters 
(which are collectively denoted as $\epsilon$) 
we have formulated a perturbative framework which we have 
dubbed as the ``$\epsilon$ perturbation theory''.
Within this framework we have calculated the $S$ matrix 
elements to order $\epsilon^2$ and derived the NSI second-order 
formula of the oscillation probability in all channels. 
It allows us to estimate size of the contribution of the particular NSI 
element $\varepsilon_{\alpha \beta}$ ($\alpha, \beta = e, \mu, \tau$) to 
the particular oscillation probability 
$P(\nu_\kappa \rightarrow \nu_\omega )$ ($\kappa, \omega = e, \mu, \tau$), 
as tabulated in Table~\ref{order}. 
To complete the table (and for other reasons) we have also calculated the 
oscillation probability in the $\nu_e$ related channels to third order in $\epsilon$, 
which is given in Appendix~\ref{third-order-formula}. 
We have given a global overview of neutrino oscillation with NSI and 
hope that the table serves as a ``handbook'' for hunting NSI effects 
in neutrino propagation.

Thanks to the NSI second-order formula we have discussed, 
for the first time, the way how the SI and the NSI parameters 
can be determined simultaneously. We found that measurement 
of all the relevant NSI and SI parameters is extremely demanding; 
While all the NSI elements in $\nu_e$ related sector can in principle 
be determined together with $\theta_{13}$ and $\delta$, it requires 
$\nu_e \rightarrow \nu_\mu$, $\nu_\mu \rightarrow \nu_e$, 
$\nu_e \rightarrow \nu_\tau$, and their CP conjugate channels 
if we do it by the rate only measurement. 
We have also proven to the accuracy of 
$\epsilon^2$ that, if we restrict to the rate only analysis, 
all the NSI elements in $\nu_\mu - \nu_\tau$ sector cannot be 
determined even if we prepare $\nu_\tau$ beam.

Clearly, the right strategy is to pursue the appropriate experimental 
setup which enables us the spectrum analysis to determine several 
coefficients at the same time. 
The capability of spectrum analysis with good resolution 
would be a mandatory requirement for future facilities which aim 
at searching for effects of NSI at least as one of their objectives. 
To our knowledge, the leading candidate for such setup is the two-detector 
setup at $L \simeq 3000$ km and $L \simeq 7000$ km in neutrino 
factory with use of the golden channel \cite{golden}, 
which are proven to be powerful in resolving the conventional 
parameter degeneracy \cite{intrinsicD,huber-winter}. 
In a previous paper, it was shown that the setting is also powerful 
in resolving the $\theta_{13}$-NSI (and probably the two-phase) 
confusion \cite{NSI-nufact}. 
It must be stressed, however, that we still do not know if the setting is 
sufficiently powerful in determining all the SI and the NSI parameters.

We have observed that the phenomenon of parameter degeneracy prevails 
in the system with NSI. Notably, it exists in an extended form of 
involving not only the SI but also the NSI parameters. 
In a concrete setting of six probabilities at monochromatic beam, 
we have uncovered  a new type of degeneracy, 
the solar-atmospheric variable exchange degeneracy. 
To have a first grasp of the nature of the parameter degeneracy 
of more conventional type, we have discussed the matter perturbation 
theory of neutrino oscillation with NSI. 
We have found that the sign-$\Delta m^2_{31}$ and the 
$\theta_{23}$ octant degeneracies are robust, and the analysis 
indicates the way how the NSI parameters are involved into the 
new form of degeneracy. 
The decoupling between degeneracies, a salient feature in the 
matter perturbative regime, is also revisited in an extended setting 
with NSI.

In our investigation we have also noticed a new feature 
of neutrino oscillation in matter in the standard three-flavor oscillation 
without NSI, that is, the matter hesitation. 
It states that the matter effect comes in into the oscillation probability only 
at the second order in $\epsilon$.
The property allows us to understand why it is so difficult to detect the 
matter effect in various long-baseline experiments, and explains 
why $\varepsilon_{ee}$ is absent from the NSI second order formula. 
Notice that the property does not hold in the $\nu_\mu - \nu_\tau$ system 
with NSI.

Of course, a number of cautions have to be made to correctly interpret 
our results; 
Many of our statements are based on the NSI second order formula 
which is reliable only if the assumptions made in formulating our 
perturbative treatment are correct. 
We do not deal with effects of NSI in production and detection of neutrinos. 
The program of complete determination of the NSI parameters mentioned 
above must be cooperated with search for NSI in production and detection 
processes.

In this paper we confined the case of relatively small $\theta_{13}$ in 
accordance to our perturbative hypothesis in (\ref{def-epsilon}). 
What happens if $\theta_{13}$ is large enough so that not only $\theta_{13}$ 
but also $\delta$ are determined by the next generation reactor/accelerator 
\cite {reactor-exp,T2K,NOVA} and upgraded superbeam 
\cite{superbeam} experiments prior NSI search?
Then, one might argue that the discussion of parameter determination 
would become much less complicated in this case. 
We argue that this is not quite correct. 
As we have seen in Sec.~\ref{formula} the NSI and the SI parameters appear 
in the oscillation probability in a tightly coupled way. 
Hence, determination of the former with size of 
$\varepsilon_{\alpha, \beta} \sim 10^{-2}$ requires 
simultaneous determination of the latter with {\em accuracy} of the similar order. 
Therefore, prior determination of $\theta_{13}$ and $\delta$, 
unless extremely precise ones, would not alter the necessity of 
simultaneous determination of SI and NSI parameters. 
However, we note that knowing the neutrino mass hierarchy 
would greatly help by decreasing the ambiguities which arise from 
the degeneracy.


\appendix

\section{$S$ Matrix Elements for Neutrino Oscillation with NSI}
\label{Smatrix-element}

Using the formalism described in Sec.~\ref{perturbation} with 
the double-tilde basis (\ref{hamiltonian2}) 
it is straightforward to compute the $S$ matrix elements 
for neutrino oscillations with NSI. 
Omitting calculations we just present the results of the $S$ matrix elements: 
The notations used below are: 
$\Delta \equiv \frac{ \Delta m^2_{31} }{2E}$, 
$r_{\Delta} \equiv \frac{ \Delta m^2_{21} }{ \Delta m^2_{31} } $, 
$r_{A} \equiv \frac{ a }{ \Delta m^2_{31} } $, and 
the NSI elements are in the tilde-basis (\ref{tilde-H1NSI}). 
%
%
\begin{eqnarray}
S_{ee} &=& 
\Bigl\{
1 - i \Delta L \left( s^2_{12} r_{\Delta} + r_{A} \tilde{\varepsilon}_{e e} \right) 
\Bigr\}
e^{- i r_{A} \Delta L} 
\nonumber \\
&+& 
s^2_{13} ( i r_{A} \Delta L ) e^{- i r_{A} \Delta L} 
- s^2_{13} \frac{ 1 + r_{A} }{ 1 - r_{A} } 
\left( e^{- i r_{A} \Delta L } - e^{- i \Delta L } \right) 
\nonumber \\ 
&-& 
2 s_{13} \text{Re}(\tilde{\varepsilon}_{e \tau} e^{i\delta}) r_{A} 
\left[ 
i \Delta L e^{- i r_{A} \Delta L } + \frac{ 1 } { 1- r_{A} }
\left( e^{- i r_{A} \Delta L } - e^{- i \Delta L } \right) 
\right] 
\nonumber \\
&-& 
( s^2_{12} \frac{ r_{\Delta} }{ r_{A} } + \tilde{\varepsilon}_{e e} )^2
\frac{ ( r_{A} \Delta L )^2 } { 2 } e^{- i r_{A} \Delta L } 
\nonumber \\
&-& 
\vert c_{12} s_{12} \frac{ r_{\Delta} }{ r_{A} } + \tilde{\varepsilon}_{e \mu} \vert^2 
\Bigl\{
( i r_{A} \Delta L ) e^{- i r_{A} \Delta L } - 
\left( 1 - e^{- i r_{A} \Delta L } \right) 
\Bigr\} 
\nonumber \\
&+& 
\vert s_{13} e^{- i \delta } + \tilde{\varepsilon}_{e \tau} \vert^2
\left( \frac{ r_{A}^2 }{ 1- r_{A} } \right)
\left[ 
i \Delta L e^{- i r_{A} \Delta L } - \frac{ 1 } { 1- r_{A} }
\left( e^{- i r_{A} \Delta L } - e^{- i \Delta L } \right) 
\right] 
\label{See} 
\end{eqnarray}
%
\begin{eqnarray}
S_{e \mu} 
&=& - c_{23} 
\left( c_{12} s_{12} \frac{ r_{\Delta} }{ r_{A} } + \tilde{\varepsilon}_{e \mu} \right) 
\left( 1 - e^{- i r_{A} \Delta L } \right) 
- s_{23} \left( 
s_{13} e^{ - i \delta} + r_{A} \tilde{\varepsilon}_{e \tau} 
\right) 
\frac{1 }{ 1 - r_{A} } 
\left( e^{- i r_{A} \Delta L } - e^{- i \Delta L } \right) 
\nonumber \\
&-& c_{23} s_{13} \tilde{\varepsilon}_{\mu \tau}^* e^{ - i \delta} 
\Bigl\{ 
\left( 1 - e^{- i r_{A} \Delta L } \right) + 
r_{A} \left( 1 - e^{- i \Delta L } \right) 
\Bigr\}
\nonumber \\
&+& 
s_{23} 
s_{13} e^{ - i \delta} ( \tilde{\varepsilon}_{e e} - \tilde{\varepsilon}_{\tau \tau} ) 
\frac{r_{A} }{ 1 - r_{A} } 
\left( e^{- i r_{A} \Delta L } - e^{- i \Delta L } \right) 
\nonumber \\
&+& 
s_{23} s_{13} e^{ - i \delta} ( i \Delta L ) 
\Bigl\{ 
( s^2_{12} r_{\Delta} + \tilde{\varepsilon}_{e e} r_{A} ) 
e^{- i r_{A} \Delta L } - 
\tilde{\varepsilon}_{\tau \tau} r_{A} e^{- i \Delta L } 
\Bigr\} 
%
%
%
\nonumber \\
&+& 
c_{23} \left( c_{12} s_{12} \frac{ r_{\Delta} }{ r_{A} } + \tilde{\varepsilon}_{e \mu} \right) 
\left[ 
i r_{A} \Delta L 
\left\{ \left( c^2_{12} \frac{ r_{\Delta} }{ r_{A} } + \tilde{\varepsilon}_{\mu \mu} \right) 
- \left( s^2_{12} \frac{ r_{\Delta} }{ r_{A} } + \tilde{\varepsilon}_{e e} \right) 
e^{- i r_{A} \Delta L } 
\right\} 
\right.
\nonumber \\
& & \left. \hspace{64mm} 
- \left( ( c^2_{12} - s^2_{12} ) \frac{ r_{\Delta} }{ r_{A} } 
- \tilde{\varepsilon}_{e e} + \tilde{\varepsilon}_{\mu \mu} \right) 
\left( 1 - e^{- i r_{A} \Delta L } \right) 
\right] 
\nonumber \\ 
&+& 
s_{23} \left( s_{13} e^{- i \delta } + \tilde{\varepsilon}_{e \tau} \right) 
\left( \frac{ r_{A}^2 }{ 1 - r_{A} } \right) 
\left[ 
i \Delta L 
\left\{ \left( s^2_{12} \frac{ r_{\Delta} }{ r_{A} } + \tilde{\varepsilon}_{e e} \right) 
e^{- i r_{A} \Delta L } - \tilde{\varepsilon}_{\tau \tau} e^{- i \Delta L } \right\} 
\right.
\nonumber \\
& & \left. \hspace{58mm} 
- \frac{ 1 }{ 1 - r_{A} } \left( s^2_{12} \frac{ r_{\Delta} }{ r_{A} } + \tilde{\varepsilon}_{e e} - \tilde{\varepsilon}_{\tau \tau} \right) 
\left( e^{- i r_{A} \Delta L } - e^{- i \Delta L } \right) 
\right] 
\nonumber \\ 
&+& 
\Bigl\{ 
c_{23} \tilde{\varepsilon}_{\mu \tau}^{*} \left( s_{13} e^{- i \delta } + \tilde{\varepsilon}_{e \tau} \right) 
+ s_{23} \tilde{\varepsilon}_{\mu \tau} \left( c_{12} s_{12} \frac{ r_{\Delta} }{ r_{A} } + \tilde{\varepsilon}_{e \mu} \right) 
\Bigr\} 
\nonumber \\ 
&\times& 
r_{A} \left[ 
( 1 - e^{- i \Delta L } ) - 
\frac{ 1 }{ 1 - r_{A} } \left( e^{- i r_{A} \Delta L } - e^{- i \Delta L } \right) 
\right] 
\label{Semu} 
\end{eqnarray}
%
\begin{eqnarray}
S_{e \tau} 
&=& s_{23} 
\left( c_{12} s_{12} \frac{ r_{\Delta} }{ r_{A} } + \tilde{\varepsilon}_{e \mu} \right) 
\left( 1 - e^{- i r_{A} \Delta L } \right) 
- c_{23} \left( 
s_{13} e^{ - i \delta} + r_{A} \tilde{\varepsilon}_{e \tau} 
\right) 
\frac{1 }{ 1 - r_{A} } 
\left( e^{- i r_{A} \Delta L } - e^{- i \Delta L } \right) 
\nonumber \\
&+& 
s_{23} s_{13} \tilde{\varepsilon}_{\mu \tau}^* e^{ - i \delta} 
\Bigl\{ 
\left( 1 - e^{- i r_{A} \Delta L } \right) + 
r_{A} \left( 1 - e^{- i \Delta L } \right) 
\Bigr\}
\nonumber \\
&+& 
c_{23} 
s_{13} e^{ - i \delta} ( \tilde{\varepsilon}_{e e} - \tilde{\varepsilon}_{\tau \tau} ) 
\frac{r_{A} }{ 1 - r_{A} } 
\left( e^{- i r_{A} \Delta L } - e^{- i \Delta L } \right) 
\nonumber \\
&+& 
c_{23} s_{13} e^{ - i \delta} ( i \Delta L ) 
\Bigl\{ 
( s^2_{12} r_{\Delta} + \tilde{\varepsilon}_{e e} r_{A} ) 
e^{- i r_{A} \Delta L } - 
\tilde{\varepsilon}_{\tau \tau} r_{A} e^{- i \Delta L } 
\Bigr\} 
%
%
\nonumber \\
&+& 
s_{23} \left( c_{12} s_{12} \frac{ r_{\Delta} }{ r_{A} } + \tilde{\varepsilon}_{e \mu} \right) 
\left[ 
i r_{A} \Delta L 
\left\{ - \left( c^2_{12} \frac{ r_{\Delta} }{ r_{A} } + \tilde{\varepsilon}_{\mu \mu} \right) 
+ \left( s^2_{12} \frac{ r_{\Delta} }{ r_{A} } + \tilde{\varepsilon}_{e e} \right) 
e^{- i r_{A} \Delta L } 
\right\} 
\right.
\nonumber \\
& & \left. \hspace{64mm} 
+ \left( ( c^2_{12} - s^2_{12} ) \frac{ r_{\Delta} }{ r_{A} } 
- \tilde{\varepsilon}_{e e} + \tilde{\varepsilon}_{\mu \mu} \right) 
\left( 1 - e^{- i r_{A} \Delta L } \right) 
\right] 
\nonumber \\ 
&+& 
c_{23} \left( s_{13} e^{- i \delta } + \tilde{\varepsilon}_{e \tau} \right) 
\left( \frac{ r_{A}^2 }{ 1 - r_{A} } \right) 
\left[ 
i \Delta L 
\left\{ \left( s^2_{12} \frac{ r_{\Delta} }{ r_{A} } + \tilde{\varepsilon}_{e e} \right) 
e^{- i r_{A} \Delta L } - \tilde{\varepsilon}_{\tau \tau} e^{- i \Delta L } \right\} 
\right.
\nonumber \\
& & \left. \hspace{58mm} 
- \frac{ 1 }{ 1 - r_{A} } \left( s^2_{12} \frac{ r_{\Delta} }{ r_{A} } + \tilde{\varepsilon}_{e e} - \tilde{\varepsilon}_{\tau \tau} \right) 
\left( e^{- i r_{A} \Delta L } - e^{- i \Delta L } \right) 
\right] 
\nonumber \\ 
&+& 
\Bigl\{ 
- s_{23} \tilde{\varepsilon}_{\mu \tau}^{*} \left( s_{13} e^{- i \delta } + \tilde{\varepsilon}_{e \tau} \right) 
+ c_{23} \tilde{\varepsilon}_{\mu \tau} \left( c_{12} s_{12} \frac{ r_{\Delta} }{ r_{A} } + \tilde{\varepsilon}_{e \mu} \right) 
\Bigr\} 
\nonumber \\ 
&\times& 
r_{A} \left[ 
( 1 - e^{- i \Delta L } ) - 
\frac{ 1 }{ 1 - r_{A} } \left( e^{- i r_{A} \Delta L } - e^{- i \Delta L } \right) 
\right] 
\label{Setau} 
\end{eqnarray}
%
\begin{eqnarray}
S_{\mu \mu} 
&=& 
c^2_{23} \Bigl\{ 
1 - i ( c^2_{12} r_{\Delta} + \tilde{\varepsilon}_{\mu \mu} r_{A} ) \Delta L 
\Big\} + 
s^2_{23} 
\left( 
1 - i \tilde{\varepsilon}_{\tau \tau} 
r_{A} \Delta L 
\right) 
e^{- i \Delta L } 
\nonumber \\
&-& 
2 c_{23} s_{23} 
\text{ Re} ( \tilde{\varepsilon}_{\mu \tau} ) 
r_{A} \left( 1 - e^{- i \Delta L } \right) 
\nonumber \\
&-& s^2_{23} s^2_{13} 
\left[ 
( i r_{A} \Delta L ) e^{- i \Delta L } - 
\frac{ 1 + r_{A} }{ 1 - r_{A} } 
\left( e^{- i r_{A} \Delta L } - e^{- i \Delta L } \right) 
\right] 
\nonumber \\
&+& 2 s^2_{23} s_{13} 
\text{Re} ( \tilde{\varepsilon}_{e \tau} e^{ i \delta} ) 
\left[ 
( i r_{A} \Delta L ) e^{- i \Delta L } + 
\frac{ r_{A} }{ 1 - r_{A} } 
\left( e^{- i r_{A} \Delta L } - e^{- i \Delta L } \right) 
\right] 
\nonumber \\
&+& 
2 c_{23} s_{23} s_{13} 
\text{ Re} ( \tilde{\varepsilon}_{e \mu} e^{ i \delta} ) 
r_{A} \left( 1 - e^{- i \Delta L } \right) 
\nonumber \\
&+& 
2 c_{23} s_{23} s_{13} 
\Bigl\{ c_{12} s_{12} \cos \delta \frac{ r_{\Delta} }{ r_{A} } + 
\text{ Re} \left( \tilde{\varepsilon}_{e \mu} e^{ i \delta} \right) 
\Bigr\} 
\left( 1 - e^{- i r_{A} \Delta L } \right) 
\nonumber \\
%
%
&-& 
\left[ 
c^2_{23} (c^2_{12} \frac{ r_{\Delta} }{ r_{A} } + \tilde{\varepsilon}_{\mu \mu} )^2 + 
s^2_{23} \tilde{\varepsilon}_{\tau \tau}^2 e^{- i \Delta L } 
\right] 
\frac{ ( r_{A} \Delta L )^2 } { 2 } 
\nonumber \\
&+& c^2_{23} \left[ 
\vert c_{12} s_{12} \frac{ r_{\Delta} }{ r_{A} } + \tilde{\varepsilon}_{e \mu} \vert^2 
\Bigl\{ 
( i r_{A} \Delta L ) - \left( 1 - e^{- i r_{A} \Delta L } \right) 
\Bigr\} 
- \vert \tilde{\varepsilon}_{\mu \tau} \vert^2 r_{A}^2 \left( 1 - i \Delta L - e^{- i \Delta L } \right) 
\right] 
\nonumber \\
&-& 
s^2_{23} \left[ 
\vert s_{13} e^{- i \delta } + \tilde{\varepsilon}_{e \tau} \vert^2
\left( \frac{ r_{A}^2 }{ 1- r_{A} } \right)
\Bigl\{ 
i \Delta L e^{- i \Delta L } - \frac{ 1 } { 1- r_{A} }
\left( e^{- i r_{A} \Delta L } - e^{- i \Delta L } \right) 
\Bigr\} 
\right.
\nonumber \\
& & \left. \hspace{74mm} 
+ \vert \tilde{\varepsilon}_{\mu \tau} \vert^2 r_{A}^2 
\Bigl\{ i \Delta L e^{- i \Delta L } - ( 1 - e^{- i \Delta L } ) \Bigr\} 
\right] 
\nonumber \\
&+& 
2 c_{23} s_{23} 
{\text Re } ( \tilde{\varepsilon}_{\mu \tau} ) r_{A}^2 
\left[ i \Delta L 
\left( c^2_{12} \frac{ r_{\Delta} }{ r_{A} } + \tilde{\varepsilon}_{\mu \mu} - 
\tilde{\varepsilon}_{\tau \tau} e^{- i \Delta L } \right) 
\right.
\nonumber \\
& & \left. \hspace{74mm} 
- \left( c^2_{12} \frac{ r_{\Delta} }{ r_{A} } + \tilde{\varepsilon}_{\mu \mu} - 
\tilde{\varepsilon}_{\tau \tau} \right) 
\left( 1 - e^{- i \Delta L } \right) 
\right] 
\nonumber \\
&+& 
2 c_{23} s_{23} 
{\text Re } \left\{
\left( c_{12} s_{12} \frac{ r_{\Delta} }{ r_{A} } + \tilde{\varepsilon}_{e \mu}^{*} \right) 
\left( s_{13} e^{- i \delta } + \tilde{\varepsilon}_{e \tau} \right) 
\right\} 
\nonumber \\ 
&\times&
r_{A} \left[ 
( 1 - e^{- i \Delta L } ) - 
\frac{ 1 }{ 1 - r_{A} } \left( e^{- i r_{A} \Delta L } - e^{- i \Delta L } \right) 
\right] 
\label{Smumu} 
\end{eqnarray}
%
\begin{eqnarray}
S_{\tau \tau} &=& 
s^2_{23} \Bigl\{ 
1 - i ( c^2_{12} r_{\Delta} + \tilde{\varepsilon}_{\mu \mu} r_{A} ) \Delta L 
\Big\} + 
c^2_{23} 
\left( 
1 - i \tilde{\varepsilon}_{\tau \tau} 
r_{A} \Delta L 
\right) 
e^{- i \Delta L } 
\nonumber \\
&+& 
2 c_{23} s_{23} 
\text{ Re} ( \tilde{\varepsilon}_{\mu \tau} ) 
r_{A} \left( 1 - e^{- i \Delta L } \right) 
\nonumber \\
&-& c^2_{23} s^2_{13} 
\left[ 
( i r_{A} \Delta L ) e^{- i \Delta L } - 
\frac{ 1 + r_{A} }{ 1 - r_{A} } 
\left( e^{- i r_{A} \Delta L } - e^{- i \Delta L } \right) 
\right] 
\nonumber \\
&+& 2 c^2_{23} s_{13} 
\text{Re} ( \tilde{\varepsilon}_{e \tau} e^{ i \delta} ) 
\left[ 
( i r_{A} \Delta L ) e^{- i \Delta L } + 
\frac{ r_{A} }{ 1 - r_{A} } 
\left( e^{- i r_{A} \Delta L } - e^{- i \Delta L } \right) 
\right] 
\nonumber \\
&-& 
2 c_{23} s_{23} s_{13} 
\text{ Re} ( \tilde{\varepsilon}_{e \mu} e^{ i \delta} ) 
r_{A} \left( 1 - e^{- i \Delta L } \right) 
\nonumber \\
&-& 
2 c_{23} s_{23} s_{13} 
\Bigl\{ c_{12} s_{12} \cos \delta \frac{ r_{\Delta} }{ r_{A} } + 
\text{ Re}\left( \tilde{\varepsilon}_{e \mu} e^{ i \delta} \right) 
\Bigr\} 
\left( 1 - e^{- i r_{A} \Delta L } \right) 
%
%
\nonumber \\
&-& 
\left[ 
s^2_{23} (c^2_{12} \frac{ r_{\Delta} }{ r_{A} } + \tilde{\varepsilon}_{\mu \mu} )^2 + 
c^2_{23} \tilde{\varepsilon}_{\tau \tau}^2 e^{- i \Delta L } 
\right] 
\frac{ ( r_{A} \Delta L )^2 } { 2 } 
\nonumber \\
&+& s^2_{23} \left[ 
\vert c_{12} s_{12} \frac{ r_{\Delta} }{ r_{A} } + \tilde{\varepsilon}_{e \mu} \vert^2 
\Bigl\{ 
( i r_{A} \Delta L ) - \left( 1 - e^{- i r_{A} \Delta L } \right) 
\Bigr\} 
- \vert \tilde{\varepsilon}_{\mu \tau} \vert^2 r_{A}^2 \left( 1 - e^{- i \Delta L } \right) 
\right] 
\nonumber \\
&-& 
c^2_{23} \left[ 
\vert s_{13} e^{- i \delta } + \tilde{\varepsilon}_{e \tau} \vert^2
\left( \frac{ r_{A}^2 }{ 1- r_{A} } \right)
\Bigl\{ 
i \Delta L e^{- i \Delta L } - \frac{ 1 } { 1- r_{A} }
\left( e^{- i r_{A} \Delta L } - e^{- i \Delta L } \right) 
\Bigr\} 
\right.
\nonumber \\
& & \left. \hspace{54mm} 
+ \vert \tilde{\varepsilon}_{\mu \tau} \vert^2 r_{A}^2 
\Bigl\{ i \Delta L e^{- i \Delta L } - ( 1 - i \Delta L - e^{- i \Delta L } ) \Bigr\} 
\right] 
\nonumber \\
&-& 
2 c_{23} s_{23} 
{\text Re } ( \tilde{\varepsilon}_{\mu \tau} ) r_{A}^2 
\left[ i \Delta L 
\left( c^2_{12} \frac{ r_{\Delta} }{ r_{A} } + \tilde{\varepsilon}_{\mu \mu} - 
\tilde{\varepsilon}_{\tau \tau} e^{- i \Delta L } \right) 
\right.
\nonumber \\
& & \left. \hspace{62mm} 
- \left( c^2_{12} \frac{ r_{\Delta} }{ r_{A} } + \tilde{\varepsilon}_{\mu \mu} - 
\tilde{\varepsilon}_{\tau \tau} \right) 
\left( 1 - e^{- i \Delta L } \right) 
\right] 
\nonumber \\
&-& 
2 c_{23} s_{23} 
{\text Re } \left\{
\left( c_{12} s_{12} \frac{ r_{\Delta} }{ r_{A} } + \tilde{\varepsilon}_{e \mu}^{*} \right) 
\left( s_{13} e^{- i \delta } + \tilde{\varepsilon}_{e \tau} \right) 
\right\} 
\nonumber \\ 
&\times& 
r_{A} \left[ 
( 1 - e^{- i \Delta L } ) - 
\frac{ 1 }{ 1 - r_{A} } \left( e^{- i r_{A} \Delta L } - e^{- i \Delta L } \right) 
\right] 
\label{Stautau} 
\end{eqnarray}
%
\begin{eqnarray}
S_{\mu \tau} &=& 
- c_{23} s_{23} \Bigl\{ 
1 - i ( c^2_{12} r_{\Delta} + \tilde{\varepsilon}_{\mu \mu} r_{A} ) \Delta L 
\Big\} + 
c_{23} s_{23} 
\left( 
1 - i \tilde{\varepsilon}_{\tau \tau} 
r_{A} \Delta L 
\right) 
e^{- i \Delta L } 
\nonumber \\
&-& 
\Bigl\{ 
( c^2_{23} - s^2_{23} ) \text{Re} ( \tilde{\varepsilon}_{\mu \tau}) - i \text{ Im} ( \tilde{\varepsilon}_{\mu \tau}) 
\Bigr\} 
r_{A} \left( 1 - e^{- i \Delta L } \right) 
\nonumber \\
&-& c_{23} s_{23} s^2_{13} 
\left[ 
( i r_{A} \Delta L ) e^{- i \Delta L } - 
\frac{ 1 + r_{A} }{ 1 - r_{A} } 
\left( e^{- i r_{A} \Delta L } - e^{- i \Delta L } \right) 
\right] 
\nonumber \\
&+& 2 c_{23} s_{23} s_{13} 
\text{Re} ( \tilde{\varepsilon}_{e \tau} e^{ i \delta} ) 
\left[ 
( i r_{A} \Delta L ) e^{- i \Delta L } + 
\frac{ r_{A} }{ 1 - r_{A} } 
\left( e^{- i r_{A} \Delta L } - e^{- i \Delta L } \right) 
\right] 
\nonumber \\
&+& 
s_{13} \Bigl\{ 
( c^2_{23} - s^2_{23} ) \text{Re} ( \tilde{\varepsilon}_{e \mu} e^{ i \delta} ) + i \text{ Im} ( \tilde{\varepsilon}_{e \mu} e^{ i \delta} ) 
\Bigr\} 
r_{A} \left( 1 - e^{- i \Delta L } \right) 
\nonumber \\
&+& 
s_{13} \left[ 
(c^2_{23} - s^2_{23} ) \text{Re} 
\Bigl\{ e^{ i \delta} 
\left( c_{12} s_{12} \frac{ r_{\Delta} }{ r_{A} } + \tilde{\varepsilon}_{e \mu} \right) 
\Bigr\} 
- i \text{ Im} \Bigl\{ e^{ i \delta} 
\left( c_{12} s_{12} \frac{ r_{\Delta} }{ r_{A} } + \tilde{\varepsilon}_{e \mu} \right) 
\Bigr\} 
\right] 
\left( 1 - e^{- i r_{A} \Delta L } \right) 
%
%
\nonumber \\
&+& 
r_{A}^2 
\left( c^2_{23} \tilde{\varepsilon}_{\mu \tau} - s^2_{23} \tilde{\varepsilon}_{\mu \tau}^{*} \right) 
\left[ 
i \Delta L 
\left( c^2_{12} \frac{ r_{\Delta} }{ r_{A} } + \tilde{\varepsilon}_{\mu \mu} - \tilde{\varepsilon}_{\tau \tau} e^{- i \Delta L } \right) 
\right.
\nonumber \\
& & \left. \hspace{74mm} 
- \left( c^2_{12} \frac{ r_{\Delta} }{ r_{A} } + \tilde{\varepsilon}_{\mu \mu} - \tilde{\varepsilon}_{\tau \tau} \right) 
\left( 1 - e^{- i \Delta L } \right) 
\right] 
\nonumber \\
&+& 
\left[ 
c^2_{23} 
\left( c_{12} s_{12} \frac{ r_{\Delta} }{ r_{A} } + \tilde{\varepsilon}_{e \mu}^{*} \right) 
\left( s_{13} e^{- i \delta } + \tilde{\varepsilon}_{e \tau} \right) - 
s^2_{23} 
\left( c_{12} s_{12} \frac{ r_{\Delta} }{ r_{A} } + \tilde{\varepsilon}_{e \mu} \right) 
\left( s_{13} e^{ i \delta } + \tilde{\varepsilon}_{e \tau}^{*} \right) 
\right] 
\nonumber \\
&\times& 
r_{A} \left[ 
( 1 - e^{- i \Delta L } ) - 
\frac{ 1 }{ 1 - r_{A} } \left( e^{- i r_{A} \Delta L } - e^{- i \Delta L } \right) 
\right] 
\nonumber \\
&+& c_{23} s_{23} 
\left[ 
\left( c^2_{12} \frac{ r_{\Delta} }{ r_{A} } + \tilde{\varepsilon}_{\mu \mu} \right)^2 - \tilde{\varepsilon}_{\tau \tau}^2 e^{- i \Delta L } 
\right] 
\frac{ ( r_{A} \Delta L )^2 } { 2 } 
\nonumber \\
&-&
c_{23} s_{23} \vert s_{13} e^{- i \delta } + \tilde{\varepsilon}_{e \tau} \vert^2 
\left( \frac{ r_{A}^2 }{ 1 - r_{A} } \right) 
\left[ 
i \Delta L e^{- i \Delta L } - 
\frac{ 1 }{ 1 - r_{A} } 
\left( e^{- i r_{A} \Delta L } - e^{- i \Delta L } \right) 
\right] 
\nonumber \\
&-&
c_{23} s_{23} \vert c_{12} s_{12} \frac{ r_{\Delta} }{ r_{A} } + \tilde{\varepsilon}_{e \mu} \vert^2 
\Bigl\{ i r_{A} \Delta L - \left( 1 - e^{- i r_{A} \Delta L } \right) \Bigr\} 
\nonumber \\
&-&
c_{23} s_{23} \vert \tilde{\varepsilon}_{\mu \tau} \vert^2 r_{A}^2 
\Bigl\{ i \Delta L ( 1 + e^{- i \Delta L } ) - 2 \left( 1 - e^{- i r_{A} \Delta L } \right) \Bigr\}
\label{Smutau} 
\end{eqnarray}
%

The other $S$ matrix elements are given by either the T-conjugate relations 
\begin{eqnarray}
S_{\mu e} (\delta, \phi_{\alpha \beta} ) &=& 
S_{e \mu} (- \delta, - \phi_{\alpha \beta} ),
\nonumber \\
S_{\tau e} (\delta, \phi_{\alpha \beta} ) &=& 
S_{e \tau} (- \delta, - \phi_{\alpha \beta} ),
\nonumber \\
S_{\tau \mu} (\delta, \phi_{\alpha \beta} ) &=& 
S_{\mu \tau} (- \delta, - \phi_{\alpha \beta} ),
\label{Tconjugate}
\end{eqnarray}
or by the CP-conjugate relations for antineutrino channels 
\begin{eqnarray}
\bar{ S }_{e \mu} (\delta, \phi_{\alpha \beta}, a ) &=& 
S_{e \mu} (- \delta, - \phi_{\alpha \beta}, -a ),
\nonumber \\
\bar{ S }_{e \tau} (\delta, \phi_{\alpha \beta}, a ) &=& 
S_{e \tau} (- \delta, - \phi_{\alpha \beta}, -a ),
\nonumber \\
\bar{ S }_{\mu \tau} (\delta, \phi_{\alpha \beta}, a ) &=& 
S_{\mu \tau} (- \delta, - \phi_{\alpha \beta}, -a ). 
\label{CPconjugate}
\end{eqnarray}

\newpage

\section{NSI second-order Probability Formulas}
\label{2nd-order-formula}

In this Appendix we give the explicit expressions of the oscillation 
probabilities to second order in $\epsilon$ in all channels, except for those 
which can be readily obtained by the extended transformation 
(\ref{transformation}).

\subsection{Oscillation probability in the $\nu_e$-related sector}
\label{nu-e-sector}

We present here the explicit forms of $P(\nu _e \rightarrow \nu _e )$ and 
$P(\nu _e \rightarrow \nu _\mu )$ for completeness and possible convenience 
of the readers considering importance of the appearance channels. 
%
\begin{eqnarray}
&& \hspace{-10mm}
P(\nu _e\rightarrow \nu _e) = \ 1-4\biggl |\ci \si \frac{\mn}{a}+\cn \varepsilon_{e\mu }-\sn \varepsilon_{e\tau }\biggr |^2\sin ^2\frac{aL}{4E} 
\nonumber \\
&-& 4\biggl |\st e^{-i\delta }\frac{\mt}{a}+\sn \varepsilon_{e\mu }+\cn \varepsilon_{e\tau } \biggr |^2\biggl (\frac{a}{\mt -a}\biggr )^2\sin ^2\frac{\mt -a}{4E}L, 
\label{PeeB}
\end{eqnarray}
%
\begin{eqnarray}
&& \hspace{-10mm}
P(\nu _e \rightarrow \nu _\mu ) = \ 4\cn ^2\biggl |\ci \si \frac{\mn}{a}+\cn \varepsilon_{e\mu }-\sn \varepsilon_{e\tau }\biggr |^2\sin ^2\frac{aL}{4E} 
\nonumber \\
&+&4\sn ^2\biggl |\st e^{-i\delta }\frac{\mt}{a}+\sn \varepsilon_{e\mu }+\cn \varepsilon_{e\tau } \biggr |^2\biggl (\frac{a}{\mt -a}\biggr )^2\sin ^2\frac{\mt -a}{4E}L 
\nonumber \\
&+&8\cn \sn \text{ Re } \biggl[ (\ci \si \frac{\mn}{a}+\cn \varepsilon_{e\mu }-\sn \varepsilon_{e\tau })(\st e^{i\delta }\frac{\mt}{a}+\sn \varepsilon_{e\mu }^*+\cn \varepsilon_{e\tau }^*)\biggr ] 
\nonumber \\
&&\hspace{56mm} \times \frac{a}{\mt -a}\sin \frac{aL}{4E}\cos \frac{\mt L}{4E}\sin \frac{\mt -a}{4E}L 
\nonumber \\
&+&8\cn \sn \text{ Im } \biggl[ (\ci \si \frac{\mn}{a}+\cn \varepsilon_{e\mu }-\sn \varepsilon_{e\tau })(\st e^{i\delta }\frac{\mt}{a}+\sn \varepsilon_{e\mu }^*+\cn \varepsilon_{e\tau }^*)\biggr ] 
\nonumber \\
&&\hspace{56mm}\times \frac{a}{\mt -a}\sin \frac{aL}{4E}\sin \frac{\mt L}{4E}\sin \frac{\mt -a}{4E}L. 
\label{PemuB}
\end{eqnarray}
%
$P(\nu _e \rightarrow \nu _\tau )$ can be obtained from 
$P(\nu _e \rightarrow \nu _\mu )$ by the transformation (\ref{transformation}). 
Or, the simpler way of remembering the operation is to do transformation 
$c_{23} \rightarrow - s_{23}$ and $s_{23} \rightarrow c_{23}$ in 
$P(\nu _e \rightarrow \nu _\mu )$, but undoing any transformation in the 
generalized atmospheric and the solar variables defined in 
(\ref{def-Theta-Xi}), the pieces bracketed in the real and imaginary parts in 
(\ref{PemuB}). (See also (\ref{def-recapit}).

\subsection{Oscillation probability in the $\nu_{\mu} - \nu_{\tau}$ sector}

For compact expressions of the oscillation probabilities in the 
$\nu_{\mu} - \nu_{\tau}$ sector, we define the simplified notations 
which involve $\varepsilon$'s in the 
$\nu_\mu - \nu_\tau$ sector as well as $\varepsilon_{ee}$.\footnote{
For readers who want to see the fully explicit expressions of all the oscillation probabilities, we refer the first arXiv version of this paper \cite{KMU-arXiv-v1}. }
%
Together with the ones already defined in Sec.~\ref{nu_e-sector}, 
they are as follows: 
\begin{eqnarray} 
\Theta_{\pm} &\equiv& 
\st \frac{\mt}{a} + (s_{23} \varepsilon_{e \mu} + c_{23}  \varepsilon_{e \tau} ) e^{i \delta} 
\equiv \vert \Theta_{\pm} \vert e^{i \theta_{\pm} }, 
\nonumber \\ 
\Xi &\equiv& 
\left( \ci \si \frac{\mn}{a}+ c_{23} \varepsilon_{e \mu} - s_{23} \varepsilon_{e \tau} \right) e^{i \delta} 
\equiv \vert \Xi \vert e^{i \xi}, 
\nonumber \\
\mathcal{E} &\equiv& 
\cn \sn (\varepsilon _{\mu \mu}-\varepsilon _{\tau \tau})+\cn ^2\varepsilon _{\mu \tau}-\sn ^2\varepsilon _{\mu \tau}^* 
\equiv \vert \mathcal{E} \vert e^{i \phi},
\nonumber \\
\mathcal{S}_{1} &\equiv& 
 (\cn ^2-\sn ^2)(\varepsilon _{\tau \tau}-\varepsilon _{\mu \mu})+2\cn \sn (\varepsilon _{\mu \tau}+\varepsilon _{\mu \tau}^*)-\ci ^2\frac{\mn}{a}, 
\nonumber \\
\mathcal{S}_{2} &\equiv& 
(\varepsilon _{\mu \mu}-\varepsilon _{ee})+\cn ^2(\varepsilon _{\tau \tau}-\varepsilon _{\mu \mu})+\cn \sn (\varepsilon _{\mu \tau}+\varepsilon _{\mu \tau}^*)-\si ^2\frac{\mn}{a}, 
\nonumber \\
\mathcal{S}_{3} &\equiv& 
(\varepsilon _{\mu \mu}-\varepsilon _{ee})+\sn ^2(\varepsilon _{\tau \tau}-\varepsilon _{\mu \mu})-\cn \sn (\varepsilon _{\mu \tau}+\varepsilon _{\mu \tau}^*)+(\ci ^2-\si ^2)\frac{\mn}{a}. 
\label{def-recapit}
\end{eqnarray}
Notice that $\mathcal{S}_{1}$, $\mathcal{S}_{2}$,  and $\mathcal{S}_{3}$ 
are not independent, 
$\mathcal{S}_{1} = \mathcal{S}_{2} - \mathcal{S}_{3}$. 
We also note that $\Theta_{\pm}$, $\Xi$, and $\mathcal{E}$ are 
complex numbers while the others are real.

To present the oscillation probabilities in the $\nu_{\mu} - \nu_{\tau}$ sector,
we start by recapitulating the decomposition formula 
(\ref{Pmutau-decomposition}) in Sec.~\ref{mu-tau-sector}: 
\begin{eqnarray} 
P(\nu _\alpha \rightarrow \nu _\beta; 
\varepsilon_{e \mu }, \varepsilon_{e \tau }, \varepsilon_{\mu \mu }, \varepsilon_{\mu \tau }, \varepsilon_{\tau \tau }) &=& 
P(\nu _\alpha \rightarrow \nu _\beta; \text{2 flavor in vacuum}) 
\nonumber \\
&+& 
P(\nu _\alpha \rightarrow \nu _\beta; 
\varepsilon_{e \mu }, \varepsilon_{e \tau }) 
\nonumber \\
&+& 
P(\nu _\alpha \rightarrow \nu _\beta; 
\varepsilon_{\mu \mu }, \varepsilon_{\mu \tau }, \varepsilon_{\tau \tau }) 
\label{Pmutau-decomposition2}
\end{eqnarray}
where $\alpha$ and $\beta$ denote one of $\mu$ and $\tau$. 
The first term in (\ref{Pmutau-decomposition2}) has a form that it appears 
in the two flavor oscillation in vacuum: 
%
\begin{eqnarray}
P(\nu _\mu \rightarrow \nu _\mu; \text{2 flavor in vacuum}) &=&
P(\nu _\tau \rightarrow \nu _\tau; \text{2 flavor in vacuum}) 
\nonumber \\ 
&=& 1- 4\cn ^2\sn ^2\sin ^2\frac{\mt L}{4E}, 
\nonumber \\ 
P(\nu _\mu \rightarrow \nu _\tau; \text{2 flavor in vacuum}) &=& 
4\cn ^2\sn ^2\sin ^2\frac{\mt L}{4E}. 
\label{Pmutau-vac}
\end{eqnarray}

We have shown in Sec.~\ref{mu-tau-sector} 
that the third term in (\ref{Pmutau-decomposition2}) in the 
$\nu_{\mu} \rightarrow \nu_{\mu}$, 
$\nu_{\tau} \rightarrow \nu_{\tau}$, and 
$\nu_{\mu} \rightarrow \nu_{\tau}$ channels are given by the single equation\footnote{
To second order in $\epsilon$ the sensitivity to $\varepsilon_{\mu \mu}$ 
and $\varepsilon_{\tau \tau}$ is through the form 
$\varepsilon_{\mu \mu} - \varepsilon_{\tau \tau}$, and 
hence no sensitivity to the individual $\varepsilon$'s. 
Generally, the diagonal $\varepsilon$'s appear in a form of difference 
in the oscillation probabilities as one can observe in the third-order 
formula given in Appendix~\ref{third-order-formula}. 
It must be the case because the over-all phase is an unobservable. 
}
%
\begin{eqnarray}
&& \hspace{-5mm}
P(\nu _\mu \rightarrow \nu _\mu; \varepsilon_{\mu \mu }, \varepsilon_{\mu \tau }, \varepsilon_{\tau \tau }) 
\nonumber \\
&=& 
P(\nu_{\tau} \rightarrow \nu_{\tau}; \varepsilon_{\mu \mu }, \varepsilon_{\mu \tau }, \varepsilon_{\tau \tau }) = 
- P(\nu_{\mu} \rightarrow \nu_{\tau}; \varepsilon_{\mu \mu }, \varepsilon_{\mu \tau }, \varepsilon_{\tau \tau }) 
\nonumber \\
&=& 2\cn ^2\sn ^2 
\left(  \st ^2\frac{\mt}{a} - \mathcal{S}_{1} \right) 
\left( \frac{aL}{2E} \right) \sin \frac{\mt L}{2E} 
- \cn ^2\sn ^2 
\mathcal{S}_{1}^2
\biggl (\frac{aL}{2E}\biggr )^2\cos \frac{\mt L}{2E} 
\nonumber \\
&+& 8\cn \sn (\cn ^2-\sn ^2) 
\left [ 
\ci \si \st \cos \delta \left( \frac{\mn}{a} \right)  - \vert \mathcal{E} \vert \cos \phi  
\right ] 
\left( \frac{a}{\mt} \right) \sin ^2\frac{\mt L}{4E} 
\nonumber \\
&-& 4\cn \sn (\cn ^2-\sn ^2)  \mathcal{S}_{1} 
\vert \mathcal{E} \vert \cos \phi 
 \biggl (\frac{a}{\mt}\biggr )
\left[ 
\left( \frac{aL}{2E} \right) \sin \frac{\mt L}{2E}-2 \biggl (\frac{a}{\mt}\biggr ) \sin ^2\frac{\mt L}{4E} 
\right] 
\nonumber \\
&-& 4\cn ^2\sn ^2 
\vert \mathcal{E} \vert^2
\left( \frac{a}{\mt}\frac{aL}{2E} \right) \sin \frac{\mt L}{2E} 
\nonumber \\
&-& 4 \vert \mathcal{E} \vert^2 
\biggl [ 
(\cn ^2-\sn ^2)^2 - 4 \cn ^2\sn ^2 \cos^2 \phi 
\biggr ] 
\biggl (\frac{a}{\mt}\biggr )^2\sin ^2\frac{\mt L}{4E}. 
\label{Pmumu-mu-tau}
\end{eqnarray}
%
The second term in (\ref{Pmutau-decomposition2}) is given in 
$\nu_{\mu} \rightarrow \nu_{\mu}$ 
channel as 
%
\begin{eqnarray}
&& \hspace{-5mm}
P(\nu _\mu \rightarrow \nu _\mu; \varepsilon_{e \mu }, \varepsilon_{e \tau }) 
= - 4\cn ^2 \vert \Xi \vert^2 
\sin ^2\frac{aL}{4E} 
- 2\cn ^2\sn ^2 \vert \Xi \vert^2 
\left( \frac{aL}{2E} \right) \sin \frac{\mt L}{2E} 
\nonumber \\
&+& 8\cn ^2\sn ^2  \vert \Xi \vert^2 
\sin \frac{aL}{4E}\sin \frac{\mt L}{4E}\cos \frac{\mt -a}{4E}L 
\nonumber \\
&-& 4\sn ^2 \vert \Theta_{\pm} \vert^2  
\biggl (\frac{a}{\mt -a}\biggr )^2\sin ^2\frac{\mt -a}{4E}L 
- 2\cn ^2\sn ^2  \vert \Theta_{\pm} \vert^2 
\left( \frac{a}{\mt -a}\frac{aL}{2E} \right) \sin \frac{\mt L}{2E} 
\nonumber \\
&+& 8\cn ^2\sn ^2   \vert \Theta_{\pm} \vert^2 
\biggl (\frac{a}{\mt -a}\biggr )^2\cos \frac{aL}{4E}\sin \frac{\mt L}{4E}\sin \frac{\mt -a}{4E}L 
\nonumber \\
&+& 8\cn \sn 
\vert \Xi \vert  \vert \Theta_{\pm} \vert \cos (\xi - \theta_{\pm}) 
\left( \frac{a}{\mt -a} \right) 
\nonumber \\
&&\hspace{-2mm} \times \biggl [ \cn ^2\sin ^2\frac{aL}{4E}+\sn ^2\sin ^2\frac{\mt -a}{4E}L-\sn ^2\sin ^2\frac{\mt L}{4E}-(\cn ^2-\sn ^2)\frac{a}{\mt}\sin ^2\frac{\mt L}{4E}\biggr ].
\nonumber \\
\label{Pmumu-emu-etau}
\end{eqnarray}
%
The subscript $\pm$ in this and the following equations 
denotes the normal and the inverted mass hierarchies, which corresponds 
to the positive and negative values of $\Delta m^2_{31}$.
Notice again that 
$P(\nu _\tau \rightarrow \nu _\tau; \varepsilon_{e \mu }, \varepsilon_{e \tau }) $
can be obtained from 
$P(\nu _\mu \rightarrow \nu _\mu; \varepsilon_{e \mu }, \varepsilon_{e \tau }) $
by the extended transformation (\ref{transformation}), or by the operation 
described at the end of the previous subsection.

Finally, the second term in the oscillation probability in the 
$\nu_{\mu} \rightarrow \nu_{\tau}$ channel is given by 
%
\begin{eqnarray}
&& \hspace{-5mm}
P(\nu _\mu \rightarrow \nu _\tau; \varepsilon_{e \mu }, \varepsilon_{e \tau }) 
\nonumber \\
&=& 4 \cn ^2\sn ^2 \vert \Xi \vert^2 
\left( \frac{aL}{4E} \right) \sin \frac{\mt L}{2E} 
- 8\cn ^2\sn ^2 \vert \Xi \vert^2 
\sin \frac{aL}{4E}\sin \frac{\mt L}{4E}\cos \frac{\mt -a}{4E}L 
\nonumber \\
&+& 4\cn ^2\sn ^2 \vert \Theta_{\pm} \vert^2 
\left( \frac{a}{\mt -a} \right) \left( \frac{aL}{4E} \right) \sin \frac{\mt L}{2E} 
\nonumber \\
&-& 8\cn ^2\sn ^2 \vert \Theta_{\pm} \vert^2 
\biggl (\frac{a}{\mt -a}\biggr )^2\cos \frac{aL}{4E}\sin \frac{\mt L}{4E}\sin \frac{\mt -a}{4E}L 
\nonumber \\
&+& 8\cn \sn (\cn ^2-\sn ^2) \vert \Xi \vert  \vert \Theta_{\pm} \vert \cos (\xi - \theta_{\pm}) 
\left( \frac{a}{\mt -a} \right) 
\left ( \frac{a}{\mt} \right) \sin ^2\frac{\mt L}{4E} 
\nonumber \\
&+& 8\cn \sn \vert \Xi \vert  \vert \Theta_{\pm} \vert 
\left( \frac{a}{\mt -a} \right) 
\sin \frac{aL}{4E}\sin \frac{\mt L}{4E} 
\nonumber \\
&& \hspace{14mm}  \times 
\left[ 
\sn ^2 \cos \left( \xi - \theta_{\pm} - \frac{\mt -a}{4E}L \right)  
- \cn ^2 \cos \left( \xi - \theta_{\pm} + \frac{\mt -a}{4E}L \right)  
\right]. 
\label{Pmutau-emu-etau}
\end{eqnarray}

\section{NSI Third-Order Formula}
\label{third-order-formula}

We present here the third-order formula for the oscillation probability with NSI. 
Though utility of such lengthy formula may be subject to doubt we can offer 
at least three arguments to justify the presentation of the formula in this Appendix.  
Firstly, the third-order formula is needed to complete Table~\ref{order}.  
Secondly, by turning off all the NSI elements 
one can obtain the SI third-order formula for the oscillation probabilities 
with SI only, which is valid to order $\epsilon^3$.  
To our knowledge, such formula has never been derived in the literature. 
Utility of the SI third-order formula for theoretical analysis may be obvious 
if the sensitivity to the oscillation probability reaches to the level of 
$\sim10^{-5}$, which is smaller than terms of order $\epsilon^2$. 
In fact, it appears to be the case in some of the future facilities according 
to the analysis in \cite{ISS-report}. 
Thirdly, once the sensitivity to the oscillation probability comes down to 
$\sim10^{-5}$, a complete treatment of neutrino oscillation probability 
must include NSI elements up to the same order as $\theta_{13}$ and 
$\Delta m^2_{21} / \Delta m^2_{31} $, as far as our ansatz (\ref{def-epsilon}) 
in formulating the $\epsilon$ perturbation theory is correct. 
Thus, we believe that the NSI third-order formula has a good chance to be useful.

In presenting the third-order probability formula we restrict ourselves 
to the $\nu_{e}$ related channel, 
and only present $P(\nu_e \rightarrow \nu_\mu)$ here 
because from which 
$P(\nu_e \rightarrow \nu_\tau)$ can be obtained by the extended 
transformation (\ref{transformation}). Then, 
$P(\nu_e \rightarrow \nu_e)$ can be readily calculated by using the 
unitarity relation. 
The NSI third-order formula for $P(\nu_e \rightarrow \nu_\mu)$ reads 
%
\begin{eqnarray}
&& 
\hspace{-8mm} 
P(\nu _e \rightarrow \nu _\mu ) 
= \ 4\cn ^2
\vert \Xi \vert^2 
\sin ^2\frac{aL}{4E} 
+ 4\sn ^2
\vert \Theta_{\pm} \vert^2 
\biggl (\frac{a}{\mt -a}\biggr )^2
\sin ^2\frac{\mt -a}{4E}L 
\nonumber \\
&+&8\cn \sn 
\vert \Xi  \vert  \vert \Theta_{\pm} \vert 
\cos \left( \xi - \theta_{\pm} -  \frac{\mt L}{4E} \right)  
\biggl (\frac{a}{\mt -a}\biggr )\sin \frac{aL}{4E} \sin \frac{\mt -a}{4E}L 
\nonumber \\
&-&8\cn \sn \si ^2\st 
\vert \Xi  \vert 
\cos \left( \xi - \frac{\mt L}{4E} \right)  
\left( \frac{\mn}{a} \right) 
\biggl (\frac{a}{\mt -a}\biggr )\sin \frac{aL}{4E} \sin \frac{\mt -a}{4E}L 
 \nonumber \\
&-&8\sn ^2 \si ^2\st 
\vert \Theta_{\pm} \vert \cos \theta_{\pm}
\left( \frac{\mn}{a} \right)
\biggl (\frac{a}{\mt -a}\biggr )^2\sin ^2\frac{\mt -a}{4E}L 
 \nonumber \\
&+&4\sn ^2 
\vert \Theta_{\pm} \vert^2 
\mathcal{S}_{2} 
\biggl (\frac{a}{\mt -a}\biggr )^2\left [ \biggl (\frac{aL}{4E} \biggr ) \sin \frac{\mt -a}{2E}L -2\biggl (\frac{a}{\mt -a}\biggr )\sin ^2\frac{\mt -a}{4E}L \right ] 
\nonumber \\
&+&4\cn ^2 
\vert \Xi \vert^2 
\mathcal{S}_{3} 
\left [ 2\sin ^2\frac{aL}{4E}-\biggl ( \frac{aL}{4E}\biggr )\sin \frac{aL}{2E}\right ] 
\nonumber \\
&+&8\cn \sn \vert \Xi \vert^2 
\vert \mathcal{E} \vert 
\cos \left( \phi + \frac{\mt -a}{4E}L \right) 
\biggl ( \frac{a^2}{\mt (\mt-a)}\biggr ) 
  \sin \frac{aL}{4E} \sin \frac{\mt L}{4E} 
 \nonumber \\
&-&8\cn \sn \vert \Xi \vert^2 
\vert \mathcal{E}  \vert \cos \phi 
\biggl (\frac{a}{\mt -a}\biggr )  \sin^2 \frac{aL}{4E} 
 \nonumber \\
&+&8\cn \sn \vert \Theta_{\pm} \vert^2 
\vert \mathcal{E}  \vert \cos \phi 
\biggl (\frac{a}{\mt -a}\biggr )^2  
\sin^2 \frac{\mt -a}{4E}L 
\nonumber \\
&-&8\cn \sn \vert \Theta_{\pm} \vert^2
\vert \mathcal{E} \vert 
\cos \left( \phi + \frac{aL}{4E} \right)  
\biggl ( \frac{a^2}{\mt (\mt-a)}\biggr ) 
\sin \frac{\mt L}{4E}  \sin \frac{\mt -a}{4E}L 
 \nonumber \\
&-&4\cn \sn 
\vert \Xi \vert  \vert  \Theta_{\pm} \vert 
\biggl (\frac{a}{\mt -a}\biggr )\biggl (\frac{aL}{4E}\biggr ) 
\nonumber \\
&&\hspace{-6mm}
\times 
\left [ 
\mathcal{S}_{1} \sin \left( \xi - \theta_{\pm} - \frac{\mt L}{2E}   \right) 
+ \mathcal{S}_{3}  \sin \left( \xi - \theta_{\pm} - \frac{aL}{2E}  \right) 
- \mathcal{S}_{2} \sin \left( \xi - \theta_{\pm} - \frac{\mt -a}{2E}L  \right) 
\right ] 
\nonumber \\
&-&8\cn \sn 
\vert \Xi \vert  \vert  \Theta_{\pm} \vert 
\cos \left( \xi - \theta_{\pm} -  \frac{\mt L}{4E} \right)  
\nonumber \\
&& \hspace{36mm}  \times 
\left [ \biggl (\frac{a}{\mt -a}\biggr ) 
\mathcal{S}_{2} 
- \mathcal{S}_{3} \right ] 
\biggl (\frac{a}{\mt -a}\biggr )\sin \frac{aL}{4E} \sin \frac{\mt -a}{4E}L 
\nonumber \\
&-&8\sn ^2 
\vert \Xi \vert  \vert  \Theta_{\pm} \vert  \vert \mathcal{E} \vert 
\cos \left( \xi + \phi - \theta_{\pm} - \frac{aL}{4E}  \right) 
\biggl ( \frac{a^2}{\mt (\mt-a)}\biggr ) 
\sin \frac{\mt L}{4E} \sin \frac{\mt -a}{4E}L 
\nonumber \\
&+& 8\cn ^2 
\vert \Xi \vert  \vert  \Theta_{\pm} \vert  \vert \mathcal{E} \vert 
\cos \left( \xi + \phi - \theta_{\pm} - \frac{\mt -a}{4E}L  \right) 
\biggl ( \frac{a^2}{\mt (\mt-a)}\biggr ) 
\sin \frac{aL}{4E} 
\sin \frac{\mt L}{4E} 
\nonumber \\
&+&8 \vert \Xi \vert  \vert  \Theta_{\pm} \vert  \vert \mathcal{E} \vert 
\cos \left( \xi + \phi - \theta_{\pm}  \right) 
\biggl (\frac{a}{\mt -a}\biggr ) 
\left[
\sn ^2  \biggl (\frac{a}{\mt -a}\biggr ) \sin^2 \frac{\mt -a}{4E}L -  \cn ^2  \sin^2 \frac{aL}{4E} 
\right].
\nonumber \\
\label{Pemu-3rd}
\end{eqnarray}

\section{Intrinsic Degeneracy in Vacuum}
\label{intrinsic-vac}

We re-examine the problem of intrinsic degeneracy in vacuum. 
For simplicity, we focus on the channel 
$ \nu_{\mu} \rightarrow \nu_{\rm e}$. 
We use a simplified notation $s_{13} \equiv s$ below. 
The neutrino and anti-neutrino oscillation probabilities in vacuum 
are given by 
%
\begin{eqnarray}
P( \nu_{\mu} \rightarrow \nu_{\rm e} ) &=& 
X s^2 + 
\left( Y_{c} \cos \delta - Y_{s} \sin \delta \right) s + P_{\odot}
\nonumber \\
P( \bar{\nu}_{\mu} \rightarrow \bar{\nu}_{\rm e} ) &=&
X s^2 + 
\left( Y_{c} \cos \delta + Y_{s} \sin \delta \right) s + P_{\odot}
\label{Pmue_vac}
\end{eqnarray}
%
where $X$, $Y$'s, etc. are defined with simplified symbol 
$\Delta_{ji} \equiv \frac{ \Delta m^2_{ji} L} {4 E} $ as 
\begin{eqnarray} 
X &\equiv& 4 s^2_{23} \sin^2 \Delta_{31}, 
\nonumber \\
Y_{c} &\equiv& 
\sin 2\theta_{12} \sin 2\theta_{23}
\Delta_{21} \sin 2 \Delta_{31}, 
\nonumber \\
Y_{s} &\equiv& 
2 \sin 2\theta_{12} \sin 2\theta_{23}
\Delta_{21} \sin^2 \Delta_{31}, 
\nonumber \\
P_{\odot} &\equiv&
\sin^2{2\theta_{12}} c^2_{23} 
\Delta_{21}^2. 
\label{definition}
\end{eqnarray}

Let us denote two set of intrinsic degenerate solutions as 
$(s_{1}, \delta_{1})$ and $(s_{2}, \delta_{2})$. 
They satisfy 
\begin{eqnarray} 
P - P_{\odot} &=& 
X s_{1}^2 + \left( Y_{c} \cos \delta_{1} - Y_{s} \sin \delta_{1} \right) s_{1} 
\nonumber \\
P - P_{\odot} &=& 
X s_{2}^2 + \left( Y_{c} \cos \delta_{2} - Y_{s} \sin \delta_{2} \right) s_{2} 
\label{def-nu}
\end{eqnarray}
and 
\begin{eqnarray} 
\bar{P} - P_{\odot} &=& 
X s_{1}^2 + \left( Y_{c} \cos \delta_{1} +Y_{s} \sin \delta_{1} \right) s_{1} 
\nonumber \\
\bar{P} - P_{\odot} &=& 
X s_{2}^2 + \left( Y_{c} \cos \delta_{2} + Y_{s} \sin \delta_{2} \right) s_{2} 
\label{def-nubar}
\end{eqnarray}
By subtracting two equations in (\ref{def-nu}) and (\ref{def-nubar}) 
respectively, we obtain 
\begin{eqnarray} 
X ( s_{1}^2 - s_{2}^2) + Y_{c} ( s_{1} \cos \delta_{1} - s_{2} \cos \delta_{2} ) 
- Y_{s} ( s_{1} \sin \delta_{1} - s_{2} \sin \delta_{2} ) &=& 0,
\nonumber \\
X ( s_{1}^2 - s_{2}^2) + Y_{c} ( s_{1} \cos \delta_{1} - s_{2} \cos \delta_{2} ) 
+ Y_{s} ( s_{1} \sin \delta_{1} - s_{2} \sin \delta_{2} ) &=& 0. 
\label{reduct1}
\end{eqnarray}
They further simplifies to 
\begin{eqnarray} 
&& s_{1} \sin \delta_{1} - s_{2} \sin \delta_{2} = 0,
\label{reduct2a}
\\
%
&&X ( s_{1}^2 - s_{2}^2) + Y_{c} ( s_{1} \cos \delta_{1} - s_{2} \cos \delta_{2} ) 
= 0. 
\label{reduct2b}
\end{eqnarray}
Equation (\ref{reduct2a}) can be solved as 
\begin{eqnarray} 
s_{2} \cos \delta_{2} = \pm 
\sqrt{ s_{2}^2 - s_{1}^2 \sin^2 \delta_{1} }
\label{reduct3}
\end{eqnarray}
which can be inserted to (\ref{reduct2b}) to yield the 
(formally quartic but actually) quadratic equation for $s_{2}$. 
Now, the issue here is to choose the correct sign in (\ref{reduct3}). 
One can show that by using (\ref{reduct2b}) if $Y_{c} > 0$ ($Y_{c} < 0$), 
minus (plus) sign has to be chosen.

These equations can be easily solved for $(s_{2}, \delta_{2})$ 
for given values of $(s_{1}, \delta_{1})$ as inputs:
\begin{eqnarray} 
s_{2} &=& \sqrt { s_{1}^2 + 2 \left( \frac{Y_{c}}{X} \right) s_{1} \cos \delta_{1} + 
\left( \frac{Y_{c}}{X} \right)^2 } 
\nonumber \\
\sin \delta_{2} &=& \frac{ s_{1} }{ s_{2} } \sin \delta_{1} 
\nonumber \\
\cos \delta_{2} &=& \mp \frac{ 1 }{ s_{2} } 
\left(
s_{1} \cos \delta_{1} + \frac{Y_{c}}{X} 
\right)
\label{solution}
\end{eqnarray}
where the sign $\mp$ for $\cos \delta_{2}$ are for $Y_{c} = \pm |Y_{c}|$, 
and $s_{2}$ in the solution of $\delta$ is meant to be the 
$s_{2}$ solution given in the first line in (\ref{solution}). 
By using 
\begin{eqnarray} 
\frac{Y_{c}}{X} = \sin 2\theta_{12} \cot \theta_{23} \Delta_{21} \cot \Delta_{31} 
\label{Y/X}
\end{eqnarray}
$s_{2}$ can be written as 
\begin{eqnarray} 
s_{2} &=& \sqrt { s_{1}^2 + 
2 \sin 2\theta_{12} \cot \theta_{23} \Delta_{21} \cot \Delta_{31} 
s_{1} \cos \delta_{1} + 
\left( \sin 2\theta_{12} \cot \theta_{23} \Delta_{21} \cot \Delta_{31} \right)^2 } 
\label{solution-s}
\end{eqnarray}
Similarly, $\cos \delta$ is given as 
\begin{eqnarray} 
\cos \delta_{2} &=& \mp \frac{ 1 }{ s_{2} } 
\left(
s_{1} \cos \delta_{1} + 
\sin 2\theta_{12} \cot \theta_{23} \Delta_{21} \cot \Delta_{31} 
\right)
\label{solution-delta}
\end{eqnarray}

By further expanding (\ref{solution}) by $\frac{Y_{c}}{X} $, assuming it small, 
the Burguet-Castell {\it et al.} solution \cite{intrinsicD,MNP2} is reproduced; 
\begin{eqnarray} 
s_{2} \simeq s_{1} + \frac{Y_{c}}{X} \cos \delta_{1} 
\label{app-solution}
\end{eqnarray}

\begin{acknowledgments}
Two of the authors (H.M. and S.U.) thank Hiroshi Numokawa and 
Renata Zukanovich Funchal for useful discussions and for sharing 
various knowledges of neutrino oscillation with NSI through fruitful 
collaborations. 
They are grateful to Belen Gavela and Andrea Donini for illuminating 
discussions and for hospitality at Departamento de F\'\i sica Te\'orica 
and Instituto de F\'\i sica Te\'orica, Universidad Aut\'onoma de Madrid, 
where this work was completed. Their visits were supported by 
the JSPS-CSIC (Japan-Spain) Bilateral Joint Projects. 
H.M. thanks Stephen Parke for critical discussions on systems with NSI, 
and Theoretical Physics Department of Fermilab for hospitality in the 
summer 2008. 
This work was supported in part by KAKENHI, Grant-in-Aid for
Scientific Research No. 19340062, and Grant-in-Aid for JSPS Fellows 
No. 209677, Japan Society for the Promotion of Science.

\end{acknowledgments}

\end{document}